\documentclass[a4paper,10.5pt]{article} 
\usepackage{graphicx}  
\usepackage{dcolumn}   
\usepackage{bm}        
\usepackage{amsmath,amssymb,amsxtra,bm,epsfig}   
\usepackage{float}
\usepackage[dvipsnames]{xcolor}
\usepackage{xspace}
\usepackage{multirow}
\usepackage{pifont}
\usepackage[normalem]{ulem}
\usepackage[flushleft]{threeparttable}
\usepackage{array,booktabs,makecell}
\usepackage[]{changes}
\usepackage{caption}
\usepackage{jheppub}
\usepackage{adjustbox,lipsum}

\usepackage{hyperref}
\usepackage{comment}
\usepackage[font=scriptsize]{subcaption}

\begin{document}
	
	
	\title{\boldmath Common origin of $\theta_{13}$ and dark matter within the\\ flavor symmetric scoto-seesaw framework
		}

\author[a,b]{Joy Ganguly,}
\author[c]{Janusz Gluza,}
\author[c,1]{Biswajit Karmakar\note{Corresponding author.}}


\affiliation[a]{Department of Physics, Indian Institute of Technology Hyderabad, Telenagana, India}
\affiliation[b]{Department of BSH, University of Engineering and Management, Kolkata, India}
\affiliation[c]{Institute of Physics, University of Silesia,  Katowice, Poland}

\emailAdd{joyganguly.hep2022@gmail.com}
\emailAdd{janusz.gluza@us.edu.pl}
\emailAdd{biswajit.karmakar@us.edu.pl}

\abstract{To understand the observed pattern of neutrino masses and  mixing  as well as to account for the dark matter we propose a hybrid scoto-seesaw model based on the $A_4$ discrete flavor symmetry.  In this setup, including at least two heavy right-handed neutrinos is essential to employ the discrete flavor symmetry that mimics once popular tribimaximal  neutrino mixing at the leading order via type-I seesaw.  The scotogenic contribution then acts as a critical  deviation to  reproduce  the observed value of the reactor mixing angle $\theta_{13}$ (within the trimaximal mixing scheme) and to accommodate potential  dark matter candidates, pointing towards a common origin of $\theta_{13}$  and dark matter. The model  predicts the atmospheric angle to be in the upper  octant,  excludes some regions on the Dirac CP phase, and restricts  the Majorana phases too. Further, normal and inverted mass hierarchies can be distinguished for specific values of the relative  phases associated with the complex light neutrino mass matrix.  Owing to the considered flavor symmetry,  contributions coming from the scotogenic mechanism  towards the lepton flavor violating decays such as  $\mu \rightarrow e \gamma$, $\tau \rightarrow e \gamma$ vanish, and  a lower limit on the second right-handed neutrino mass can be obtained.  Prediction for the effective mass parameter appearing in the  neutrinoless double beta decay falls  within the sensitivity of future experiments such as  LEGEND-1k and nEXO.}

\maketitle

\section{Introduction}
The discovery of neutrino oscillation \cite{Pontecorvo:1967fh,SNO:2001kpb,Super-Kamiokande:1998kpq,Gonzalez-Garcia:2007dlo,deSalas:2020pgw} suggests that at least two neutrinos are massive but having very small masses with respect  to the charged leptons and quarks. This situation opens up a  window for interpretations which go beyond the Standard Model (SM) of particle physics. On the other hand, it is also well established  that the neutrino flavor mixing is significantly large compared  to the quark mixing which also demands an extension of the SM (either by particle content or symmetry extension). Another extraordinary problem in particle physics as of today is the nature of dark matter (DM), whose relic abundance is precisely measured by the WMAP~\cite{WMAP:2012nax} and PLANCK~\cite{Aghanim:2018eyx} satellite experiments  and  the existence of such DM is strongly supported by the gravitational lensing, galactic rotation curve and large scale structure of the Universe~\cite{Bertone:2004pz} as well. However, the SM of particle physics fails to provide an appropriate  candidate for DM. Now, from  a theoretical perspective, the origin of tiny  neutrino masses  can be well understood within the framework of various seesaw mechanisms. The simplest one is  the type-I seesaw mechanism \cite{Minkowski:1977sc, Gell-Mann:1979vob, Mohapatra:1979ia, Schechter:1981cv, Schechter:1980gr} where usually three singlet right-handed neutrinos are added to the SM;  then three left-handed neutrinos can be massive. This  “complete” sequential  seesaw scenario is denoted by ({\scriptsize $\#$}$\nu_L$,{\scriptsize $\#$}$N_R$)=(3,3) where {\scriptsize $\#$} $\nu_L$, {\scriptsize $\#$}$N_R$ denote the number of generations of left-handed and right-handed neutrinos respectively.  However, as first pointed out in~~\cite{Schechter:1980gr},  the number of right-handed neutrinos added to the SM is not fixed as they do not carry any anomaly~\cite{King:2015sfk}. A simpler version of the type-I seesaw is the minimal seesaw   which further reduces the number of free parameters~\cite{Xing:2020ald,King:1999mb,King:2002nf}. In the (3,2) seesaw mechanism only two right-handed neutrinos are introduced to obtain viable neutrino masses.   However, the flavor structure of  the relevant lepton mass matrices  still remains undetermined. 

The flavor structures of the lepton mass matrices and hence the   observed non-trivial pattern of the lepton flavor mixing can be examined by incorporating  non-Abelian discrete flavor symmetries to the SM. For this purpose  discrete symmetry groups such as $A_4, S_4, A_5, \Delta(27)$ are often used~ \cite{Morisi:2012fg}.  For a general overview and  implementation of discrete symmetries in neutrino physics see~\cite{Ishimori:2010au, Altarelli:2010gt, Feruglio:2019ybq, King:2013eh, Petcov:2017ggy, Xing:2020ijf,Chauhan:2022gkz}. Such discrete flavor symmetries can explain various fixed mixing schemes such as bi-maximal (BM), golden ratio (GR) and hexagonal (HG)~\cite{Vissani:1997pa, Barger:1998ta,Datta:2003qg, Kajiyama:2007gx,Albright:2010ap} mixings, with the most popular one being the tri-bimaximal (TBM) mixing~\cite{Harrison:2002er, Harrison:2002kp}. In the TBM mixing scheme the solar and atmospheric mixing angles take  values   $\sin^2\theta_{12}=1/3$ and  $\sin^2\theta_{23}=1/2$ respectively,  whereas the reactor mixing angle is fixed at  $\sin^2\theta_{13}=0$. Therefore, a discrete flavor symmetric construction with two right-handed neutrinos is a very economical scenario to explain neutrino masses and TBM mixing simultaneously~\cite{Chang:2004wy,Park:2011zt,Zhao:2011pv}. Neutrino model building with two right-handed neutrinos in various limiting cases can be found in \cite{King:1999cm}. Among  various discrete groups employed for this purpose, $A_4$ is the most popular one~\cite{Ma:2001dn, Ma:2004zv, Ma:2002yp, Altarelli:2005yp, Altarelli:2005yx,Babu:2002dz}. This symmetry was initially proposed as an underlying family symmetry for the quark sector.  It is a discrete group of even permutations of four objects with three inequivalent one-dimensional representations (1, $1'$, and $1''$ ) and a three-dimensional representation (3). Interestingly, the three generations (or flavors) of right-handed charged lepton singlets can fit into three inequivalent one-dimensional representations. Conversely, $A_4$ is the smallest group with a three-dimensional irreducible  representation. Then three SM lepton doublets can transform together as a triplet under $A_4$ \cite{Altarelli:2005yx, Altarelli:2005yp, Ma:2002yp}. So far, so good; however,  in the last decade, the  reactor neutrino mixing
angle $\theta_{13}$ is decisively measured~\cite{Abe:2011fz, An:2012eh, Ahn:2012nd, T2K:2013ppw, MINOS:2013utc} to be adequately  large ($\sim 9^{\circ}$) and hence  the era of fixed patterns (such as BM, TBM, GR, HG)  of the lepton mixing matrix is over.  To generate non-zero $\theta_{13}$ various approaches are considered either by additional contributions to the neutrino sector or considering  additional corrections from   the charged lepton sector or including corrections to vacuum alignments of the flavons  etc. \cite{Adhikary:2008au,Brahmachari:2008fn, King:2009qt, Branco:2009by,AristizabalSierra:2009ex,Morisi:2009sc, Ahn:2011yj, Shimizu:2011xg, Ganguly:2020riw, Ganguly:2021nqx,Ahn:2011if, Antusch:2011ic, Borah:2019ldn, Ding:2011gt, King:2011ab,Mukherjee:2015axj, deMedeirosVarzielas:2010ppv, Ahn:2012cga, Ahn:2012tv,BenTov:2012tg, Branco:2012vs, Borah:2017qdu, CarcamoHernandez:2013yiy, Bhattacharya:2016rqj,Barry:2010zk, Chen:2012st, Karmakar:2016cvb, Zhao:2014yaa, Antusch:2013wn, Borah:2013upa, Ding:2013bpa,Vien:2021eog, Ahn:2013mva, AristizabalSierra:2014zeq, Vien:2014pta, Vien:2020dlk, CarcamoHernandez:2015rmj, Holthausen:2012wz, Pramanick:2015qga, Kalita:2015jaa, King:2013hj, Nomura:2016nfi, Borah:2013jia, Memenga:2013vc, Karmakar:2014dva,Puyam:2022mej}. As consequence various descendants of the fixed mixing schemes have emerged. For example, even if the TBM mixing is obsolete now, two of its successors are still compatible with data. These mixing schemes are known as trimaximal mixing (TM) mixings~\cite{Albright:2010ap,Xing:2006ms,King:2011ab,Grimus:2008tt} which  preserves the first (second) column of the TBM mixing matrix and are called ${\rm TM}_1$ (${\rm TM}_2$) mixing, respectively.

In this work, we consider a  scotogenic contribution~\cite{Ma:2006km} to the underlying TBM mixing scheme establishing  a common origin of the nonzero $\theta_{13}$ and cosmological DM. Many radiative models account for the tiny  neutrino masses~\cite{Zee:1980ai, Cheng:1980qt, Restrepo:2013aga, Babu:1988ki, Cai:2017jrq} and   perhaps the simplest one is the scotogenic model which also naturally accommodates potential candidate for dark matter~\cite{Ma:2006km}.  In the present article, we explore  the idea of combining seesaw and scotogenic (termed as scoto-seesaw) model \cite{Rojas:2018wym} to explain neutrino mass and mixing in a consistent way with well-motivated $A_4$ discrete flavor symmetry. {\it  In our proposal,  the nature of cosmological dark matter and  reactor mixing angle $\theta_{13}$ share a unified origin}.  Our full model comprises of (3,2) seesaw and then combines it with the scotogenic mechanism, which predicts three neutrinos to be massive. The Yukawa structure of the scotogenic contribution is formulated by adding flavon fields which transform non-trivially under the flavor symmetry. With this, we show that our model can successfully explain the lepton mixing with non-zero reactor angle $\theta_{13}$ and includes leptonic CP violation. Due to the flavor symmetric construction, the model is extremely  predictive in nature and offers many interesting  results involving neutrino mass hierarchy, octant of the atmospheric mixing angle $\theta_{23}$ and restricts the  Dirac CP phase.  In addition, we constrain the absolute neutrino masses and Majorana CP phases. On the other hand, the model can  be falsiﬁed by null results of   future  neutrinoless double beta decay experiments. Interestingly, due to the specific flavor structure of the Yukawa couplings   as a consequence of the flavor symmetry,  the scotogenic part does not contribute in the lepton flavor violating decays such as $\mu \rightarrow e \gamma$ and only right-handed neutrinos contribute in such decays.  We begin with a  minimal type-I seesaw assisted by the $A_4$ discrete flavor symmetry, which helps reproduce the TBM mixing. The model also contains additional $Z_N$ symmetries to forbid unwanted contributions in the lepton sector, and an inherent  $\mathcal{Z}_2$ symmetry also ensures the stability of the dark matter. Thanks to the considered symmetry,  the charged lepton mass matrix as well as the heavy right-handed Majorana neutrino mass matrix, are diagonal to start with. Therefore, the structure of the $3\times2$ Dirac Yukawa matrix turns out to be solely responsible for generating the TBM mixing with two right-handed neutrinos. Now,  the inclusion of the scotogenic contribution to the neutrino mass helps in reproducing the ${\rm TM}_2$ mixing and generates the observed value of the reactor mixing angle $\theta_{13}$. It also naturally incorporates  dark matter candidates  (three potential dark matter candidates, such as the dark fermion and real and imaginary components of the scalar field involved in the scotogenic contribution) into the picture. 

The rest of the paper is organized as follows. In Section \ref{sec:scoto-seesaw} we first describe the minimal scoto-seesaw model. Then in Section \ref{sec:our model} we present the  $A_4$ flavor symmetric  scoto-seesaw model with two right-handed neutrinos  and describe the construction of the model based on the symmetries of the framework. In Section \ref{sec:nuetrino mass and mixing}  we present the correlation among the  parameters involved in our analysis.  We carry out the complete  analysis for various limiting and general  cases and present their predictions in Section \ref{sec:numerical analysis}. Then in Section \ref{sec:pheno} we mention various phenomenological implications of undertaken analysis and finally conclude in Section \ref{sec:conc}. We also included in the Appendix a short note on $A_4$ multiplication rules used in our analysis.

\section{Minimal scoto-seesaw model}\label{sec:scoto-seesaw}
In Ref.~\cite{Rojas:2018wym}, the (3,1) scenario of  seesaw mechanism~\cite{Schechter:1980gr, Schechter:1981bd, Schechter:1981cv} and the scotogenic model \cite{Ma:2006km} are combined to propose a minimal  scoto-seesaw model. This model consists of only one right-handed neutrino $N_R$, one singlet fermion $f$, and one extra scalar doublet $\eta$. In addition to these particles, one $Z_2$ symmetry is introduced, which is responsible for the stability of the dark matter. All the standard model fields, $N_R$, are even under the $Z_2$ symmetry, while the dark sector consists of one fermion $f$ and scalar field $\eta$ which are odd under $Z_2$. This proposal generates the atmospheric neutrino mass scale at the tree level with the conventional $(3,1)$ seesaw term with $N_R$, and the solar neutrino mass scale is generated at a one-loop level, as a result, the hierarchy between the atmospheric and solar scale is maintained\footnote{ Earlier such hierarchy of the atmospheric and solar neutrino mass scales (and associated mixing) was explained with a type-I seesaw mechanism where the right-handed neutrinos contribute hierarchically and implemented within the frameworks of sequential dominance~\cite{King:1998jw, King:1999cm, King:1999mb, King:2002nf,Antusch:2010tf,Antusch:2004gf} and constrained sequential dominance ~ \cite{King:2005bj, Antusch:2011ic}}. With this field content, the lepton Yukawa and mass terms can be written as
\begin{eqnarray}\label{eq:scoto-seesaw Lag}
		{\mathcal{L}}=-Y_{N}^k \bar{L}^ki\sigma_2 H^* N_R+\frac{1}{2}M_R \bar{N}_R^cN_R+ Y_{f}^k \bar{L}^ki\sigma_2 \eta^* f +\frac{1}{2}M_f \bar{f}^c f + h.c..
\end{eqnarray}
where $L^k$ are the lepton doublets. The scalars $H=(H^+,H^0)^T$ and $\eta=(\eta^+,\eta^0)$ are the $SU(2)$ doublets. $Y_N$ and $Y_f$ are complex $3\times 1$ Yukawa coupling matrices, and $M_{R,f}$ are the mass matrices for $N_R$ and $f$. The total neutrino mass reads \cite{Rojas:2018wym}
\begin{eqnarray}\label{eq:scoto-seesaw mass matrix}
	M_{\nu}^{ij}=-\frac{v^2}{M_N}Y_N^i Y_N^j+{\mathcal{F}}(m_{\eta_R},m_{\eta_I},M_f)Y_{f}^i Y_f^j.
\end{eqnarray}
Here the first term is due to the tree-level seesaw mechanism while the second term originates from the scotogenic correction with
\begin{eqnarray}\label{eq: loop function F}
	\mathcal{F}(m_{\eta_R},m_{\eta_I},M_f)=\frac{1}{32 \pi^2}\Big[\frac{m_{\eta_R}^2 \log\Big(M_f^2/m_{\eta_R}^2\Big)}{M_f^2-m_{\eta_R}^2}-\frac{m_{\eta_I}^2 \log\Big(M_f^2/m_{\eta_I}^2\Big)}{M_f^2-m_{\eta_I}^2}\Big],
\end{eqnarray}
where $m_{\eta_R}$ and $m_{\eta_I}$ are the masses of the neutral component of $\eta$~\cite{Mandal:2021yph}.  However, this model predicts one massless neutrino and demands extension to explain all the neutrino oscillation data. Hence, to understand the observed pattern of neutrino masses and  mixing, in the next section we will present a modified scoto-seesaw model where two right-handed neutrinos and $A_4$ discrete flavor symmetry will reproduce non-zero $\theta_{13}$ mixing angle and will account for the dark matter content.  In Ref.~\cite{Barreiros:2020gxu}, the authors already discussed the model with two right-handed neutrinos in the context of the scoto-seesaw scenario but with the $Z_8$ symmetry. As we will see, the set-up based on the $A_4$ flavor symmetry is substantially different in construction from the $Z_8$ scenario, resulting in a completely different texture of the Yukawa couplings and mass matrices.  As mentioned earlier, the $A_4$ flavor symmetry is well motivated in reproducing the TBM mixing scheme  and such symmetry can arise in various ways, such as starting from a continuous group~\cite{Koide:2007sr,Adulpravitchai:2009kd,Luhn:2011ip,Merle:2011vy,Rachlin:2017rvm,King:2018fke} or  superstring theory in compactified extra
dimensions~\cite{Altarelli:2005yp,Altarelli:2005yx,Burrows:2009pi,King:2017guk,deAdelhartToorop:2011re,Feruglio:2017spp,deAnda:2018ecu,Novichkov:2018yse,Criado:2018thu,Kobayashi:2018wkl,Penedo:2018nmg,Ding:2019xna}. We will show that due to the presence of the $A_4$ symmetry with diagonal structure of the charged leptons and heavy right-handed neutrinos, the TBM mixing can be generated at the leading order in the context of the minimal type-I seesaw. Subsequently, the scotogenic contribution acts as a crucial deviation from TBM mixing to generate non-zero $\theta_{13}$ (reproducing the ${\rm TM}_2$ mixing scheme) as well as providing essential dark matter candidates, thus unifying the origin of $\theta_{13}$ and dark matter. The $A_4$ symmetry assists us to obtain  analytic expressions for neutrino masses and mixing angles as well as  yields interesting correlations among the oscillation parameters with distinctive predictions; see section \ref{sec:nuetrino mass and mixing}. In ~\cite{Barreiros:2020gxu}, the CP symmetry is spontaneously broken by the complex vacuum expectation value of the singlet field whereas in our analysis the source of CP violation is due to the complex couplings and relative values of CP phases  determine the hierarchy of the masses.

\section{Scoto-seesaw with flavor $A_4$ symmetry: the FSS model}\label{sec:our model}
The model which we propose is a hybrid scoto-seesaw framework with usual scotogenic fermion $f$ and scalar doublet $\eta$, supported additionally by the $A_4$ discrete flavor symmetry and two right-handed neutrinos $N_{R_{1,2}}$. To obtain the flavor structure of the Yukawa couplings the  flavons $\phi_s$, $\phi_a$, $\phi_T$, $\xi$ are introduced. The inclusion of  flavon fields (SM gauge singlets) is a characteristic feature of models with discrete flavor symmetries~\cite{Ishimori:2010au, Altarelli:2010gt, Feruglio:2019ybq, King:2013eh, Petcov:2017ggy, Xing:2020ijf,Chauhan:2022gkz}. In a similar manner, 
we also incorporate additional $Z_N$ discrete symmetries which forbid the exchange of flavon fields eliminating unwanted terms~\cite{Altarelli:2010gt, Feruglio:2019ybq, King:2013eh, Petcov:2017ggy, Xing:2020ijf, Chauhan:2022gkz,Borah:2017dmk,Borah:2018nvu,Borah:2018gjk}. In what follows, we will call the whole framework the Flavor-Scoto-Seesaw (FSS) model where each element of the FSS model's construction is well motivated towards understanding a common origin of $\theta_{13}$ and DM. As we employ   the $A_4$ discrete symmetry, compared to Ref. ~\cite{Barreiros:2020gxu} and the $Z_8$ choice, we have a fuller symmetry with larger particle content. This is the price we pay to predict the ${\rm TM}_2$ structure of neutrino masses and mixing. The role of each of $Z_N$ auxiliary symmetries will be explained in detail as we proceed.  Interestingly,  the model  contains an intrinsic   $\mathcal{Z}_2$  symmetry under which both $f$ and $\eta$ are odd. The stability of the dark matter is ensured by this  $\mathcal{Z}_2$ symmetry.  In Table \ref{table:A4 table},   we  present transformation properties of all the fields content of our model under the complete discrete $A_4$ flavor symmetry. The desired mass matrix structure will be obtained when the flavons get a vacuum expectation value (VEV) in a suitable direction.
	\begin{table}[h!]
		\begin{center}
			\begin{tabular}{cccccccc|cccc}
				\hline
				Fields & $e_R$, $\mu_R$, $\tau_R$ & $L_{\alpha}$ & $H$ & $N_{R_1}$& $N_{R_2}$ & $f$& $\eta$ & $\phi_s$& $\phi_a$ & $\phi_T$ & $\xi$ \\
				\hline
				$A_4$ & $1$ , $1^{\prime\prime}$ , $1^{\prime}$ & $3$ & $1$ & $1$ & $1$ & $1$ & $1$ & $3$ &$3$& 3 & $1^{\prime\prime}$ \\
				$Z_4$ &  $-i$ & $-i$ & $1$ & $-1$ & $1$ & $1$ & $1$ & $i$ & $-i$ & $1$& $-1$ \\
				$Z_3$ &  $\omega$ & $\omega$ &$\omega^2$ & 1& 1& 1 & $\omega^2$ & 1 & 1 & $\omega$ & $1$ \\
				$Z_2$ & $-1$ & $1$ & $1$ & $1$ & $-1$ & $-1$ & $1$ & $1$ & $-1$ & $-1$& $-1$ \\
			    \hline
			\end{tabular}
		\caption{Field contents and transformation under the symmetries of our model. The second right-handed neutrino and the flavons field in second block of the table are introduced to  implement the  $A_4$ symmetry. }
		\label{table:A4 table}
		\end{center}
	\end{table}
The VEV alignment  considered here is widely used \cite{Altarelli:2005yx, King:2005bj, Zhao:2011pv}  and can be realized in a natural way by analyzing the complete scalar potential \cite{Altarelli:2010gt, Karmakar:2016cvb,  He:2006dk, Lin:2008aj, King:2005bj}. Here, the low energy scalar potential is identical to the potential presented in  \cite{Rojas:2018wym}, and for brevity we omit it here.  With the  fields content of Table \ref{table:A4 table}, the charged lepton Lagrangian can be described by 
\begin{eqnarray}\label{eq:charged lepton Lag}
	\mathcal{L}_l=\frac{y_e}{\Lambda}(\bar{L}\phi_T)H e_R + \frac{y_{\mu}}{\Lambda}(\bar{L}\phi_T)H \mu_R + \frac{y_{\tau}}{\Lambda}(\bar{L}\phi_T)H \tau_R + h.c.,
\end{eqnarray}
to the leading order, where $\Lambda$ is the cut-off scale of our model and $y_e$, $y_{\mu}$ and $y_{\tau}$ are coupling constants. As the SM lepton doublet, $L$ transforms as a triplet under $A_4$, the involvement of another triplet is essential as seen in the above Lagrangian.   Terms in the first parenthesis of Eq. (\ref{eq:charged lepton Lag}) represent the product of two $A_4$ triplets which results a true singlet after contracting with $A_4$ singlets  $e_R$, $\mu_R$ and $\tau_R$ (charged as $1$, $1^{\prime\prime}$ and $1^{\prime}$ respectively). The multiplication rule of $A_4$ symmetry is summarized in the Appendix and a detailed discussion on $A_4$ symmetry can be found in \cite{ Altarelli:2010gt, Ishimori:2010au}. Now, when the flavon $\phi_T$ gets VEV in the direction $\langle \phi_T \rangle =(v_T,0,0)^T$ \cite{Altarelli:2005yx} and also the Higgs field gets VEV $\langle H \rangle =v$, the charged lepton mass matrix will be a diagonal form
\begin{eqnarray}
	M_l=\frac{v_T}{\Lambda} v\begin{pmatrix}
		y_e & 0 & 0 \\
		0 & y_{\mu} & 0 \\
		0 & 0 & y_{\tau}
	\end{pmatrix}.
\end{eqnarray}
The Lagrangian in the neutrino sector constitutes two parts: a type-I seesaw contribution with two right-handed neutrinos $N_{R_1}$ and $N_{R_2}$ and another is a scotogenic contribution with a scalar field $\eta$ and a fermionic field $f$. The Lagrangian that generates neutrino mass at a tree level by the type-I seesaw mechanism in our model can be written as
\begin{eqnarray}\label{eq:Lag seesaw}
	\mathcal{L}_N = \frac{y_{N_1}}{\Lambda}(\bar{L}\phi_s)\tilde{H} N_{R_1}+\frac{y_{N_2}}{\Lambda}(\bar{L}\phi_a)\tilde{H} N_{R_2}+\frac{1}{2}M_{N_1} \bar{N}_{R_1}^c N_{R_1}+\frac{1}{2}M_{N_2} \bar{N}_{R_2}^c N_{R_2}+ h.c.,
\end{eqnarray} 
where $y_{N_{1,2}}$ are the corresponding couplings and $M_{N_{1,2}}$ are the Majorana masses of right-handed neutrinos. To get the flavor structure, we assume that the flavon fields get VEVs along $\langle \phi_s \rangle=(0,v_s,-v_s)$, $\langle \phi_a \rangle=(v_a,v_a,v_a)$ ~\cite{Antusch:2011ic, King:2005bj}. With these flavon vevs, the Dirac mass matrix will appear from the first two terms of  Eq. (\ref{eq:Lag seesaw}) while the Majorana matrix which follows from the next two terms of the Lagrangian of Eq. (\ref{eq:Lag seesaw}) can be found as follows 
\begin{eqnarray}\label{eq:dirac and majorana mass matrix}
	M_D=\frac{v}{\Lambda}\begin{pmatrix}
		0 & y_{N_2}v_a \\
		-y_{N_1}v_s & y_{N_2}v_a \\
		y_{N_1}v_s & y_{N_2}v_a
	\end{pmatrix}=v Y_{N},\quad
M_{R}=\begin{pmatrix}
	M_{N_1} & 0 \\
	0 & M_{N_2}
\end{pmatrix}.
\end{eqnarray}
The $Z_4$ symmetry in Table \ref{table:A4 table}, under which $N_{R_1}$ is odd and $N_{R_2}$ even and the $Z_2$ symmetry  under which $N_{R_1}$ is even and $N_{R_2}$ odd ensures the diagonal structure of $M_{R}$ as obtained in Eq. \ref{eq:dirac and majorana mass matrix}.  The VEV alignment considered here is widely used in the context of form dominance~\cite{Chen:2009um}, sequential dominance~\cite{Antusch:2004gf}, constrained sequential dominance~~\cite{Antusch:2011ic} etc., to obtain the textures of the Dirac and Majorana mass matrices.  Now using  the type-I seesaw formula the light neutrino mass matrix at the leading order can be written as 
\begin{eqnarray}\label{eq:seesaw formula}
	(M_{\nu})_{\rm TREE}=-M_D M_R^{-1}M_D^T.
\end{eqnarray}
With the structure  of  $M_D$ and $M_R$ obtained in  Eq. (\ref{eq:dirac and majorana mass matrix}), the light neutrino mass matrix is given by 
\begin{eqnarray}\label{eq:seesaw mass matrix}
	({M_{\nu}})_{\rm TREE}=-\begin{pmatrix}
	B & B & B \\
	B & A+B & -A+B \\
	B & -A+B & A+B
	\end{pmatrix},\quad A=\frac{v^2 v_s^2 y_{N_1}^2}{\Lambda^2 M_{N_1}},\quad B=\frac{v^2 v_a^2 y_{N_2}^2}{\Lambda^2 M_{N_2}}.
\end{eqnarray}
The above mass matrix for  light neutrinos obtained from type-I seesaw is incapable to  generate  non-zero $\theta_{13}$ (charged lepton mass matrix being diagonal).    As neutrino oscillation data established adequately large   $\theta_{13}$, we include a scotogenic contribution  to our model to explain correct neutrino mixing  which also naturally incorporates few potential DM candidates. The scotogenic contribution in our model with the fermion  $f$ and scalar field $\eta$ can be written as
\begin{eqnarray}\label{eq:scoto Lag}
	\mathcal{L}_S=\frac{y_s}{\Lambda^2}(\bar{L}\phi_s)\xi i\sigma_2 \eta^* f+ \frac{1}{2}M_f \bar{f}^c f+h.c.,
\end{eqnarray}
where $y_s$ is the coupling and $M_f$ is the mass of $f$. Owing to the considered symmetry in Table \ref{table:A4 table}, the leading order contribution $\bar{L} i\sigma_2 \eta^* f$ is disallowed,  as it is not invariant under $A_4$ (and $Z_4$, $Z_2$) symmetry.  The SM lepton doublet being a triplet under $A_4$, just like charged lepton sector, involvement of  another $A_4$ triplet (here $\phi_s$) is essential.  As a consequence of  $Z_{4,2}$ symmetry involvement of the $A_4$ flavons $\phi_s$ and $\xi$ is necessary in the first term of Eq. (\ref{eq:scoto Lag}). The VEV of $\phi_s$ (mentioned below) and the non-trivial $A_4$ singlet $\xi$ (provides appropriate $A_4$ contraction) crucially dictates the structure of  the scotogenic contribution and helps in breaking of TBM mixing~\cite{Karmakar:2014dva, Karmakar:2015jza, Karmakar:2016cvb, Bhattacharya:2016lts,Bhattacharya:2016rqj}.   Therefore the contribution in the effective neutrino mass matrix originated from the scotogenic radiative
corrections is given by~\cite{Ma:2006km, Rojas:2018wym,Barreiros:2020gxu} 
\begin{eqnarray}\label{eq:scoto mass formula}
		(M_{\nu})_{\rm LOOP}=\mathcal{F}(m_{\eta_R},m_{\eta_I},M_f)M_f Y_f^i Y_f^j. 
\end{eqnarray}
Once the flavons  $\phi_s$ and $\xi$ acquire VEVs in the direction $\langle \phi_s \rangle=(0,v_s,-v_s)$ and $\langle \xi \rangle=v_{\xi}$ respectively the associated couplings can be written as 
\begin{eqnarray}\label{eq: Yukawa for scoto}
   Y_F=(Y_F^e,Y_F^{\mu},Y_F^{\tau})^T=(y_s \frac{v_s}{\Lambda}\frac{v_{\xi}}{\Lambda},0,-y_s \frac{v_s}{\Lambda}\frac{v_{\xi}}{\Lambda})^T. 
\end{eqnarray}
Therefore, the corresponding mass matrix takes  the form
\begin{eqnarray}\label{eq:scoto mass matrix}
	(M_{\nu})_{\rm LOOP}=C\begin{pmatrix}
		1 & 0 & -1 \\
		0 & 0 & 0 \\
		-1 & 0 & 1
	\end{pmatrix}, \quad C={\mathcal{F}(m_{\eta_R},m_{\eta_I},M_f)}y_s^2 \frac{v_s^2 v_{\xi}^2}{\Lambda^4}.
\end{eqnarray}
Here ${\mathcal{F}(m_{\eta_R},m_{\eta_I},M_f)}$ is the loop function given in Eq. (\ref{eq: loop function F}). Finally, combining the seesaw and scotogenic contributions,  the effective light neutrino mass matrix is the addition of the two mass matrices given in Eq. (\ref{eq:seesaw mass matrix}) and Eq. (\ref{eq:scoto mass matrix}) and reads  
\begin{eqnarray}\label{eq:total mass matrix}
	M_{\nu}&=&(M_{\nu})_{\rm TREE}+(M_{\nu})_{\rm LOOP}\nonumber \\
&=&\begin{pmatrix}\label{eq:full mass matrix}
		-B+C & -B & -B-C \\
		-B & -(A+B) & A-B \\
		-B-C & A-B & -A-B+C
	\end{pmatrix}.
\end{eqnarray}

In the present context, neutrino masses are obtained through a combination of the type-I seesaw and scotogenic mechanisms. Now, there could also be operators like $LHLH/\Lambda$, which can also contribute to the light neutrino mass. In our model, this term is not invariant under  the $Z_4$ symmetry mentioned in Table \ref{table:A4 table}. Any contributions coming from $LHLH (\phi_a, \phi_s, \phi_T, \xi) / \Lambda^2$ is also disallowed due to the considered discrete symmetries $Z_4,Z_3$ and $Z_2$.  For the scotogenic contribution,  the coupling $\bar{L}i\sigma_2 \eta^* f$ is allowed only at $1/\Lambda^2$ level with the involvement of the flavons $\phi_s,\xi$,  see Eq. (\ref{eq:scoto Lag}).  In this sector, any higher-order contributions in the bare mass term of $f$ can be absorbed in the leading order contribution. On the other hand for charged lepton sector,  the leading contribution only appears at dimension-5 due to the considered $A_4$ symmetry. There are next-to-leading order  corrections present in this sector coming from $(\bar{L}\phi_s\phi_a)H \alpha_R/\Lambda^2$, where $\alpha_{R}$ is the associated right-handed charged lepton. Thanks to the VEV alignment of the flavons $\phi_s$ and $\phi_a$ this term essentially vanishes following the $A_4$ multiplication rules mentioned in the appendix.  For right-handed Majorana neutrinos, the non-vanishing next-to-leading order corrections up-to $\mathcal{O}$ $(1/\Lambda^2)$ in the mass matrix arise from the following terms:
\begin{eqnarray}\label{eq:NLORH}
	\delta\mathcal{L}_{M_R} &=&\frac{1}{\Lambda} \left(\bar{N}_{R_1}^c N_{R_1}+\bar{N}_{R_2}^c N_{R_2}\right) \left(\phi^{\dagger}_a\phi_a+\phi^{\dagger}_s\phi_s\right)\nonumber \\
	&+&  \frac{1}{\Lambda}(\bar{N}_{R_1}^c N_{R_2}+\bar{N}_{R_2}^c N_{R_1})(\phi^{\dagger}_s\phi_a+\phi_s\phi^{\dagger}_a+\phi^{\dagger}_s\phi^{\dagger}_a)+\frac{1}{\Lambda^2}(\bar{N}_{R_1}^c N_{R_2}+\bar{N}_{R_2}^c N_{R_1})\xi^3. 
\end{eqnarray}
Here in Eq. (\ref{eq:NLORH}), the first term represents a correction to the diagonal entry which can be absorbed in the leading order $M_R$. The second term represents off-diagonal entries of the right-handed neutrino mass matrix at  $\mathcal{O}$ $(1/\Lambda)$ which also vanishes due to the specific VEV direction of $\phi_s$ and $\phi_a$. The last term in Eq. (\ref{eq:NLORH}) represents    off-diagonal entries at  $\mathcal{O}$ ($1/\Lambda^2$). Although this contribution is very small compared to the leading order contribution, it  can also be forbidden by considering another $Z'_2$ symmetry under which charged leptons, $f$ and the flavons $\phi_{T},\xi'$ are odd (while all other particles are even). The  Dirac Yukawa coupling is allowed at dimension-5 as given in Eq. (\ref{eq:Lag seesaw}). Here the next-to-leading order contribution at $\mathcal{O}$ $(1/\Lambda)$  can be written as $(\bar{L} \phi^{\dagger}_a \phi_T ) \tilde{H} N_{R_1}/\Lambda^2$ and $(\bar{L}\phi^{\dagger}_s\phi_T)\tilde{H} N_{R_2}/\Lambda^2$ respectively. These terms are however forbidden owing to the $Z_3$ symmetry mentioned in Table \ref{table:A4 table}. Therefore, from Table \ref{table:A4 table}, it is clear that  along with the $A_4$ symmetry the auxiliary  discrete symmetries  crucially dictate allowed structures of the fermionic mass matrices, and such symmetries are an integral part of the flavor symmetric approach to understand neutrino mixing ~\cite{Altarelli:2005yp,Altarelli:2005yx,Molinaro:2009lud,Branco:2011iw,King:2011zj,Altarelli:2010gt,Ma:2001dn}.   

\section{Neutrino masses and mixing in the FSS model}\label{sec:nuetrino mass and mixing}
From the previous discussion, we find  that the effective neutrino  mass matrix consists of two parts,  one of them is coming from the type-I seesaw mechanism given by Eq. (\ref{eq:seesaw mass matrix}) and another one  originates from the scotogenic contribution given by Eq. (\ref{eq:scoto mass matrix}). Now the mass matrix originating from type-I seesaw given in  Eq. (\ref{eq:seesaw mass matrix}) can be diagonalized by the TBM mixing matrix ($U_{TB}$) via 
\begin{eqnarray}\label{eq:UTB seesaw diag}
	U_{TB}^T (M_{\nu})_{\rm TREE}U_{TB}=\begin{pmatrix}
		0 & 0 & 0 \\
		0 &-3 B & 0 \\
		0 & 0 & -2 A
	\end{pmatrix}, 
\end{eqnarray}
where 
\begin{eqnarray}\label{eq:UTB}
	U_{TB}=\begin{pmatrix}
		\sqrt{\frac{2}{3}} & \frac{1}{\sqrt{3}} & 0 \\
		-\frac{1}{\sqrt{6}} & \frac{1}{\sqrt{3}} & -\frac{1}{\sqrt{2}}\\
		-\frac{1}{\sqrt{6}} & \frac{1}{\sqrt{3}} & \frac{1}{\sqrt{2}}
	\end{pmatrix}. 
\end{eqnarray}
Clearly, a pure type-I seesaw contribution in the present set-up predicts  $\theta_{13}=0$.  However, thanks to the scotogenic contribution, we can obtain a deviation from $\theta_{13}=0$ to be consistent with  the observed experimental value~\cite{Abe:2011fz,An:2012eh,Ahn:2012nd}. 
Therefore  considering the effective light neutrino mass matrix given in Eq. (\ref{eq:total mass matrix}) and rotating it by $U_{TB}$, $M_{\nu}$ takes the form 
\begin{eqnarray}\label{eq:UTB full mass mstrix rotation}
M_{\nu}^{\prime}&=&U_{TB}^T M_{\nu} U_{TB}\nonumber \\
&&=\frac{1}{2}\begin{pmatrix}
	3 C & 0 & -\sqrt{3}C \\
	0 & -6 B & 0 \\
	-\sqrt{3}C & 0 & -4 A + C 	
\end{pmatrix},
\end{eqnarray}
Here we find  that the mass matrix in the tri-bimaximal basis is block diagonalized. Therefore a further rotation by a unitary matrix $U_{13}$ in the  13 plane via $M_{\nu}^{diag}=U_{13}^T M_{\nu}^{\prime}U_{13}$ takes $M'_{\nu}$ to a diagonal one. This  unitary matrix $U_{13}$  can be parametrized  as 
\begin{eqnarray}\label{eq:U13}
U_{13}=\begin{pmatrix}
	\cos\theta & 0 & \sin\theta e^{-i\phi}\\
	0 & 1 & 0 \\
	-\sin\theta e^{i\phi} & 0 & \cos\theta
\end{pmatrix}, 	
\end{eqnarray}
where $\theta$ is the rotation angle and $\phi$ is the associated phase factor. The full diagonalization relation of the mass matrix  $M_{\nu}$ can be written as
\begin{eqnarray}
	(U_{TB}U_{13})^T M_{\nu}U_{13}U_{TB}={\rm diag}(m_1 e^{i\gamma_1},m_2 e^{i\gamma_2},m_3e^{i\gamma_3}), 
\end{eqnarray}
where $m_1$, $m_2$, $m_3$ are the real and positive mass eigenvalues and $\gamma_1$, $\gamma_2$ and $\gamma_3$ are the phases extracted from the corresponding complex eigenvalues. We are now in a position to evaluate the neutrino mixing matrix $U_{\nu}$ such that $U_{\nu}^T M_{\nu}U_{\nu}={\rm diag}(m_1,m_2,m_3)$. Thus $U_{\nu}$ becomes $U_{\nu}=U_{TB}U_{13}U_m$, where $U_m={\rm diag}(1,e^{i\alpha_{21}/2},e^{i\alpha_{31}/2})$ is the Majorana phase matrix with $\alpha_{21}=\gamma_1-\gamma_2$ and $\alpha_{31}=\gamma_1-\gamma_3$, one common phase being irrelevant. Using Eq. (\ref{eq:UTB}) and Eq. (\ref{eq:U13}), the  $U_{\nu}$ mixing matrix in its explicit form can be written as  
\begin{eqnarray}\label{eq:Unufull}
    U_{\nu}=\begin{pmatrix}
    \sqrt{\frac{2}{3}}\cos\theta & \frac{1}{\sqrt{3}} & \sqrt{\frac{2}{3}}e^{i\phi}\sin\theta \\
    -\frac{\cos\theta}{\sqrt{6}} +\frac{e^{i\phi \sin\theta}}{\sqrt{2}} &\frac{1}{\sqrt{3}} &   -\frac{\cos\theta}{\sqrt{2}} -\frac{e^{i\phi \sin\theta}}{\sqrt{6}} \\
     -\frac{\cos\theta}{\sqrt{6}} -\frac{e^{i\phi \sin\theta}}{\sqrt{2}} & \frac{1}{\sqrt{3}} & \frac{\cos\theta}{\sqrt{2}} -\frac{e^{i\phi \sin\theta}}{\sqrt{6}}
    \end{pmatrix}U_m. 
\end{eqnarray}
Such deviation from the TBM mixing is well known and this particular pattern of $U_{\nu}$ is called ${\rm TM}_2$ mixing as described  earlier. This lepton mixing matrix $U_{\nu}$ can now be compared with $U_{PMNS}$ which in its standard parametrization   is given by~\cite{ParticleDataGroup:2020ssz} 
\begin{eqnarray}\label{eq:UPMNS}
	U_{PMNS}=\begin{pmatrix}
		c_{12}c_{13} & s_{12}c_{13} & s_{13}e^{-i\delta_{\rm CP}}\\
		-s_{12}c_{23}-c_{12}s_{23}s_{13}e^{i\delta_{\rm CP}} & c_{12}c_{23}-s_{12}s_{23}s_{13}e^{i\delta_{\rm CP}} & s_{23}c_{13}\\
		s_{12}s_{23}-c_{12}c_{23}s_{13}e^{i\delta_{\rm CP}} & -c_{12}s_{23}-s_{12}c_{23}s_{13}e^{i\delta_{\rm CP}} & c_{23}c_{13}
	\end{pmatrix}U_m, 
\end{eqnarray}
where $\theta_{12}, \theta_{13}$ and $\theta_{23}$ are three mixing angles, $\delta_{\rm CP}$ is the CP violating Dirac phase and $\alpha_{21}$, $\alpha_{31}$ are the Majorana phases.

The parameters $A$, $B$ and $C$  appearing in Eq. (\ref{eq:UTB full mass mstrix rotation}) are in general  complex, and without loss of generality we can  write $A=|A|e^{i\phi_A}$, $B=|B|e^{i\phi_B}$, $C=|C|e^{i\phi_C}$. Now, for calculation purpose, let us define $\alpha=|A|/|C|$ and $\beta=|B|/|C|$, and phase differences $\phi_{AC}=\phi_A-\phi_C$ and $\phi_{BC}=\phi_B-\phi_C$. As $U_{13}$ diagonalizes $M_{\nu}^{\prime}$ of Eq. (\ref{eq:UTB full mass mstrix rotation}), $\theta$ and $\phi$ can be expressed in terms of model parameters  as
\begin{eqnarray}\label{eq:phi and theta expre}
	\tan\phi=\frac{ \alpha\sin\phi_{AC}}{ 1-\alpha\cos\phi_{AC}},\quad \tan2\theta=\frac{\sqrt{3}}{\cos\phi+2\alpha\cos(\phi_{AC}+\phi)}.
\end{eqnarray}
Further  comparing $U_{\nu}=U_{TB}U_{13}U_m$ as given in Eq. (\ref{eq:Unufull}) with $U_{PMNS}$ as in Eq. (\ref{eq:UPMNS}), we find the following relations for mixing angles  and $\delta_{\rm CP}$ as a function of $\theta$ and $\phi$ as~\cite{Ma:2012ez, Hernandez:2012ra, Tanimoto:2015nfa, Shimizu:2014ria, Karmakar:2014dva}
\begin{eqnarray}\label{eq:mixing angles expres}
	&&	\sin\theta_{13} e^{-i\delta_{\rm CP}}=\sqrt{\frac{2}{3}}e^{-i\phi}\sin\theta, \quad	\tan^2\theta_{12}=\frac{1}{2-3 \sin^2\theta_{13}},  \\
	&&\tan^2\theta_{23}=\frac{\Big(1+\frac{\sin\theta_{13}\cos\phi}{\sqrt{2-3 \sin^2\theta_{13}}}\Big)^2+\frac{ \sin^2\theta_{13}\sin^2\phi}{ (2- 3\sin^2\theta_{13})}}{\Big(1-\frac{\sin\theta_{13}\cos\phi}{\sqrt{2-3 \sin^2\theta_{13}}}\Big)^2+\frac{ \sin^2\theta_{13}\sin^2\phi}{ (2- 3\sin^2\theta_{13})}}.\label{eq:mixing angles expres2}
\end{eqnarray}
The above relations show that the mixing angles are correlated  which is a characteristic feature of the considered $A_4$ discrete flavor symmetry. For $\sin\theta>0$, the relation of $\delta_{\rm CP}$ implies that $\delta_{\rm CP}=\phi$ and for $\sin\theta<0$, the same relation implies that $\delta_{\rm CP}=\phi\pm \pi$. Hence, for both  cases, we have $\tan\delta_{\rm CP}=\tan\phi$. Now, using Eq. (\ref{eq:full mass matrix}), the complex mass eigenvalues are calculated to be
\begin{eqnarray}
	m_{1,3}^c&=&-A+C\pm\sqrt{A^2+AC+C^2},\label{eq:complex mass eigenvalues1}  \\
	m_2^c&=& -3B.\label{eq:complex mass eigenvalues2}
\end{eqnarray}
The real and positive eigenvalues can be written as  
\begin{eqnarray}
&&m_1=|C|\big[(1-\alpha \cos\phi_{AC}-P)^2+(Q+\alpha \sin\phi_{AC})^2\big]^{1/2},\label{eq:real mass  m1} \\
&& m_2=|C|3 \beta,\label{eq:real mass  m2} \\
&& m_3=|C|\big[(1-\alpha \cos\phi_{AC}+P)^2+(Q-\alpha \sin\phi_{AC})^2\big]^{1/2}, \label{eq:real mass  m3}
\end{eqnarray}
where 
\begin{eqnarray}
	&& P^2=\frac{M\pm \sqrt{M^2+N^2}}{2},\quad Q^2=\frac{-M\pm\sqrt{M^2+N^2}}{2},  \\
	&& M=1+\alpha \cos\phi_{AC}+\alpha^2 \cos 2\phi_{AC},\quad N=\alpha \sin\phi_{AC}+\alpha^2 \sin 2 \phi_{AC}.
\end{eqnarray}
Following Eq. (\ref{eq:complex mass eigenvalues1}) and Eq. (\ref{eq:complex mass eigenvalues2}), the phase associated with complex mass eigenvalues $m^c_{1,2,3}$ can be written as  $\gamma_i=\phi_C+\phi_i$, where $\phi_i$ are
\begin{eqnarray}\label{eq:phi1,2,3 express}
	\phi_1=\tan^{-1}\Big(\frac{Q+\alpha \sin\phi_{AC}}{1-\alpha \cos\phi_{AC}-P}\Big),\quad\phi_2=\phi_{BC},\quad \phi_3=\tan^{-1}\Big( \frac{Q-\alpha\sin\phi_{AC}}{1-\alpha \cos\phi_{AC}+P}\Big).
\end{eqnarray}
The two Majorana phases in $U_m$ (see eq. (\ref{eq:Unufull})) therefore can be derived as 
\begin{eqnarray}
	\alpha_{21}&=&\tan^{-1}\Big(\frac{Q+\alpha \sin\phi_{AC}}{1-\alpha \cos\phi_{AC}-P}\Big)-\phi_{BC}\label{eq:majo1}, \\ \alpha_{31}&=&\tan^{-1}\Big(\frac{Q+\alpha \sin\phi_{AC}}{1-\alpha \cos\phi_{AC}-P}\Big)-\tan^{-1}\Big( \frac{Q-\alpha\sin\phi_{AC}}{1-\alpha \cos\phi_{AC}+P}\Big).\label{eq:majo2}
\end{eqnarray}
The overall phase factor $\phi_C$ appearing in $\gamma_i$ has no physical significance in computing the Majorana phases.  The mixing angles and the phases depend on the parameters $\alpha,\phi_{AC},\phi_{BC}$ whereas the light neutrino masses  depend on these parameters as well as on $\beta$ and  $|C|$ as observed in  Eq. (\ref{eq:phi and theta expre}) - Eq. (\ref{eq:majo2}). In the next section, we constrain these parameters using experimental data  for neutrino mixing angles and masses.

\begin{table}[h!]
	\begin{center}\resizebox{0.45\textwidth}{!}{
		\begin{tabular}{c||c|c}
			\hline
			parameters & best-fit & $3 \sigma$ range \\
			\hline
			\hline
		$\dfrac{\Delta m_{21}^2}{10^{-5}{\rm eV}^2}$ & 7.50 & 6.94 - 8.14 \\
			\hline 
   	$\dfrac{|\Delta m_{31}^2|}{10^{-3}{\rm eV}^2}$(NH) & 2.55 & 2.47 - 2.63 \\  \cline{2-3}
	\rule{0pt}{4ex}	$\dfrac{|\Delta m_{31}^2|}{10^{-3}{\rm eV}^2}$(IH) & 2.45 & 2.37 - 2.53 \\
			\hline
			$\sin^2 \theta_{12}/10^{-1}$ & 3.18 & 2.71 - 3.69 \\
			\hline
			$\sin^2 \theta_{13}/10^{-2}$(NH) & 2.200 & 2.000 - 2.405 \\  \cline{2-3}
			$\sin^2 \theta_{13}/10^{-2}$(IH) & 2.225 & 2.018 - 2.424 \\
			\hline
			$\sin^2 \theta_{23}/10^{-1}$(NH) & 5.74 & 4.34 - 6.10 \\  \cline{2-3}
			$\sin^2 \theta_{23}/10^{-1}$(IH) & 5.78 & 4.33 - 6.08 \\
			\hline	
		\end{tabular}}
		\caption{Global fits of three active-neutrino oscillation data taken for Ref.~\cite{deSalas:2020pgw} for NH and IH,  used in our analysis.    }
		\label{table:neutrino osc data}
	\end{center}
\end{table}


\section{Numerical analysis of the FSS model}\label{sec:numerical analysis}
In order to constrain the parameters involved in our analysis, using  Eq. (\ref{eq:real mass  m1}) - Eq. (\ref{eq:real mass  m3}), we can define  a ratio $r$  as
\begin{eqnarray}\label{eq:r}
	r=\frac{\Delta m_{21}^2}{|\Delta m_{31}^2|}, 
\end{eqnarray}
where $\Delta m_{21}^2=m_2^2-m_1^2$ and $|\Delta m_{31}^2|=|m_3^2-m_1^2|$ are the  solar and atmospheric mass squared differences. From the expressions for the mixing angles (namely, $\theta_{13}$, $\theta_{12}$ and  $\theta_{23}$) as well as the absolute neutrino masses ($m_{1,2,3}$), their sum ($\sum m_i$) and the ratio $r$ defined in Eq. (\ref{eq:r}) all depend on  the variables   $\alpha$, $\beta$, $\phi_{AC}$ and $\phi_{BC}$ as discussed in Section \ref{sec:nuetrino mass and mixing}. Over the last two decades neutrino oscillation parameters have been measured with incredible accuracy~ \cite{ deSalas:2020pgw, Esteban:2020cvm,Capozzi:2021fjo}. Therefore using the precisely determined  neutrino oscillation data on $\theta_{13}$, $\theta_{12}$, $\theta_{23}$, $r$, $\Delta m_{21}^2$ and $|\Delta m_{31}^2|$ one can constrain  four model parameters. Once the model parameters are constrained, we can further compute the  Dirac CP phase $\delta_{\rm CP}$, Majorana phases $\alpha_{21}$, $\alpha_{31}$, the sum of the absolute masses of the three light neutrinos ($\sum m_i$) as well as the effective mass parameter  appearing in the neutrinoless double beta decay ($m_{\beta\beta}$). Table \ref{table:neutrino osc data} summarizes the best fit and 3$\sigma$ ranges of the neutrino oscillation data~\cite{deSalas:2020pgw} for both normal and inverted hierarchy of light neutrino masses which are used  in  the subsequent  numerical analysis.

\begin{figure}[h!]
\begin{center}
	\includegraphics[width=.43\textwidth]{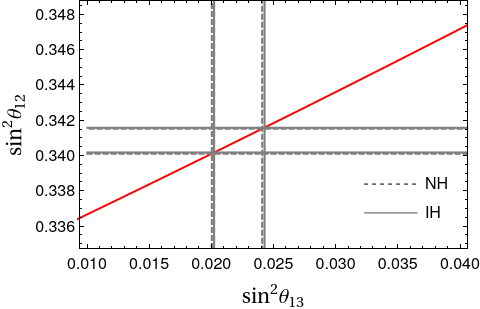}
	\caption{$\sin^2\theta_{12}$ is plotted against $\sin^2\theta_{13}$ for both NH and IH of neutrino masses. The vertical grid-lines are 3$\sigma$ allowed for $\sin^2\theta_{13}$ and horizontal grid-lines represent corresponding restriction on $\sin^2\theta_{12}$ in our analysis.  }
	\label{fig:ss12-ss13}
	\end{center}
\end{figure}

Before we proceed further, let us  point out  that the correlation between the mixing angles $\theta_{13}$ and $\theta_{12}$  given in Eq. (\ref{eq:mixing angles expres})  is a feature of the ${\rm TM}_2$ mixing mentioned above~\cite{Shimizu:2014ria,Hernandez:2012ra}.  In Fig. \ref{fig:ss12-ss13}, we have plotted this $\sin^2\theta_{12}$$-$$\sin^2\theta_{13}$ correlation  for ${\rm TM}_2$ mixing and we
find that $\sin^2\theta_{12}$ is restricted within a narrow range (between 0.3401~$\leq \sin^2\theta_{12} \leq$~0.3415) corresponding to  the $3\sigma$ ranges of $\sin^2\theta_{13}$.Now, in order to evaluate  absolute neutrino masses we also need to find the  overall factor $|C|$   appearing in the mass eigenvalues given in Eq. (\ref{eq:real mass  m1}) - Eq. (\ref{eq:real mass  m3}). Although this common factor $|C|$ cancels out when we calculate $r$ but  $|C|$ can  be calculated by fitting the solar (or atmospheric) mass-squared differences knowing the model parameters. After  $|C|$ is evaluated, we can get the estimation of  absolute neutrino masses $m_{1,2,3}$ and their sum $\sum m_i$. Similarly, substituting the estimations for  $\alpha,\beta,\phi_{AC}$ and $\phi_{BC}$  in Eq. (\ref{eq:majo1}) and Eq. (\ref{eq:majo2}) we can also quantify the Majorana phases. Knowing the neutrino mixing angles, masses, and associated  CP phases, finally in our  analysis, we will have a prediction on the effective neutrino mass parameter $m_{\beta\beta}$ characterizing  the neutrinoless double beta decay.  The effective  mass parameter $m_{\beta\beta}$ can be described as a function of the lightest neutrino mass ($m_1$ for NH and $m_3$ for IH respectively) and can be written as  \footnote{The expression for the effective mass parameter $m_{\beta\beta}$ can also be written in a symmetrical form where  only  two Majorana phases~\cite{Rodejohann:2011vc} appear instead of three phases appearing in the standard PDG parametrization \cite{Zyla:2020zbs}.}
 \begin{eqnarray}\label{eq:ndbd exp}
 \text{NH:}\quad	m_{\beta\beta}&=&\Big|m_1 c_{12}^2c_{13}^2+\sqrt{m_1^2+\Delta m_{21}^2} s_{12}^2 c_{13}^2 e^{i\alpha_{21}}+\sqrt{m_1^2+\Delta m_{31}^2} s_{13}^2 e^{i(\alpha_{31}-2\delta_{\rm CP})}\Big| ,\quad \\
 \text{IH:}\quad m_{\beta\beta} &=&\Big|\sqrt{m_3^2+\Delta m_{31}^2} c_{12}^2c_{13}^2+\sqrt{m_3^2+\Delta m_{31}^2+\Delta m_{21}^2} s_{12}^2 c_{13}^2 e^{i\alpha_{21}}+m_3 s_{13}^2 e^{i(\alpha_{31}-2\delta_{\rm CP})}\Big|\label{eq:ndbd exp2}. 
 \end{eqnarray}
Now, for a better understanding of the behaviour of the parameters involved and the model predictions, we can divide our numerical analysis into some special cases by taking some particular values of the relative  phases $\phi_{AC}$ and $\phi_{BC}$. All mixing angles in Eq.  (\ref{eq:mixing angles expres}), (\ref{eq:mixing angles expres2}) and neutrino mass eigenvalues in Eq. (\ref{eq:real mass  m1}) - Eq. (\ref{eq:real mass  m3}) and the Majorana phase $\alpha_{31}$ depend on one relative phase, namely,  $\phi_{AC}$. In contrast, the other Majorana phase $\alpha_{21}$  as well as $m_{\beta\beta}$  depend on both $\phi_{AC}$ and $\phi_{BC}$. In the following, we choose five simple special cases depending on the values of these phases, namely: $(i)$ Case I : $\phi_{AC}=0$, $\phi_{BC}=0$ , $(ii)$ Case II $\phi_{AC}=0$, $(iii)$ Case III $\phi_{AC}=\phi_{BC}$, $(iv)$ Case IV $\phi_{BC}=0$ and subsequently in $(v)$ Case V we present the  general scenario where both $\phi_{AC}$ and $\phi_{BC}$ vary between $0-2\pi$.  Below, we have explored these cases and as we proceed it will be clear that some of these cases  have the potential to distinguish the normal  and inverted hierarchy  of light neutrino masses and produce interesting predictions on neutrino parameters. 
\subsection{Case I: ${\phi_{AC}=\phi_{BC}=0}$}\label{subsec:1}
Here we make the simplest choice for the relative  phases, $i.e.,$ $\phi_{AC}=\phi_{BC}=0$. With this value, the Eq. (\ref{eq:phi and theta expre}) and (\ref{eq:mixing angles expres}) have the simple form
\begin{eqnarray}\label{eq:theta s13 exp phiac=0}
\tan2\theta=\frac{\sqrt{3}}{1+2\alpha},\quad \sin\theta_{13}=\sqrt{\frac{2}{3}}|\sin\theta|.
\end{eqnarray}
with $\tan\delta_{\rm CP}=0$.
\begin{figure}[h!]
	\begin{center}
	\includegraphics[width=0.43\textwidth]{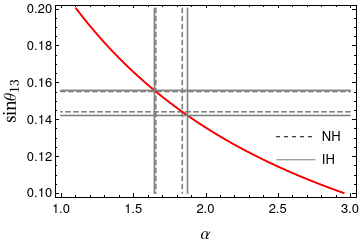}
	\includegraphics[width=0.43\textwidth]{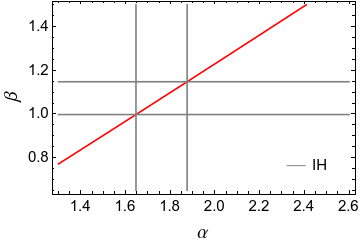}
		\caption{Left Panel: Plot for $\sin\theta_{13}$ vs $\alpha$. Horizontal grid lines are the $3\sigma$ allowed range for $\sin\theta_{13}$ (for NH and IH) whereas the vertical grid lines represent the corresponding allowed range for $\alpha$ ( $1.6557 - 1.8391$ and $ 1.6465 - 1.8754$ for NH and IH respectively). Right Panel: Contour plot for $r=0.03$ in $\alpha-\beta$ plane. The vertical grid lines represent the allowed regions for $\alpha$ from the left panel and the horizontal grid lines are the corresponding allowed regions for $\beta$. }
		\label{fig:ss13-alpha phiac=0}
	\end{center}
\end{figure}
Clearly,  $\sin\theta_{13}$ only depends on $\alpha$ and  in Fig. (\ref{fig:ss13-alpha phiac=0}) left panel, we have plotted $\sin\theta_{13}$ as a function of $\alpha$ using Eq. (\ref{eq:theta s13 exp phiac=0}). The $3\sigma$ allowed range for $\sin\theta_{13}$ (given by the area between horizontal lines) restricts $\alpha$ within $1.6465-1.8754$ ($1.6557 - 1.8391$) for IH (NH)  given by the dashed (continuous) vertical lines.  With  $\phi_{AC}=\phi_{BC}=0$  the real positive mass eigenvalues given in  Eq. (\ref{eq:real mass  m1}) - Eq. (\ref{eq:real mass  m3}) can be expressed as 
\begin{eqnarray}\label{eq:mass case 1}
m_1&=&|C|(1-\alpha-\sqrt{1+\alpha+\alpha^2}),  \\
m_2&=&|C|3\beta, \label{eq:mass case 1-2} \\
m_3&=&|C|(1-\alpha+\sqrt{1+\alpha+\alpha^2})\label{eq:mass case 1-3}. 
\end{eqnarray}
With the above mass eigenvalues, one can write the ratio of solar to atmospheric mass-squared differences as defined in Eq. (\ref{eq:r}) as
\begin{eqnarray}\label{eq:r case 1}
r=\pm \frac{9 \beta^2-(1-\alpha-\sqrt{1+\alpha+\alpha^2})^2}{(1-\alpha+\sqrt{1+\alpha+\alpha^2})^2-(1-\alpha-\sqrt{1+\alpha+\alpha^2})^2}, 
\end{eqnarray}
where the $\pm$ signs are for NH and IH respectively. When ${\phi_{AC}=\phi_{BC}=0}$, as a consequence of the considered discrete flavor symmetry, NH of light neutrino masses can not be realized with Eq. (\ref{eq:mass case 1}) - Eq. (\ref{eq:r case 1}). 
\begin{figure}[h!]
  	\begin{center}
  		\includegraphics[width=.43\textwidth]{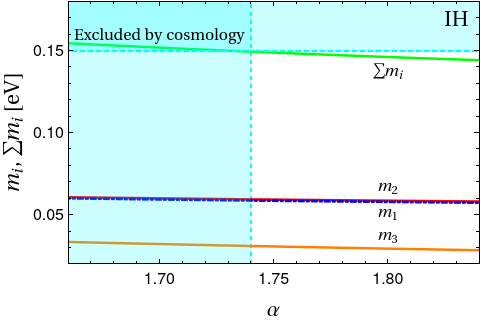}
  		\includegraphics[width=.43\textwidth]{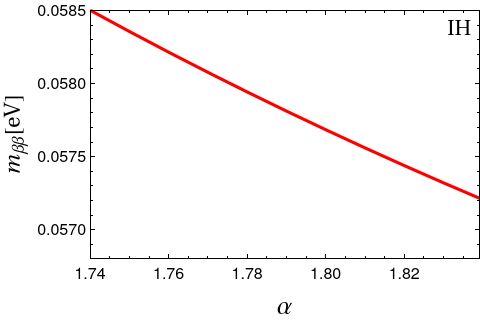}
  	\end{center}
\caption{Left Panel: Absolute neutrino masses $m_3$ (orange line), $m_2$ (blue dashed line), $m_3$ (red line) and their sum $\sum m_i$ (green line) plotted against $\alpha$. Right Panel : The prediction on  the effective mass  parameter $m_{\beta\beta}$. In both cases we have considered $\phi_{AC}=\phi_{BC}=0$.}
\label{fig:mass-case1}
\end{figure}
Hence only IH of light neutrino masses is allowed. From Eq. (\ref{eq:r case 1}), we notice that $r$ depends on both $\alpha$ and $\beta$ and in the right panel of Fig. \ref{fig:ss13-alpha phiac=0}, we have plotted this dependence in this $\alpha-\beta$ plane for the best fit value of the ratio $r~ (=0.03)$~\cite{deSalas:2020pgw} only for IH. For the 3$\sigma$ allowed range for $\alpha$ (fitting $\sin\theta_{13}$),  obtained from the left panel of Fig. \ref{fig:ss13-alpha phiac=0}, $\beta$  found to be in the range $0.997-1.14$ for IH. The  contour plot for $r$ yields a one-to-one correspondence between $\alpha$ and $\beta$ as evident from the right panel of Fig. \ref{fig:ss13-alpha phiac=0}. For example,  the best fit value of $\sin\theta_{13}$ and $r$ fixes   $\alpha:1.752,\beta:1.067$ for IH.  With the known sets of $(\alpha,\beta)$ corresponding to  3$\sigma$  range of $\sin\theta_{13}$, we can calculate $|C|$ by fitting the best fit value of the solar mass-squared difference (given in Table \ref{table:neutrino osc data}) as $|C|=(2.04-1.68)\times 10^{-2}$ eV. Then, for the allowed sets of $(\alpha,\beta,|C|)$, we can estimate the   absolute neutrino masses $m_{1,2,3}$ and  their sum $\sum m_i$. In Fig. \ref{fig:mass-case1}, we have plotted the individual neutrino masses $m_{1,2,3}$ (by  blue dashed, red and orange lines respectively)  and  their sum $\sum m_i$ (green line) against   $\alpha$ as obtained from the left panel of Fig. \ref{fig:ss13-alpha phiac=0}. Now for IH,  cosmology sets an upper limit on the sum of the masses of three light neutrinos as $\sum m_i \leq 0.15$ eV~\cite{deSalas:2020pgw} as given by the horizontal  cyan shaded region in   the left panel of Fig. $\ref{fig:mass-case1}$.  Thus  all  $(\alpha,\beta)$  found from the the right panel of Fig. \ref{fig:ss13-alpha phiac=0}  are not allowed.   The vertical dashed line in the left panel of Fig. $\ref{fig:mass-case1}$ thus represents a further restriction on $\alpha$ (and hence on $\beta$) and the lower bound on $\alpha$ shifted from 1.6567 to 1.74. As a result  the corresponding values of $\beta$ and $|C|$ will also be shifted. The final  allowed values of the model parameters are summarized in Table. \ref{table:summary spec case I}.
\begin{table}[h!]
	\begin{center} \resizebox{0.32\textwidth}{!}{
		\begin{tabular}{c|c}
			\hline \hline
			Parameters & Allowed ranges \\
			\hline \hline
			$\alpha$ & 1.74-1.875 \\
			\hline
			$\beta$ & 1.059-1.147 \\
			\hline
			$|C|$ (eV) & (1.87-1.68)$\times 10^{-2}$ \\
			\hline
			$\sum {m_i}$ (eV) & 0.1496-0.1408 \\
			\hline
			$m_{\beta\beta}$ (eV) & 0.0568-0.0585 \\
			\hline 
		\end{tabular}}
	\end{center}
\caption{The allowed ranges for $\alpha$, $\beta$, $|C|$, $\sum m_i$ and $m_{\beta\beta}$  when $\phi_{AC} = \phi_{BC}=0$.}
	\label{table:summary spec case I}
\end{table}
Substituting  $\phi_{AC}=0$ in Eq. (\ref{eq:phi1,2,3 express}) we obtain $\phi_1=\phi_3=0$ and with $\phi_{BC}=0$ we get $\phi_2=0$. Altogether substituting these in Eq. (\ref{eq:majo1}) and Eq. (\ref{eq:majo2}), the Majorana phases are found to be zero. Thus for vanishing values of the relative phase $\phi_{AC}=0$ and $\phi_{BC}=0$, the Dirac and Majorana phases also vanishes. With these values of the phases and known sets of $(\alpha, \beta,  |C|)$ we can now finally  estimate the effective mass parameter $m_{\beta\beta}$ appearing in the neutrinoless double beta decay. In the right panel of Fig. \ref{fig:mass-case1}, we have plotted the prediction for $m_{\beta\beta}$ as a function of $\alpha$ with  $\phi_{AC}=\phi_{BC}=0$.  The finding for the sum of the absolute neutrino masses and the effective mass parameter as obtained from Fig. \ref{fig:mass-case1} are also summarized in Table \ref{table:summary spec case I}. 

\subsection{Case II :~$\phi_{AC}=0$}\label{subsec:2}
In the second case, we consider $\phi_{AC}=0$ but do not make any choice of $\phi_{BC}$. We already know that neutrino mixing angles and neutrino masses both depend on the phase $\phi_{AC}$ alone, which can be understood from the general expressions of Eq. (\ref{eq:phi and theta expre}) -  Eq. (\ref{eq:mixing angles expres2}) and Eq. (\ref{eq:real mass  m1}) - Eq. (\ref{eq:real mass  m3}). The Majorana phases $\alpha_{21}$ and hence neutrinoless double beta decay effective mass parameter $m_{\beta\beta}$ depend on both $\phi_{AC}$ and $\phi_{BC}$. Now, with the choice of $\phi_{AC}=0$, the simplified expressions for $\theta$ and $\sin\theta_{13}$ are same as in Eq. (\ref{eq:theta s13 exp phiac=0}). The real positive mass  eigenvalues and the ratio of the mass-squared differences  $r$ will take the form as in Eqs. (\ref{eq:mass case 1}) - (\ref{eq:r case 1}) respectively. Hence, with these expressions, the limits on $\alpha$, $\beta$, $|C|$, $m_i$ and $\Sigma m_i$ will same as in Case I ($\phi_{AC} = \phi_{BC} = 0$) which are summarized in Table \ref{table:summary spec case I}. Thus in this case too only IH of light neutrino masses is allowed.  The only difference between Case I and Case II is that  in the latter case $\phi_{BC}$ is free. As $\phi_{BC}$ only appears for Majorana phases, which can be understood just by looking at the Eq. (\ref{eq:phi1,2,3 express}), the prediction on Majorana phases will be different from the previous case and hence  prediction on $m_{\beta\beta}$ will change accordingly.  For $\phi_{AC}=0$, $\phi_1$ and $\phi_3$ are zero (obtained from Eq. (\ref{eq:phi1,2,3 express})) as mentioned in Case I. Hence, one of the Majorana phase $\alpha_{31}$ vanishes  but due to non-zero $\phi_2$, the other Majorana phase  ($\alpha_{21}$) found to be $\alpha_{21}=-\phi_{BC}$. Hence, we will calculate the neutrinoless double beta decay effective mass parameter $m_{\beta\beta}$ using Eq. (\ref{eq:ndbd exp}) by choosing different values for $\phi_{BC}$. 
\begin{table}[h!]
    \centering \resizebox{0.8\textwidth}{!}{
    \begin{tabular}{|c|c|c|c|c|c|c|}
    \hline
     $\phi_{BC}$&${\pi}/{6}$    & ${\pi}/{3}$ & ${\pi}/{2}$ & ${2\pi}/{3}$ & ${5\pi}/{6}$ & $\pi$   \\
     \hline
     $m_{\beta\beta}$ (eV)&  0.057-0.055  & 0.051-0.050 & 0.043-0.042 & 0.034-0.033 & 0.024-0.023 & 0.019-0.018 \\
     \hline
    \end{tabular}}
    \caption{Prediction on  $m_{\beta\beta}$ depending on different values of $\phi_{BC}$ in the Case II with $\phi_{AC}=0$.}
    \label{tab:case II}
\end{table}
In Table \ref{tab:case II} where we have summarized the ranges of $m_{\beta\beta}$ for different values of $\phi_{BC}$ such as  $\phi_{BC}={\pi}/{6}$, ${\pi}/{3}$, ${\pi}/{2}$, ${2\pi}/{3}$, ${5\pi}/{6}$ and $\pi$. Prediction on $m_{\beta\beta}$ in this case for $\phi_{BC}=0-\pi$ recurs the same value again for $\phi_{BC}=\pi-2\pi$.    From Case I and Case II, one can say that only IH is allowed with $\phi_{AC}=0$ (with $\phi_{BC}=0$ or arbitrary). 

\subsection{Case III: $\phi_{AC}=\phi_{BC}=\phi_x$}\label{subsec:3}
\begin{figure}[h!]
	\begin{center}
		\includegraphics[width=0.43\textwidth]{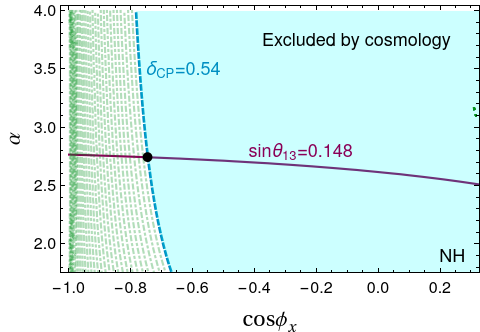}
	\end{center}
\caption{The contour plot for $\delta_{\rm CP}=0.78$ (blue dots) and best fit of $\sin\theta_{13}$ (red) for NH of neutrino masses. The region between red dots represents the $3\sigma$ region of $\sin\theta_{13}$ value of neutrino oscillation data for NH of neutrino masses. The intersection point represents the value of $\alpha$ and $\cos\phi_{x}$.}
\label{fig:alpha-cosphix}
\end{figure}

In this case, we consider the scenario when the relative phases $\phi_{AC}$ and $\phi_{BC}$ both are equal, say,  $\phi_{AC}=\phi_{BC}=\phi_x$. Hence, the general expressions for the rotation angle $\theta$ and associated phase $\phi$ appearing in the unitary rotation matrix $U_{13}$ as given in Eq. (\ref{eq:phi and theta expre}) can be rewritten as 
\begin{eqnarray}\label{eq:case 3 theta phi}
    \tan\phi=\frac{\alpha \sin\phi_{x}}{1-\alpha \cos\phi_{x}},\quad \tan2\theta=\frac{\sqrt{3}}{\cos\phi+2 \alpha \cos(\phi_x + \phi)}.
\end{eqnarray}
Thus we can substitute $\phi$ in the second equation above to evaluate $\sin\theta_{13}$ using Eq. (\ref{eq:mixing angles expres}). Furthermore as  $\tan\delta_{\rm CP}=\tan\phi$,  we find that  a particular  value of  $\delta_{\rm CP}$, $\sin\theta_{13}$ both depends on $\phi_{x}$ and $\alpha$.  In Fig. \ref{fig:alpha-cosphix} we provide contour plots for $\sin\theta_{13}=0.148$ and $\delta_{\rm CP}=0.541 $ ( or $31 ^{\circ}$) denoted by purple   and blue dotted lines respectively. The intersection between $\sin\theta_{13}$ and $\delta_{\rm CP}$ contours indicate the simultaneous satisfaction of them and Fig. \ref{fig:alpha-cosphix} it is indicated by a black dot  with which a pair of $\cos\phi_{x}$ and $\alpha$ are attached. Similar intersections of the green dot-dashed lines ($\delta_{\rm CP} \leq 0.541$) with the purple line represent other pairs  of $\cos\phi_{x}$ and $\alpha$.  
In this case the mass eigenvalues $m_1$, $m_2$ and $m_3$ can be written as
\begin{eqnarray}\label{eq:case3m1}
    m_1&=&|C|\big[(1-\alpha \cos\phi_x -P)^2+(Q+\alpha \sin\phi_x)^2\big]^{1/2}, \\
    m_2&=& |C|3\beta,  \label{eq:case3m2}\\
    m_3&=& |C| \big[(1-\alpha \cos\phi_x + P)^2+(Q-\alpha \sin\phi_x)^2 \big]^{1/2}\label{eq:case3m3}, 
\end{eqnarray}
and using these equations we can also evaluate the ratio of the mass-squared differences  $r$ using Eq. (\ref{eq:r}).  One can then compute $\beta$ with $r=0.03$ and the corresponding value for $|C|$ for each pair of  $\alpha$ and $\cos\phi_x$ obtained from intersecting points in Fig. \ref{fig:alpha-cosphix}. With this flavor structure of the  mass eigenvalues given in Eq. (\ref{eq:case3m1}) - Eq. (\ref{eq:case3m3}),  the inverted hierarchy of neutrino mass is not possible with $\phi_{AC}=\phi_{BC}$. For each set of $\alpha,\beta,\phi_x$ and $|C|$ estimated above, we can predict the  sum of the absolute neutrino mass $\sum m_i$ and in Table \ref{table:summary spec case III} we present few such representative values. From Eq. (\ref{eq:phi1,2,3 express}) - Eq. (\ref{eq:majo2}), we find that the Majorana phases $\alpha_{21}$ and $\alpha_{31}$ are defined as $\alpha_{21}=\phi_1-\phi_2$ and $\alpha_{31} = \phi_1-\phi_3$ where $\phi_1$, $\phi_2$ and $\phi_3$ in the present case can be written as \begin{eqnarray}\label{eq:phi1,2,3 case 3}
    \phi_1=\tan^{-1}\Big(\frac{Q+\alpha \sin\phi_x}{1-\alpha \cos\phi_x-P}\Big),\quad  \phi_2=\phi_x,\quad \phi_3=\tan^{-1}\Big( \frac{Q-\alpha \sin \phi_x}{1-\alpha \cos\phi_x +P}\Big). 
\end{eqnarray}
Substituting $\alpha$ and $\cos\phi_x$ obtained  from Fig. \ref{fig:alpha-cosphix} in the above equations we can  compute the Majorana phases and estimate the effective mass parameter appearing in the neutrinoless double beta decay as defined in Eq. (\ref{eq:ndbd exp}).
\begin{table}[h!]
	\begin{center}\resizebox{0.8\textwidth}{!}{
		\begin{tabular}{c c c c c c c| |c}
			\hline \hline 
			$\delta_{\rm CP}$ & $\alpha$ &  $\cos\phi_x$ & $\beta$ & $|C|$~(eV)& $\sum m_i $ (eV) & $m_{\beta\beta}$~(eV) & $m_{\beta\beta}\big|_{\phi_{BC}=0}$~(eV)\\
			\hline \hline 
			0.087 & 2.765 & -0.9848 & 0.5721 & 0.0084 & 0.0773 & 0.00589 & 0.0128 \\
			0.174 & 2.765 & -0.9659 &  0.5895 & 0.00858 & 0.0792 & 0.0077& 0.0148\\
	    	0.262 & 2.765 & -0.9327 & 0.6356 & 0.008956 & 0.0804 & 0.0100 & 0.0141\\	
            0.349 &  2.759 & -0.8853 & 0.7089 & 0.0096 & 0.0922 & 0.0144& 0.0108\\
 	    	0.436 & 2.754 & -0.8332 & 0.7849 & 0.0104 & 0.1026 & 0.0188 & 0.0188\\
	    	0.523 & 2.754 & -0.7669 & 0.872& 0.0117 & 0.1181 & 0.0252 & 0.0259\\
	    	\hline
			0.541 & 2.754 & -0.748 & 0.894 & 0.0122 & 0.123 & 0.0257 & 0.0251\\
			\hline 
		\end{tabular}}
	\end{center}
	\caption{The allowed ranges of parameters satisfying neutrino  data for various values of $\delta_{\rm CP}$ with $\phi_{AC}=\phi_{BC}$. The last column represents prediction for $m_{\beta\beta}$ for  $ \phi_{AC}=0 \div 2\pi$ and $\phi_{BC}=0$. 
	}
	\label{table:summary spec case III}
\end{table}
These predictions for $m_{\beta\beta}$ are listed in the last column of Table \ref{table:summary spec case III}. From the results obtained form Fig. \ref{fig:alpha-cosphix} and Table \ref{table:summary spec case III}, we find that all values for  $\delta_{\rm CP}$ are not compatible with the cosmological bound on the sum of neutrino masses $\sum m_i$ (satisfying correct neutrino oscillation data). The region  $\delta_{\rm  CP} \leq 0.54~ (31^{\circ})$ given by the cyan shaded region is disallowed due to the cosmological upper bound $\sum m_i \leq 0.12$ eV for NH~\cite{deSalas:2020pgw}. Thus the green dot-dashed region in Fig. \ref{fig:alpha-cosphix} with $0<\delta_{\rm CP}\leq \pi/6$ satisfy   both neutrino oscillation data and the cosmological bound on the sum of neutrino masses.  In  Table \ref{table:summary spec case III} we have summarized constraints on $\alpha$, $\beta$, $\cos\phi_x$ and $|C|$ for a specific  value of $\delta_{\rm CP}$ as well as predictions on $\sum m_i$ and $m_{\beta\beta}$. Here we also find that predictions for the  parameters   given in Table \ref{table:summary spec case III} repeats again for $\pi<\delta_{\rm CP}\leq 7\pi/6$.  From the results summarized in the Table \ref{table:summary spec case III}  one can find that with increase of  ($\beta$, $\cos\phi_x$, $|C|$),   $\alpha$ decreases  whereas both $\sum m_i$, $m_{\beta\beta}$ increases  with the increase of $\delta_{\rm CP}$.

\subsection{Case IV: $\phi_{BC}=0$}\label{subsec:4}
In this case we consider $\phi_{BC}=0$ while  $\phi_{AC}$ is varied arbitrarily between $0-2\pi$. This scenario is very similar to the previous case.  with $\phi_{AC}=\phi_{BC}=\phi_x$. From Eq. (\ref{eq:phi1,2,3 express})$-$(\ref{eq:majo2}) we find that $\phi_{BC}=0$ only effects the Majorana phases  and hence the  effective mass parameter appearing in the neutrinoless double beta decay.    The conclusion drawn in Case III for absolute values of  light neutrino masses and their   hierarchy will be identical. For simplicity and resembles to the previous case, we consider $\phi_{AC}=\phi_x$.  With this, the expressions for rotation angle $\theta$ and the phase $\phi$ appearing in $U_{13}$ are already given in Eq. (\ref{eq:case 3 theta phi}). Using Eq. (\ref{eq:mixing angles expres}) we also find that $\delta_{\rm CP}$ and $\sin\theta_{13}$ are both function of  $\alpha$ and $\phi_x$. For  a particular value of $\delta_{\rm CP}$ we can again compute the pairs of $\alpha$ and $\phi_x$  from the intersections of contour plots for the chosen value of $\delta_{\rm CP}$ and best-fit value of $\sin\theta_{13}$ as given in  Fig. \ref{fig:alpha-cosphix}.  Similar to Case III, by fitting $r=0.03$, we can get $\beta$ (and subsequently $|C|$), where the expression of $r$ in Eq. (\ref{eq:r}) involves neutrino masses those are given in Eq. (\ref{eq:case3m1}) - Eq. (\ref{eq:case3m3}). Therefore the results of Table \ref{table:summary spec case III} will be same for this case up-to the sixth column for $\sum m_i$. The change will occur in the seventh column of Table \ref{table:summary spec case III} as here in Case IV, we have $\phi_{BC}=0$. Although, with this choice, the expression for $\phi_1$ and $\phi_3$ will be the same as Eq. (\ref{eq:phi1,2,3 case 3}) but unlike the previous case we have $\phi_2=0$ here. The two Majorana phases follow the relation $\alpha_{21}=\phi_1-\phi_2$,  $\alpha_{31}=\phi_1-\phi_3$ hence $\alpha_{31}$ will be identical  and $\alpha_{21}$ will be different from Case III.  As a result, $m_{\beta\beta}$ will be different compared to the previous scenario. Hence in the last column of  Table \ref{table:summary spec case III} we  append the prediction of $m_{\beta\beta}$  for a few allowed values for  $\delta_{\rm CP}$. Note that here also the allowed ranges of the  Dirac CP phase are  $0<\delta_{\rm CP}\leq \pi/6$ and $\pi<\delta_{\rm CP}\leq 7\pi/6$ respectively which satisfy   both neutrino oscillation data and the cosmological bound on the sum of absolute neutrino masses. A point to remember is that this case only satisfies  NH of neutrino mass and IH is not allowed. 
\begin{figure}[h]
	\begin{center}
		\includegraphics[width=.43\textwidth]{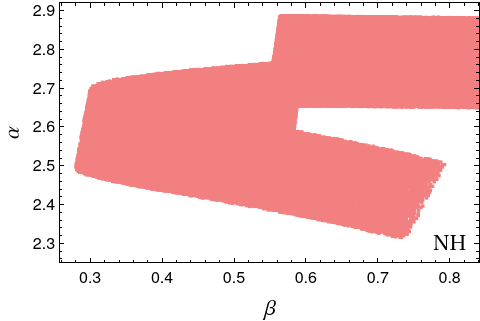}
		\includegraphics[width=.43\textwidth]{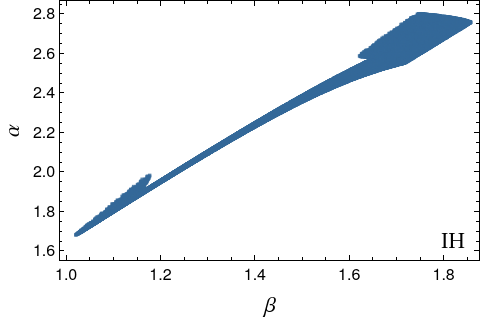}
\end{center}
\caption{The allowed regions for $\alpha$ and $\beta$ for the $3\sigma$ ranges for $\sin^2\theta_{13}$ and $r$ for NH (left panel, light red shaded region) and IH (right panel, blue shaded region) of neutrino masses where $\phi_{AC}$ and $\phi_{BC}$ are varied from $0$ to $2\pi$.}
\label{fig:alpha-beta}
\end{figure}


\subsection{General case}\label{subsec:5}
In the above cases, we have analyzed neutrino mixing for various limiting values for the relative phases associated with our study.   Now, we will carry out a full  numerical analysis for the most general case where we vary both $\phi_{AC}$ and $\phi_{BC}$ to their entire range from $0$ to $2\pi$. Then, ~ Eq. (\ref{eq:phi and theta expre}) - Eq. (\ref{eq:mixing angles expres2}) will be used for the calculation of mixing angles. On the other hand, the ratio of the mass-squared differences $r$ defined in Eq. (\ref{eq:r}) can also be calculated using the general expression for the mass eigenvalues of Eq. (\ref{eq:real mass  m1}) - Eq. (\ref{eq:real mass  m3}). 
\begin{figure}[h]
	\begin{center}
		\includegraphics[width=.43\textwidth]{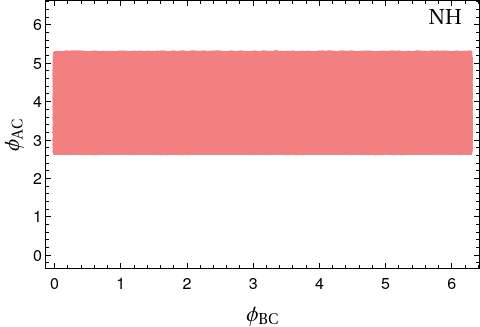}
		\includegraphics[width=.43\textwidth]{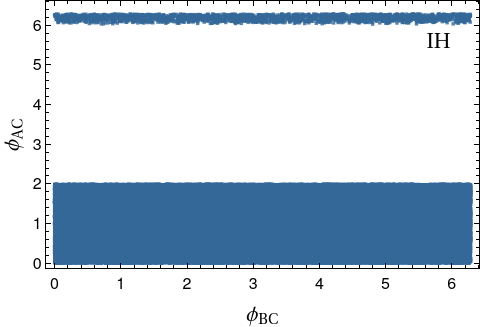}
	\end{center}
\caption{The allowed ranges for $\phi_{AC}$ and $\phi_{BC}$ for the correct value of $\sin^2\theta_{13}$ and $r$ along with for both NH (left panel) and IH (right panel) of neutrino masses.}
\label{fig:phiAC-phiBC}
\end{figure}
As explained earlier, in order to evaluate the absolute neutrino masses, we also need to evaluate the common factor $|C|$ associated with each mass eigenvalue. Here we obtain $|C|$ by fitting the solar mass-squared difference taken from~\cite{deSalas:2020pgw}. In our analysis for this most general case,  we have also included the bound coming from cosmological observations on the sum of absolute  neutrino masses as  $\sum m_i<0.12$ eV for NH and $\sum m_i<0.15$ eV for IH~\cite{deSalas:2020pgw}. Using the $3\sigma$ allowed range for the neutrino oscillation data~\cite{deSalas:2020pgw} given in Table \ref{table:neutrino osc data}, we vary both  $\phi_{AC}$ and $\phi_{BC}$ between  0 to $2\pi$. In Fig. \ref{fig:alpha-beta} we have plotted  the allowed region in  the $\alpha-\beta$ plane for NH (left panel, light red shaded region) and IH (right panel, blue shaded region) respectively.  Here we find that for NH (IH), the allowed ranges for $\alpha$ vary between $2.33\leq \alpha \leq 2.9$ ($1.65\leq \alpha \leq 2.8$). On the other hand,  allowed ranges for $\beta$ are restricted by $0.3 \leq \beta \leq 1$  for NH whereas for IH we have $1.9\geq \beta \geq 1$. This implies $\beta$ values less than 1 are favored for NH whereas values greater than 1 are favored for IH.    Similarly, in Fig. \ref{fig:phiAC-phiBC} we have plotted the allowed region in  $\phi_{AC}-\phi_{BC}$ plane for NH (left panel, light red shaded region) an IH (right panel, blue shaded region) respectively.  Here $\phi_{BC}$ between $0-2\pi$ is compatible for  both of the hierarchies,  however, two distinct regions for $\phi_{AC}$ are allowed, namely, $2.61\leq \phi_{AC} \leq 5.38$ for NH and $\phi_{AC} \leq 2$ $\&$ $6\leq \phi_{AC} \leq 2\pi$ for IH respectively. Clearly, the full range of $\phi_{BC}$  is allowed because it is not sensitive to low energy masses, and mixing and only appears in one of the Majorana phase $\alpha_{21}$. Thus the values of $\phi_{AC}$ crucially dictate the hierarchy of light neutrino masses.   As an artifact of the considered flavor symmetry,  combining the results from Fig. \ref{fig:alpha-beta} and \ref{fig:phiAC-phiBC}, we can conclude that with $\beta \leq 1$ and $2.61\leq \phi_{AC} \leq 5.38$ one can reproduce NH whereas to obtain IH we need $\beta \geq 1$ and $0 \leq \phi_{AC} \leq 2$ (or $6\leq \phi_{AC} \leq 2\pi$). 

\begin{figure}[h]
	\begin{center}
		\includegraphics[width=.43\textwidth]{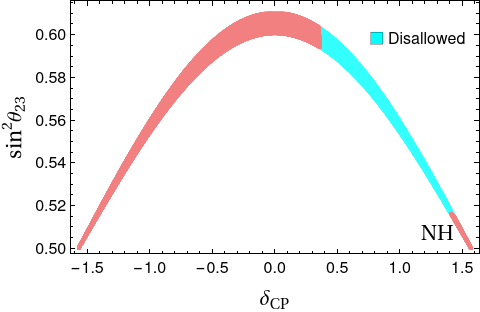}
		\includegraphics[width=.43\textwidth]{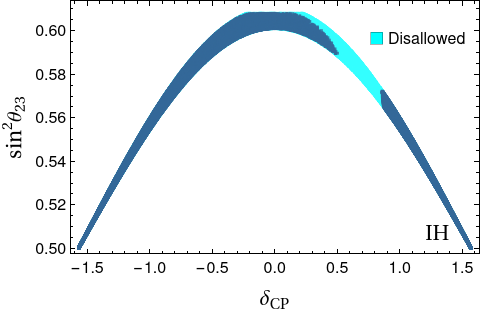}
	\end{center}
	\caption{ $\sin^2\theta_{23} - \delta_{\rm CP}$ correlation for NH (left panel) and IH (right panel).
	The cyan patch represents the disallowed region coming from the constraints on light neutrino masses. }\label{fig:ss23-dCP}
\end{figure}

Now with the allowed values for $\alpha,\beta,\phi_{AC}$ and $\phi_{BC}$ obtained from Fig. \ref{fig:alpha-beta} and \ref{fig:phiAC-phiBC}, we are now equipped to study the correlation between  neutrino mixing parameters and predictions associated with the phases and masses.  Due to the presence of the $A_4$ discrete flavor symmetry, this model  yields an interesting correlation among the observables   appearing in neutrino mixing. Following Eq. (\ref{eq:phi and theta expre}) - Eq. (\ref{eq:mixing angles expres2}), we find one  such important correlation between the atmospheric mixing angle $\theta_{23}$ and Dirac CP phase $\delta_{\rm CP}$. This correlation  is very crucial because still there are some unsettled issues  with the measurement of these two oscillation parameters such as $(a)$ octant of $\theta_{23}$, i.e., $\theta_{23} < 45^{\circ}$ (lower octant, LO) or $\theta_{23} > 45^{\circ}$ (higher octant, HO) and $(b)$ magnitude of Dirac CP phase $\delta_{\rm CP}$.  The   $\delta_{\rm CP} - \theta_{23}$ correlation obtained here is plotted in Fig. \ref{fig:ss23-dCP} and given by  light red (blue) shaded region for NH (IH) in the left (right) panel and  shades some light on the above mentioned  unsettled issues. It is evident from  Fig. \ref{fig:ss23-dCP} that for both hierarchies only higher octant of $\theta_{23}$ is  favoured ($i.e. ~\theta_{23} \geq 45^{\circ}$) in our   analysis. Furthermore, the cyan patch in both of the panels represents the disallowed region for $\delta_{\rm CP}$ in order to satisfy the limits on light neutrino masses~\cite{deSalas:2020pgw}.  From Fig. \ref{fig:ss23-dCP} the  allowed regions for Dirac CP phase $\delta_{\rm CP}$ are given by $-1.57 \leq \delta_{\rm CP} \leq 1.37 $ and $1.4 \leq \delta_{\rm CP} \leq 1.57 $ for NH, whereas for IH the predictions are $-1.57 \leq \delta_{\rm CP} \leq 0.5 $ and $0.86 \leq \delta_{\rm CP} \leq 1.57 $.  The disallowed region for $\delta_{\rm CP}$ is small in IH of neutrino masses compared to the NH as  cosmology puts a tighter constraint on  NH compared to IH ~ \cite{deSalas:2020pgw}.
\begin{figure}[h]
	\begin{center}
		\includegraphics[width=0.43\textwidth]{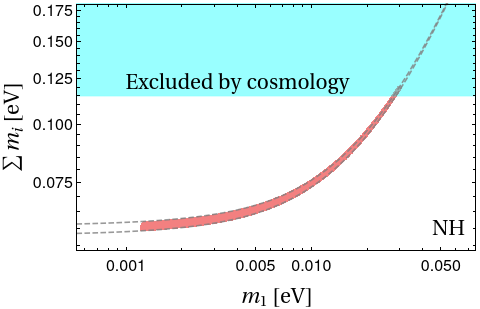}
		\includegraphics[width=0.43\textwidth]{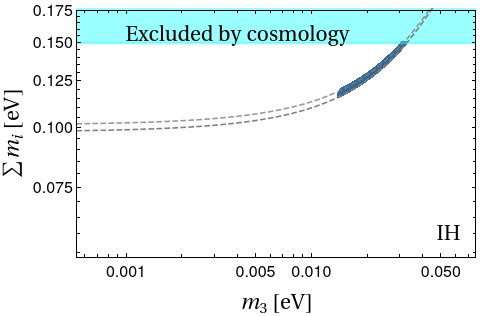}
	\end{center}
\caption{ Sum of  absolute neutrino masses $\sum m_i$ plotted against lightest neutrino mass for both NH (left panel, light red shaded region) and IH (right panel, blue shaded region). The area between the dashed lines represents $3\sigma$ allowed range  and the cyan patch represents the area excluded by cosmology.}
\label{fig:mtotal}
\end{figure}

\begin{figure}[h!]
	\begin{center}
		\includegraphics[width=.43\textwidth]{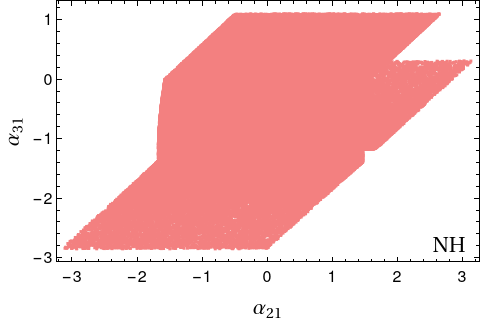}
		\includegraphics[width=.43\textwidth]{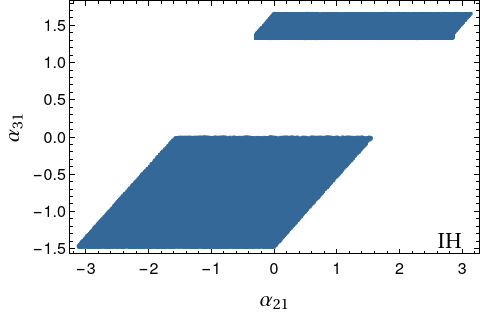}
	\end{center}
\caption{The correlation plot between two Majorana phases $\alpha_{21}$ and $\alpha_{31}$  for NH (left panel) and IH (right panel)  for $3\sigma$ allowed range of neutrino oscillation data~\cite{deSalas:2020pgw}.}
\label{fig:alpha21-alpha31}
\end{figure}

Again, with the permitted values of $\alpha$, $\beta$, $\phi_{AC}$ and $\phi_{BC}$, we have the predictions on light neutrino masses which can be understood from the correlation plot of $\sum m_i$ vs lightest neutrino ($m_1$ for NH and $m_3$ for IH) mass for both hierarchies  as given in Fig. \ref{fig:mtotal}. Here also the allowed regions are given by light red (blue) shaded region for NH (IH) in the left (right) panel. The horizontal cyan region in each plot represents the disallowed regions mentioned earlier in this subsection.  Clearly, this framework  predicts that the  lightest  neutrino mass
can take smaller values for NH ($m_{\rm lightest} \geq 0.0012$ eV), compared to the IH  scenario ($m_{\rm lightest} \geq 0.014$ eV). 
\begin{figure}[h!]
	\begin{center}
	\includegraphics[width=1\textwidth]{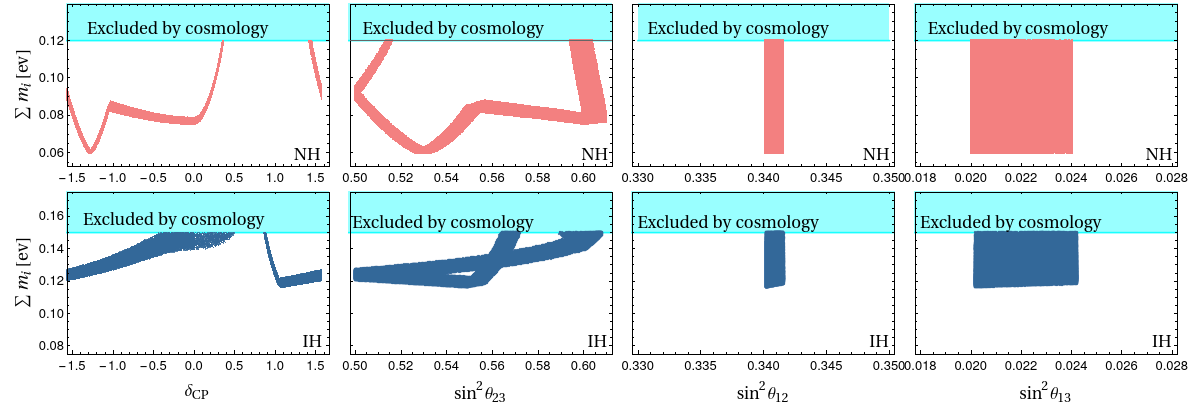}
	\end{center}
\caption{The correlation plot between $\sum m_i$ and  $\delta_{\rm CP}$, $\sin^2_{23}$, $\sin^2_{12}$, $\sin^2_{13}$. The upper panels are schematics for NH whereas    the lower panel is for IH and  the cyan patch represents the area excluded by cosmology. }
\label{fig:summ-others}
\end{figure}
In Fig. \ref{fig:ss23-dCP}, we showed  the $\sin^2\theta_{23} - \delta_{\rm CP}$ correlation which is generic feature for the TM$_2$ mixing. Now to elucidate the additional predictions which go beyond    TM$_2$ mixing in FSS, we  present a few additional  correlations among neutrino masses and mixing. Hence  in Fig. \ref{fig:summ-others} we plot the correlation between the sum of absolute neutrino masses
$\sum m_i$ and  other observables such as $\delta_{\rm CP}$, $\sin^2\theta_{23}$, $\sin^2\theta_{12}$ and $\sin^2\theta_{13}$ in the FSS framework. Here the upper panel with light red shaded regions represents the allowed parameter space for NH whereas the lower panel with blue shaded regions represents the allowed parameter space for IH and the cyan patch represents the area excluded by cosmology. Here from the $\sum m_i - \delta_{\rm CP}$ (first column of Fig. \ref{fig:summ-others}) it is clear that \emph{the bound on the absolute neutrino masses disallows some regions of the Dirac CP phase and $\sum m_i$- $\sin^2\theta_{23}$, $\sin^2\theta_{12}$ correlations are characteristics signature of this model.} From the third column of Fig. \ref{fig:summ-others} we find that $\sin^2\theta_{12}$ in FSS is restricted within a narrow range (the plot corresponds to the 3$\sigma$ allowed range). 
The Majorana phases in our analysis, for the most general case, can be evaluated using the expressions given in Eq. (\ref{eq:majo1}) and (\ref{eq:majo2}) with  the  allowed regions of $\alpha$, $\phi_{AC}$, $\phi_{BC}$ given in Fig. \ref{fig:alpha-beta} and \ref{fig:phiAC-phiBC}. Thus we can  constrain the Majorana phases using the low-energy neutrino oscillation data.  In Fig. \ref{fig:alpha21-alpha31},  present a correlation plot  in the $\alpha_{21} - \alpha_{31}$  plane  for NH (left panel, light red shaded region) and IH (right panel, blue shaded region) respectively with  $3\sigma$ allowed ranges of neutrino oscillation data~\cite{deSalas:2020pgw}. 
\begin{figure}[h!]
	\begin{center}
    \includegraphics[width=.43\textwidth]{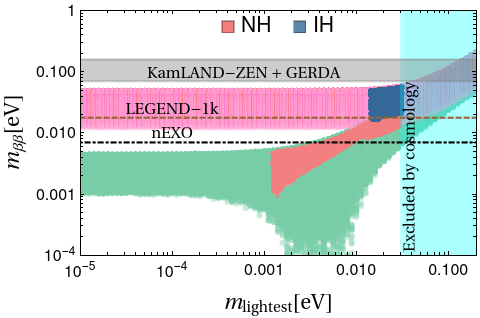}
	\end{center}
\caption{Neutrinoless double beta decay effective mass parameter $m_{\beta\beta}$ is plotted as a function of  the lightest neutrino mass  for the case of both NH (light red shaded region) and IH (blue shaded region). Green and magenta regions are three neutrino-allowed regions when all the parameters are varied within their 3$\sigma$ range. The gray-shaded region, brown dashed, and black dotted-dash lines stand for experimental limits by KamLAND-Zen+GERDA, LEGEND-1k, and nEXO respectively. The vertical cyan area represents a disallowed region by cosmology. }
\label{fig:mee-beta}
\end{figure}

Finally, with the estimation for neutrino masses and phases in hand, we are now able to plot  the effective mass parameter characterizing neutrinoless double beta decay ($m_{\beta\beta}$) given  in Eq. (\ref{eq:ndbd exp}) and Eq. (\ref{eq:ndbd exp2}). In Fig. \ref{fig:mee-beta}, we have   plotted  $m_{\beta\beta}$ against the lightest neutrino mass for both NH ($m_1$) and IH ($m_3$) respectively by light red  and blue shaded  regions respectively. The predictions for $m_{\beta\beta}$ are $1-30$ meV for NH and 16-60 meV for IH.  Here the green and magenta shaded regions represent $3\sigma$ allowed regions for the $m_{\beta\beta}$ predictions for NH and IH respectively. The vertical cyan-shaded regions represent the cosmological upper limit on the sum of absolute neutrino masses ($\sum m_{i}$).  The gray shaded region represents the upper limit for $m_{\beta\beta}$ by combined analysis of KamLAND-Zen~\cite{KamLAND-Zen:2016pfg} and GERDA~\cite{GERDA:2018pmc}   experiments and predictions for $m_{\beta \beta}$ in our model  fall within this upper limit.  
\begin{figure}[h!]
	\begin{center}
	\includegraphics[width=1\textwidth]{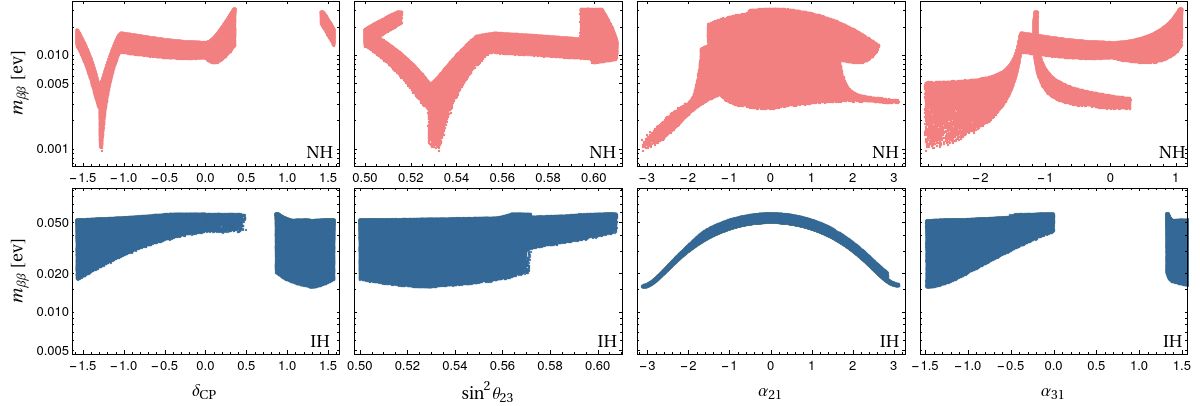}
	\end{center}
\caption{The correlation plot between $\sum m_i$ and  $\delta_{\rm CP}$, $\sin^2_{23}$, $\alpha_{21}$, $\alpha_{31}$. The upper panels are schematics for NH whereas    the lower panel is for IH. }
\label{fig:mbb-others}
\end{figure}
\begin{table}[h]
       \begin{center} \resizebox{0.9\textwidth}{!}{
		\begin{tabular}{l ||c |c |c |c |c |c| c}	
		\hline \hline
          \rule{0pt}{4ex}		Cases &  ~NH~$~$ & ~IH~~~~~$~$ & $\delta_{\rm CP}$ & $\alpha_{21}$ & $\alpha_{31}$ & $\sum m_i$(eV) & $m_{\beta\beta}$(eV)   \\ 
		\hline\hline
            Case I       & \ding{55}& \checkmark& 0,$\pi$& 0  &0  & (0.1408, 0.1496) & (0.057$, $0.059)                   \\
			\hline
             Case II  & \ding{55} & \checkmark& 0,$\pi$ &(0,2$\pi$) & 0 & (0.1408-0.1496)& (0.019-0.054)                \\
			\hline
			\multirow{2}{*}{Case III} & \multirow{2}{*}{\checkmark} & \multirow{2}{*}{\ding{55}} & (0-0.78) & \multirow{2}{*}{(-2.362,-1.26)} & \multirow{2}{*}{(0.61,1.20)} & \multirow{2}{*}{(0.0773,0.12)} & \multirow{2}{*}{(0.0059,0.026)} \\ 
			&                   &                   & (3.14,3.66) &                   &                   &                   &                  \\ \hline
			\multirow{2}{*}{Case IV} & \multirow{2}{*}{\checkmark} & \multirow{2}{*}{\ding{55}} &(0,0.78)  & \multirow{2}{*}{(0.60,1.84)} & \multirow{2}{*}{(0.61,1.20)} & \multirow{2}{*}{(0.0773,0.12)} & \multirow{2}{*}{(0.013,0.025)} \\
			&                   &                   & (3.14,3.66) &                   &                   &                   &                   \\
			\hline
			\multirow{2}{*}{General Case} & \multirow{2}{*}{\checkmark} & \multirow{2}{*}{\ding{55}} &(-1.5,0.6)  & \multirow{2}{*}{(-3,3)} & \multirow{2}{*}{(-2.85,1.13)} & \multirow{2}{*}{(0.06,0.12)} & \multirow{2}{*}{(0.001,0.03)} \\
			&                   &                   & (1.38,1.57) &                   &                   &                   &                   \\
			\hline
			\multirow{2}{*}{General Case} & \multirow{2}{*}{\ding{55}} & \multirow{2}{*}{\checkmark} & (-1.5,0.6) & \multirow{2}{*}{(-3,3)} &    (-1.5,-0.01)               & \multirow{2}{*}{(0.115,0.15)} & \multirow{2}{*}{(0.016,0.06)} \\
			&                   &                   & (0.8,1.56) &                   &                   (1.33,1.7)&                   &             \\
			\hline
		\end{tabular}}
		\caption{Summary of the different cases of our numerical analysis. The \checkmark  and \ding{55} symbols stand for allowed and disallowed regimes. The numbers in the parenthesis represent the allowed ranges in each scenario. }\label{tab:allsummary}
		\end{center}
	\end{table}
In this plot the brown dashed and black dotted-dash lines stand for future sensitivities of the LEGEND-1k~\cite{LEGEND:2021bnm} and nEXO~\cite{nEXO:2021ujk} experiments respectively. Thus these near-future experiments have the potential to almost entirely  falsify the IH prediction and probe  a major part prediction for $m_{\beta\beta}$ for NH of light neutrino mass. Guided by the symmetry construction, the model also sets a lower limit on the effective mass parameter  as $m_{\beta\beta} \geq 1$ meV for NH.  Similar to Fig. \ref{fig:summ-others}, to obtain additional predictive correlations among neutrino masses and mixing, we plot a few more schematics. Thus in Fig. \ref{fig:mbb-others} we present the dependence of $m_{\beta\beta}$ on $\sin^2\theta_{23},\delta_{\rm CP}, \alpha_{21}$ and $\alpha_{31}$ for the FSS framework. Here the upper panel with light red shaded regions represents the allowed parameter space for NH. The lower panel with blue shaded regions represents the allowed parameter space for IH. Together with Fig. \ref{fig:summ-others}, the correlations between $\sum m_{\beta\beta}$ - $\delta_{\rm CP}$, $\sin^2_{23}$, $\alpha_{21}$, $\alpha_{31}$ presented in Fig. \ref{fig:mbb-others} are typical features of FSS. The fate of the present model crucially depends on these correlations.   

To sum up the full numerical analysis, we present Table \ref{tab:allsummary}, where we give a summary of all the results including both special cases and general cases. In our analysis, we divided our special case into four categories depending on the values of the input relative phases $\phi_{AC}$ and $\phi_{BC}$. In Case I, where $\phi_{AC}=\phi_{BC}=0$, only IH of neutrino masses are allowed and $\delta_{CP}$ = $0$ or $\pi$, and two Majorana phases are coming out to be zero. In case II, where $\phi_{AC}=0$ but $\phi_{BC}\neq 0$, IH is predicted and  the main difference occurs in the prediction of Majorana phases. As a result, the prediction on $m_{\beta\beta}$ is different from Case I. In both Case III and Case IV where $\phi_{AC}=\phi_{BC}\neq 0$ and $\phi_{AC}\neq 0, \phi_{BC}=0$, respectively, NH of neutrino masses are predicted. The allowed regions of $\delta_{\rm CP}$ are given in Table \ref{tab:allsummary} which are the same for both these cases. The prediction on the Majorana phase $\alpha_{31}$ is the same for both of these cases  whereas the prediction on the $\alpha_{21}$ and hence $m_{\beta\beta}$ are different in both cases. Finally, as a most general case study, in Case V we vary $\phi_{AC}$ and $\phi_{BC}$ arbitrarily within its full range.  The analysis constrains $\delta_{\rm CP}$ into two particular regions. We also find the values of the parameters $\beta$ and the phase $\phi_{AC}$ plays a crucial role in determining the neutrino mass hierarchy with distinct limits on neutrino masses for each hierarchy. 

\section{Phenomenological implications for the FSS model}\label{sec:pheno}

Owing to the flavor symmetry of the model, charged lepton sector's Yukawa couplings   are diagonal so the flavors are  conserved. But there are sources of the lepton flavor violation arising  outside the charged lepton sector from both the Yukawa couplings $y_N$ and $y_s$ associated with the seesaw and scotogenic contributions, respectively. These Yukawa interactions lead to lepton flavor violating processes such as $l_{\alpha}\rightarrow l_{\beta} \gamma$, $l_{\alpha}\rightarrow 3 l_{\beta}$ ($\alpha,\beta=e,\mu,\tau$) etc. For related studies on lepton flavor violation in a pure scotogenic model  see~\cite{Toma:2013zsa,Vicente:2014wga,Hagedorn:2018spx}. Studies of such lepton flavor-violating processes in our framework depend heavily on the  proposed symmetry configuration as described below.  

In our framework, the branching ratios of the  $l_{\alpha}\rightarrow l_{\beta}\gamma$ decays for the scotogenic contribution can be written as~\cite{Rojas:2018wym,Toma:2013zsa}
\begin{eqnarray}
    {\rm Br}(l_{\alpha} \rightarrow l_{\beta}\gamma) \approx \frac{3 \pi\tilde{\alpha}}{64 G_F^2}|Y_F^{\beta *}Y_{F}^{\alpha}|^2 \frac{1}{m_{\eta^+}^4}\Bigg(F\Bigg(\frac{M_f^2}{m_{\eta^+}^2}\Bigg)\Bigg)^2 {\rm Br}(l_{\alpha} \rightarrow l_{\beta}\nu_{\alpha}\bar{\nu}_{\beta}).
\end{eqnarray}
Here $G_F$ is the Fermi constant, $\tilde{\alpha}=e^2/4\pi$ is the fine structure constant. $Y_F$ is the Yukawa coupling matrix coming from the scotogenic contribution given in Eq. (\ref{eq: Yukawa for scoto}). The expression for the function $F$ is given by
\begin{eqnarray}
    F(x)=\frac{1-6x-3x^2+2 x^3-6 x^2 {\rm log}x}{6(1-x)^4}.
\end{eqnarray}
In our discussion,  considered discrete symmetries dictate the structure of the associated Yukawa couplings. Due to the specific VEV alignment of the $A_4$ triplet flavon $\phi_s$ and its contraction (following the multiplication rules given in the appendix) with the non-trivial $A_4$ singlet $\xi$ (charged as $1'$ ), we find  $Y_F^{\mu}=0$ as given in Eq. (\ref{eq: Yukawa for scoto}).  Therefore owing to the $A_4$ symmetry, the scotogenic part alone yields  a vanishing contribution in the lepton flavor violating decays for $\mu \rightarrow e\gamma$ and $\tau \rightarrow \mu\gamma$. The only non-vanishing contribution arising in the $l_{\alpha} \rightarrow l_{\beta}\gamma$ decays originates from  the $\tau \rightarrow e\gamma$ decay and  the branching fraction can be written as
\begin{eqnarray}\label{eq:scoto tau decay e gamma}
    {\rm Br}(\tau \rightarrow e\gamma) &\approx& \frac{3 \pi\tilde{\alpha}}{64 G_F^2}|-y_s y_s^*\epsilon^4|^2 \frac{1}{m_{\eta^+}^4}\Bigg(F\Bigg(\frac{M_f^2}{m_{\eta^+}^2}\Bigg)\Bigg)^2 {\rm Br}(\tau \rightarrow e\nu_{\tau}\bar{\nu}_{e}), \\
    &=& \frac{3 \pi\alpha}{64 G_F^2}\Bigg(\frac{|C|}{\mathcal{F}(m_{\eta_R},m_{\eta_I},M_f)}\Bigg)^2 \frac{1}{m_{\eta^+}^4}\Bigg(F\Bigg(\frac{M_f^2}{m_{\eta^+}^2}\Bigg)\Bigg)^2 {\rm Br}(\tau \rightarrow e\nu_{\tau}\bar{\nu}_{e})
\end{eqnarray}
In the above $\epsilon=v_f/\Lambda$ where we assume flavons VEVs to be equal, $i.e.$, $v_{\xi}=v_{s,a}=v_f$. There is also possible another type of the flavor violating decay   $l_{\alpha}\rightarrow 3 l_{\beta}$ ($l_{\alpha}\rightarrow l_{\beta}\bar{l}_{\beta}l_{\beta}$) and the corresponding branching ratio  is given by~\cite{Toma:2013zsa}
\begin{eqnarray}
    {\rm Br}(l_{\alpha}\rightarrow 3 l_{\beta})\approx \frac{3\tilde{\alpha}^2}{512 G_F^2}|Y_F^{\beta ^*}Y_F^{\alpha}|^2 \frac{1}{m_{\eta^+}^4}\mathcal{G}\Big(\frac{m_{\alpha}}{m_{\beta}}\Big)\Bigg(F\Bigg(\frac{M_f^2}{m_{\eta^+}^2}\Bigg)\Bigg)^2 {\rm Br}(l_{\alpha} \rightarrow l_{\beta}\nu_{\alpha}\bar{\nu}_{\beta}).
\end{eqnarray}
where
\begin{eqnarray}
    \mathcal{G}\Big(\frac{m_{\alpha}}{m_{\beta}}\Big)=\Big(\frac{16}{3}{\rm log}\Big(\frac{m_{\alpha}}{m_{\beta}}\Big)-\frac{22}{3}\Big).
\end{eqnarray}
Since in FSS we have $Y_F^{\mu}=0$, then  branching fractions for $\mu \rightarrow 3e$ and $\tau \rightarrow 3\mu$ decays coming through the scotogenic contribution also vanishes. The only non-vanishing contribution originates from  the $\tau \rightarrow 3e$  decay, and the branching fraction  can be written as 
\begin{eqnarray}\label{eq:scoto tau decay 3e}
 {\rm Br}(\tau\rightarrow 3 e)&\approx & \frac{3\tilde{\alpha}^2}{512 G_F^2}  |-y_s y_s^*\epsilon^4|^2 \frac{1}{m_{\eta^+}^4}\mathcal{G}\Big(\frac{m_{\tau}}{m_{e}}\Big)\Bigg(F\Bigg(\frac{M_f^2}{m_{\eta^+}^2}\Bigg)\Bigg)^2 {\rm Br}(\tau \rightarrow e\nu_{\tau}\bar{\nu}_{e}). 
\end{eqnarray}
\begin{figure}
    \centering
    \includegraphics[width=0.42\textwidth]{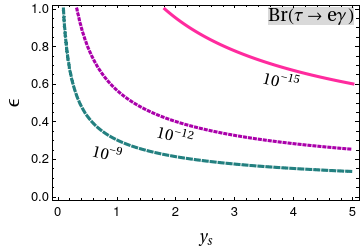}
    \includegraphics[width=0.42\textwidth]{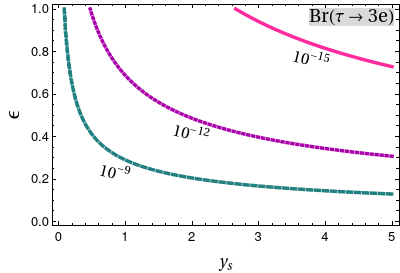}
    \caption{Contour plots for a branching fraction of $\tau \rightarrow e\gamma$ (left panel) and $\tau \rightarrow 3e$ (right panel) in the $y_s - \epsilon$ plane. The dashed, dotted, and  continuous lines represent the branching fraction to be $10^{-9}$, $10^{-12}$, and $10^{-15}$ in both panels.}
    \label{fig:ys-ep-br}
   \end{figure}
Clearly, from Eq. ({\ref{eq:scoto tau decay e gamma}}) and ({\ref{eq:scoto tau decay 3e}}) we find   with  fixed values of the mass parameters that ${\rm Br}(\tau \rightarrow e\gamma)$ and ${\rm Br}(\tau\rightarrow 3 e)$ depend upon $y_s$ as well as $\epsilon$ (the ratio of flavon VEVs $v_f$ to the cut-off scale $\Lambda$). Hence in Fig. \ref{fig:ys-ep-br} we present contour plots for the corresponding branching fractions  in the $y_s$-$\epsilon$ plane considering $m_{\eta^+} = 600$ GeV and $M_f=10$ TeV. The near future sensitivity of these two branching ratios is of the order of $10^{-9}$ \cite{Aushev:2010bq}. Therefore  we have plotted contours for  the branching fraction of $\tau \rightarrow e\gamma$  (left panel) and $\tau \rightarrow 3e$ (right panel) fixed at $10^{-9}$, $10^{-12}$ and $10^{-15}$ given by the dashed, dotted and continuous lines respectively.  The  $y_s$-$\epsilon$ correlation in Fig. \ref{fig:ys-ep-br} also helps us to estimate the ratio $\epsilon$, and we  find $\epsilon \leq 1$ since it is suppressed by the cut-off scale of the theory.

\begin{figure}[h]
    \centering
    \includegraphics[width=0.42\textwidth]{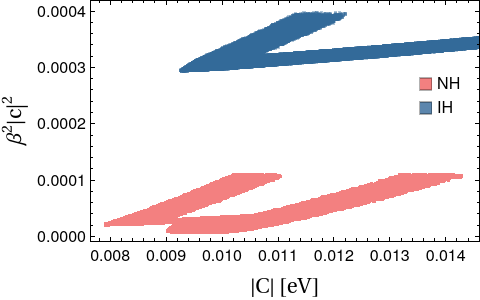}
        \includegraphics[width=0.43\textwidth]{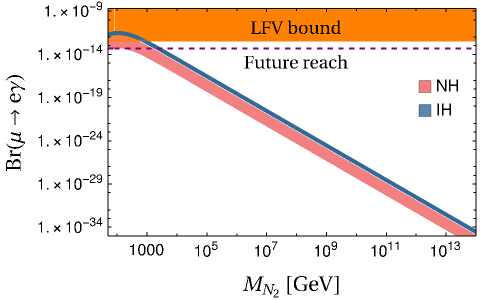}
    \caption{Left Panel : $\beta^2|C|^2$ vs $|C|$ for 3$\sigma$ allowed regions of neutrino oscillation data obtained in Fig. \ref{fig:alpha-beta} and Fig. \ref{fig:phiAC-phiBC}. Right Panel:   The branching ratio for $\mu \rightarrow e\gamma$  is plotted against the mass of the right-handed neutrino $N_2$. The horizontal shaded region and the dashed line represent current and future experimental upper limits. In both panels light red, and blue shaded patches represent regions for NH and IH respectively.}
    \label{fig:br of mu to e gamma}
\end{figure}

Now, for the type-I seesaw contribution in the lepton flavor violating decays, the decay of the form of $l_{\alpha} \rightarrow l_{\beta}\gamma$ will put a constraint on the FSS parameters. The branching ratio for such type of decay in our framework can be written as~\cite{Ilakovac:1994kj,Tommasini:1995ii,Dinh:2012bp,Bambhaniya:2016rbb,Ghosh:2017fmr} 
\begin{eqnarray}
   {\rm Br}(l_{\alpha} \rightarrow l_{\beta}\gamma)&\approx& \frac{3 \tilde{\alpha} v^4}{8 \pi } \Bigg| \sum_{i=1}^2 \mathcal{K}_{\beta i} \mathcal{K}^{\dagger}_{i\alpha} f\Bigg(\frac{M^2_{N_i}}{M_W^2}\Bigg)\Bigg|^2, \\
   &=&\sum_{i=1}^2\frac{3 \tilde{\alpha} v^4}{8 \pi M_{N_i}^4} \Bigg| (Y_N)_{\beta i} (Y_N^{\dagger})_{i\alpha} f\Bigg(\frac{M^2_{N_i}}{M_W^2}\Bigg)\Bigg|^2\label{eq:seesaw-lfv}
\end{eqnarray}
where  $\mathcal{K}=Y^{\dagger}_{N} (M^{-1}_R)^{*}$ and $Y_N=M_D/v$, as obtained from Eq. (\ref{eq:dirac and majorana mass matrix}).  Similarly to the scotogenic contribution, the VEV alignment of the $A_4$ flavon $\phi_s$ once again plays a crucial role in obtaining the estimation for the branching ratio  $l_{\alpha} \rightarrow l_{\beta}\gamma$ decays. The VEV configuration of $\phi_s$ is such that it gives rise to $(Y_N)_{e1}=0$ (see Eq. (\ref{eq:dirac and majorana mass matrix})) and the contribution associated with $N_1$ essentially vanishes for $\mu \rightarrow e\gamma$ and $\tau \rightarrow e\gamma$. Therefore only surviving contribution in these decays originates from ${N_2}$. Again due to the $A_4$ flavor symmetry we find $(Y_N)_{e2}=(Y_N)_{\mu 2}=(Y_N)_{\tau 2}$ as written in the second column of $M_D$, see Eq. (\ref{eq:dirac and majorana mass matrix}). As a result, the expression for the branching fraction for the above two decays will be the same $i.e.,$ ${\rm Br}(\mu \rightarrow e\gamma)={\rm Br}(\tau \rightarrow e\gamma)$. Out of  these two decays, the most stringent constraint comes from the $\mu \rightarrow e\gamma$ decay and in the following, we  discuss  the numerical analysis for  the same. Following Eq. (\ref{eq:seesaw-lfv}), in our framework, the branching ratio of $\mu \rightarrow e\gamma$ can be written as
\begin{eqnarray}\label{eq:seesaw mu e gamma}
 {\rm Br}(\mu \rightarrow e\gamma) = \frac{3 \tilde{\alpha} v^4}{8 \pi M_{N_2}^4} \Bigg| (y_{N_2} y_{N_2}^* \epsilon^2 f\Bigg(\frac{M^2_{N_2}}{M_W^2}\Bigg)\Bigg|^2
 =\frac{3 \tilde{\alpha} }{8 \pi M_{N_2}^2} \beta^2|C|^2 \Bigg(f\Bigg(\frac{M^2_{N_2}}{M_W^2}\Bigg)\Bigg)^2\label{eq:lfvseesawfinal}.  
\end{eqnarray}
where we have used $B=v^2y^2_{N_2}\epsilon^2/M_{N_2}$ from Eq. (\ref{eq:seesaw mass matrix}) and the definition  $\beta=|B|/ |C|$ to obtain Eq. (\ref{eq:lfvseesawfinal}). The loop function $f(x)$ in Eq. (\ref{eq:lfvseesawfinal}) can be written as  
\begin{eqnarray}
    f(x)=\frac{x(2x^3+3x^2-6x-6 x^2{\rm log}(x) +1)}{2(1-x)^4}. 
\end{eqnarray}
From Eq. (\ref{eq:lfvseesawfinal}) we find that the contribution in  ${\rm Br}(\mu \rightarrow e\gamma)$ coming from the seesaw mechanism depends on the mass of the heavy right-handed neutrino $N_2$,  $\beta$ and  $|C|$. The parameters $\beta$ and  $|C|$  are already fixed by neutrino data, as discussed in Section \ref{sec:numerical analysis}. As  ${\rm Br}(\mu \rightarrow e\gamma) \propto \beta^2|C|^2$, in Fig. \ref{fig:br of mu to e gamma} left panel we have shown the variation of $\beta^2|C|^2$ with $|C|$ for the  most general case of our analysis. In this plot, the light red and blue shaded regions represent an estimation for $\beta^2|C|^2$ as a function of $|C|$ for NH and IH light neutrino masses, respectively. Furthermore, Fig. \ref{fig:br of mu to e gamma} left panel also depicts that  $\beta^2|C|^2$  acquires higher values for IH compared to NH. This is because for NH $\beta$ is smaller ($\beta \leq 1$) compared to IH ($\beta \geq 1$) for similar values of $|C|$.  With this when we plot ${\rm Br}(\mu \rightarrow e\gamma)$ for seesaw contribution in our scenario as a function of $M_{N_2}$ in the right panel of Fig. \ref{fig:br of mu to e gamma}. Here the light red and blue shaded region represent prediction for ${\rm Br}(\mu \rightarrow e\gamma)$ for NH and IH of light neutrino mass.  As the $\beta^2|C|^2$  takes higher values for IH compared to NH, the branching ratio for the $\mu \rightarrow e \gamma$ decay is higher for IH compared to NH as seen in this figure. The horizontal orange shaded region represents current experimental limit   ${\rm Br}(\mu \rightarrow e\gamma) \leq 4.2 \times 10^{-13}$~\cite{MEG:2016leq} and the purple dashed  stands for the future reach ${\rm Br}(\mu \rightarrow e\gamma) \leq 6 \times 10^{-14}$~\cite{MEGII:2018kmf} which  puts a lower limit on the mass of $N_2$  in the range  $0.2~ {\rm TeV} \leq M_{N_2} \leq 1.6$ TeV ($1.7~ {\rm TeV} \leq M_{N_2} \leq 3$ TeV) for NH (IH) of light neutrino mass. 
Again, following Eq. (\ref{eq:seesaw-lfv}),  the branching ratio for   ${\tau} \rightarrow {\mu}\gamma$  in case of the seesaw contribution can be expressed as  
\begin{eqnarray}
   {\rm Br}({\tau} \rightarrow {\mu}\gamma)&=&\sum_{i=1}^2\frac{3 \tilde{\alpha} v^4}{8 \pi M_{N_i}^4} \Bigg| (Y_N)_{\tau i} (Y_N^{\dagger})_{i\mu} f\Bigg(\frac{M^2_{N_i}}{M_W^2}\Bigg)\Bigg|^2 ,\\
   &=&\frac{3 \tilde{\alpha} }{8 \pi M_{N_1}^2} \alpha^2|C|^2 \Bigg(f\Bigg(\frac{M^2_{N_1}}{M_W^2}\Bigg)\Bigg)^2+\frac{3 \tilde{\alpha} }{8 \pi M_{N_2}^2} \beta^2|C|^2 \Bigg(f\Bigg(\frac{M^2_{N_2}}{M_W^2}\Bigg)\Bigg)^2. 
\end{eqnarray}
\begin{figure}[h]
    \centering
    \includegraphics[width=0.42\textwidth]{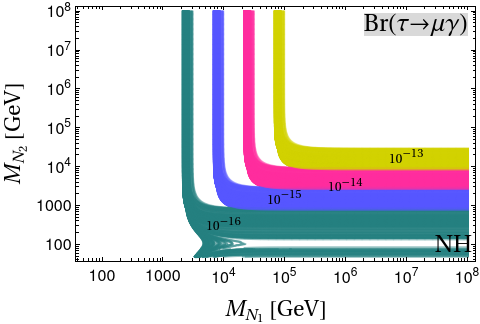}
        \includegraphics[width=0.43\textwidth]{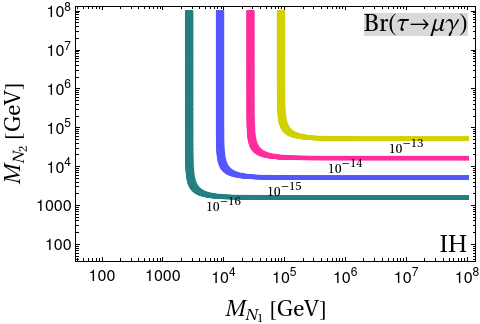}
    \caption{Contour plot for branching fraction of $\mu \rightarrow e\gamma$ (fixed at $10^{-13}$, $10^{-14}$, $10^{-15}$, $10^{-16}$ respectively) in the  $M_{N_1}$-$M_{N_2}$ plane. The left and right panels  represent allowed regions (obtained from Fig. \ref{fig:alpha-beta}) for NH and IH respectively.}
    \label{fig:br-tau-to-e-g}
\end{figure}
where we have used the relations $A={v^2  y_{N_1}^2 \epsilon^2}/{ M_{N_1}}$  and $\alpha=|A|/|C|$. Thus the contribution to  ${\rm Br}(\tau \rightarrow \mu\gamma)$ coming from the seesaw mechanism depends on the masses of the heavy right-handed neutrinos $N_{1,2}$ as well as   $\alpha,\beta$ and  $|C|$. The parameters $\alpha, \beta$, and  $|C|$  are already fixed to satisfy correct neutrino oscillation data  as discussed in Section \ref{sec:numerical analysis}. Hence in Fig. \ref{fig:br-tau-to-e-g} we have plotted different contours for   ${\rm Br}(\tau \rightarrow \mu\gamma)$ in the $M_{N_1}$-$M_{N_2}$ plane, corresponding to the $3\sigma$ allowed range of neutrino data for both NH and IH. In two panels, we have plotted the contours showing the  $M_{N_1}$-$M_{N_2}$ correlations with  branching fractions fixed at $10^{-13}$, $10^{-14}$, $10^{-15}$ and $10^{-16}$ (given by yellow, magenta, blue and green regions respectively). Here it is worth mentioning that  among $l_{\alpha} \rightarrow l_{\beta}\gamma$ decays, as a consequence of the flavor symmetric construction, only the seesaw part contributes to the $\mu \rightarrow e \gamma$ and $\tau \rightarrow  \mu \gamma$ decays. Whereas in the branching fraction of the decay $\tau \rightarrow e\gamma$,  both scotogenic and seesaw parts contribute. To understand the relative magnitude of these two contributions involved  in the $\tau \rightarrow e\gamma$ decay,  we define  a ratio  $R$ as
\begin{eqnarray}
    R=\frac{{{\rm Br}(\tau \rightarrow e\gamma)}_{\rm scoto}}{{{\rm Br}(\tau \rightarrow e\gamma)}_{\rm seesaw}}. 
\end{eqnarray}    
Now following Eq.~(\ref{eq:scoto tau decay e gamma}) and Eq.~(\ref{eq:seesaw mu e gamma}) along with Eq.~(\ref{eq:scoto mass matrix}),  in the FSS framework we find that  the ratio $R$ is proportional to $ 1/\beta^2$ for specific values  of the scoto-seesaw mass parameters.  
\begin{figure}[h]
    \centering
    \includegraphics[width=0.42\textwidth]{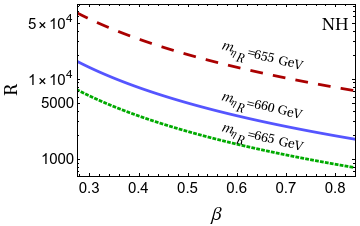}
        \includegraphics[width=0.405\textwidth]{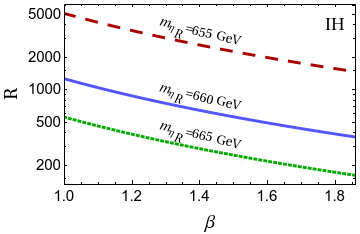}
    \caption{The ratio ($R$) of  the branching fractions for the scoto and seesaw contributions  versus $\beta$ for both NH (left panel) and IH (right panel).}
    \label{fig:ratio-br}
\end{figure}
In Fig. \ref{fig:ratio-br}, we have plotted   $R$ considering $m_{\eta^+} = 600$ GeV, $m_{\eta_I}=650 $ GeV,  $M_f=10^5$ GeV, and $M_{N_2}=10^6$ GeV for both NH (left panel) and IH (right panel) respectively. In the panels,  the brown dashed, blue continuous, and green dotted lines represent the estimation for $R$ with $m_{\eta_R}=655$ GeV, 660 GeV, and 665 GeV respectively. These lines  correspond to the allowed range for  $\beta$ obtained earlier. Since  $R \geq  10^3$  ($R \geq 2\times 10^2$) for NH (IH), \emph{we can conclude that the scotogenic part  dominates the seesaw contribution in the lepton flavor violating decay such as $\tau \rightarrow e\gamma$.}

The scotogenic contribution in our analysis offers us the opportunity to explain the nature of dark matter.  In this model, there is an inherent   dark $\mathcal{Z}_2$ symmetry that ensures the stability of the lightest dark particle, and  three feasible dark matter candidates exist. These are, namely, the fermionic dark matter $f$ and  scalar dark matter, which are real and imaginary components of $\eta$, given by $\eta_R$ and $\eta_I$, respectively.  When the lightest dark particle originates from $\eta$, it resembles the inert Higgs doublet model~\cite{Dolle:2009fn}.     Considering $\eta_R$ as the DM candidate, there exists several annihilation and co-annihilation channels  in this model which involve annihilation to quarks and leptons, SM gauge bosons, and the Higgs boson such as $\eta_R\eta_R \rightarrow ZZ,~W^{+}W^{-},~ hh, ~q\bar{q}$ etc. Collectively they all contribute to  the relic abundance of $\eta_R$. Present dark matter abundance is often expressed in terms of the relic density parameter $\Omega^2 h$ and reported to be $\Omega^2 h=0.120\pm 0.001$ at $68\%$ CL~\cite{Planck:2018vyg}. In a minimal scoto-seesaw framework~\cite{Mandal:2021yph}, it has been argued   that  correct relic density can be obtained for  three different  mass ranges for $\eta_R$. These are respectively $m_{\eta_R}<50$ GeV, $70$ GeV $<m_{\eta_R}<100$ GeV and $m_{\eta_R}>550$ GeV. The dark matter mass in the range $m_{\eta_R}<50$ is disallowed as it is in conflict with the LHC Higgs invisible decay limit~\cite{CMS:2022qva}.   The intermediate region $70$ GeV $<m_{\eta_R}<100$ GeV is not completely ruled out by the LHC and LEP data and  dark matter mass in the range $m_{\eta_R}>550$ GeV  is not affected by the collider constraints.  The phenomenology of  fermionic  dark matter is worth exploring and is beyond the scope of the current study.

\section{Conclusions}\label{sec:conc}
We have proposed the flavor-scoto-seesaw (FSS) model that establishes   a common origin of the nonzero $\theta_{13}$ and cosmological dark matter. The framework is based on the $A_4$  flavor symmetry where both type-I seesaw and scotogenic mechanisms contribute to the effective light neutrino mass.  FSS explains observed neutrino masses and mixing angles provides rich phenomenology and accommodates potential dark matter candidates. Guided by the discrete symmetry, we show that the minimal type-I seesaw first reproduces the widely popular TBM mixing,  a first-order approximation of the lepton mixing matrix. Subsequently, the scotogenic contribution acts as a requisite deviation to the TBM mixing and addresses the issue of the nature of dark matter. We also  demonstrate that the neutrino mixing pattern  exhibits  ${\rm TM}_2$ mixing scheme  (a viable descendent of the TBM mixing pattern)   within this scoto-seesaw scenario.  \\ \indent
The model which we construct here is highly predictive in nature. Using the current experimental observation on neutrino oscillation and other cosmological limits, {\it{we found that the allowed parameter space in the FSS model restricts some {of the} key  observables associated with neutrinos (the atmospheric mixing angle, Dirac and Majorana CP phases, effective mass parameter appearing in the neutrinoless double beta decay ) and crucially dictates the lepton flavor violating decays.}} To understand the behaviours of the parameters involved (namely, $\alpha, \beta, \phi_{AC}, \phi_{BC}$), we divide  the numerical analysis into a few specific cases based on the choice of the associated phases  ($\phi_{AC}, \phi_{BC}$) and then carried out the complete general numerical analysis.    These limiting cases can easily distinguish between light neutrino masses' normal and inverted hierarchy. For example, when the relative phase between the seesaw contribution associated with right-handed neutrino $N_1$ and the scotogenic contribution is considered zero ($i.e.$, $\phi_{AC}=0$ for Case I and II), only inverted hierarchy is allowed. On the other hand, when   $\phi_{AC} \neq 0$ (Case III and IV), only normal hierarchy is allowed. Considered discrete flavor symmetries play an instrumental role in producing such distinctive constraints.
Subsequently, for the most general case, we carried out the numerical analysis for all possible choices of the parameters involved and found that   normal hierarchy can be realized only with $\beta \leq 1$ and $2.61\leq \phi_{AC} \leq 5.38$  whereas to realize inverted hierarchy we need $\beta \geq 1$ and $0\leq \phi_{AC} \leq 2$ (or $6\leq \phi_{AC} \leq 2\pi$). This analysis also predicts the atmospheric mixing angle lying in the upper  octant, in good agreement with the latest global ﬁt neutrino oscillation data, and restricts  Dirac CP phase $\delta_{\rm CP}$ within  $-1.57 \leq \delta_{\rm CP} \leq 1.37 $ (and $1.4 \leq \delta_{\rm CP} \leq 1.57 $) for normal hierarchy and   $-1.57 \leq \delta_{\rm CP} \leq 0.5 $ (and $0.86 \leq \delta_{\rm CP} \leq 1.57 $) for inverted hierarchy. Along with the Dirac CP phase $\delta_{\rm CP}$, the Majorana phases also get restricted in our analysis.  Furthermore, we obtain a lower limit on the lightest neutrino mass as $m_{\rm lightest} \geq 0.0012$ eV for normal hierarchy and $m_{\rm lightest} \geq 0.014$ eV for inverted hierarchy. We have also estimated the prediction for the effective mass parameter $m_{\beta\beta}$ characterizing the neutrinoless double beta decay and found it to be in the range  $1-30$ meV for normal hierarchy and $16-60$ meV for inverted hierarchy, respectively. These values are within the reach of future neutrinoless double beta decay experiments. 
The unique  prediction on the correlation among the masses and mixing such as   $m_{\beta\beta}$-$\delta_{\rm CP}$, $\sin^2_{23}$,$\alpha_{21}$,  $\alpha_{31}$ as well as $\sum m_i$-$\delta_{\rm CP}$, $\sin^2_{23}$, etc.  are typical features of the discussed FSS  model. For example, the sum of absolute neutrino masses crucially  dictates the allowed ranges for $\delta_{\rm CP}$ mentioned above.
In the end, we also comment on the phenomenological implications, such as lepton flavor violation and the prospects of dark matter candidates in such a scenario.   As a consequence of the flavor structure, the scotogenic part of the model does not contribute to the lepton flavor violating decays such as  $\mu \rightarrow e \gamma$ and $\tau \rightarrow  \mu \gamma$, and a lower limit on the mass of the heavy right-handed neutrinos can be obtained from the constraints on the branching ratios involving such decays. On the other hand,  both scoto and seesaw parts contribute to the decay $\tau \rightarrow e \gamma$. However, the scotogenic part dominates this  rare decay  and a constraint on the  ratio of flavon VEVs  to the cut-off scale of the theory can be obtained.   A detailed discussion of the phenomenological aspects in this direction is dedicated to future investigations. 

\section*{Appendix: $A_4$ symmetry}\label{section:A4 group}
$A_4$ is a discrete group of even permutations of four
objects\footnote{For a detailed discussion on $A_4$ see Ref. \cite{Altarelli:2010gt,Ishimori:2010au}}. Geometrically, it is an invariance group of a tetrahedron. It has 12 elements which can be generated by two basic objects $S$ and $T$ which obey the following relation
\begin{eqnarray}
	S^2=T^3=(ST)^3=1
\end{eqnarray}
The $A_4$ group has three one-dimensional irreducible representations $1$,$1^{\prime}$ and $1^{\prime\prime}$ and one three dimensional irreducible representation 3. Products of the singlets and triplets are
given by
\begin{eqnarray}
	1&\otimes& 1=1; ~  
	1^{\prime}\otimes1^{\prime\prime}=1,  \\
	1^{\prime}&\otimes& 1^{\prime}=1^{\prime\prime};~  
	1^{\prime\prime}\otimes 1^{\prime\prime}=1^{\prime}, \\
	3&\otimes&3 =1\oplus1^{\prime}\oplus 1^{\prime\prime}\oplus 3_s\oplus 3_a,
\end{eqnarray}
where the subscripts $``s"$ and $``a"$ denote symmetric and antisymmetric part respectively. Writing two triplets as $(x_1,x_2,x_3)$ and $(y_1,y_2,y_3)$ respectively, their products are given by
\begin{eqnarray}
	1&\sim &x_1 y_1+x_2y_3+x_3 y_2,\\
	1^{\prime}&\sim &x_3 y_3+x_1y_2+x_2 y_1,  \\
	1^{\prime\prime}&\sim &x_2 y_2+x_1y_3+x_3 y_1,  \\
	3_s&\sim& \begin{pmatrix}
		2 x_1 y_1-x_2 y_3-x_3y_2 \\
		2 x_3 y_3-x_1 y_2-x_2 y_1 \\
		2 x_2 y_2-x_1 y_3-x_3 y_1
	\end{pmatrix} , \\
3_a &\sim & \begin{pmatrix}
       x_2 y_3-x_3 y_2 \\
        x_1 y_2-x_2 y_1 \\
        x_3 y_1-x_1 y_3
\end{pmatrix}.
\end{eqnarray}

\acknowledgments
This work has been supported in part by the Polish National Science Center (NCN) under grant 2020/37/B/ST2/02371 and the Freedom of Research (Swoboda Badań) initiative  of the University of Silesia in Katowice. BK would like to thank José W. F. Valle for useful discussions. 

\bibliography{ref1,bibliography}

\providecommand{\href}[2]{#2}\begingroup\raggedright\begin{thebibliography}{100}

\bibitem{Pontecorvo:1967fh}
B.~Pontecorvo, \emph{{Neutrino Experiments and the Problem of Conservation of
  Leptonic Charge}}, {\emph{Zh. Eksp. Teor. Fiz.} {\bfseries 53} (1967) 1717}.

\bibitem{SNO:2001kpb}
{\scshape SNO} collaboration, \emph{{Measurement of the rate of $\nu_e+d \to
  p+p+e^-$ interactions produced by $^8$B solar neutrinos at the Sudbury
  Neutrino Observatory}},
  \href{https://doi.org/10.1103/PhysRevLett.87.071301}{\emph{Phys. Rev. Lett.}
  {\bfseries 87} (2001) 071301}
  [\href{https://arxiv.org/abs/nucl-ex/0106015}{{\ttfamily nucl-ex/0106015}}].

\bibitem{Super-Kamiokande:1998kpq}
{\scshape Super-Kamiokande} collaboration, \emph{{Evidence for oscillation of
  atmospheric neutrinos}},
  \href{https://doi.org/10.1103/PhysRevLett.81.1562}{\emph{Phys. Rev. Lett.}
  {\bfseries 81} (1998) 1562}
  [\href{https://arxiv.org/abs/hep-ex/9807003}{{\ttfamily hep-ex/9807003}}].

\bibitem{Gonzalez-Garcia:2007dlo}
M.C.~Gonzalez-Garcia and M.~Maltoni, \emph{{Phenomenology with Massive
  Neutrinos}}, \href{https://doi.org/10.1016/j.physrep.2007.12.004}{\emph{Phys.
  Rept.} {\bfseries 460} (2008) 1}
  [\href{https://arxiv.org/abs/0704.1800}{{\ttfamily 0704.1800}}].

\bibitem{deSalas:2020pgw}
P.F.~de~Salas, D.V.~Forero, S.~Gariazzo, P.~Mart\'\i{}nez-Mirav\'e, O.~Mena,
  C.A.~Ternes et~al., \emph{{2020 global reassessment of the neutrino
  oscillation picture}},
  \href{https://doi.org/10.1007/JHEP02(2021)071}{\emph{JHEP} {\bfseries 02}
  (2021) 071} [\href{https://arxiv.org/abs/2006.11237}{{\ttfamily
  2006.11237}}].

\bibitem{WMAP:2012nax}
{\scshape WMAP} collaboration, \emph{{Nine-Year Wilkinson Microwave Anisotropy
  Probe (WMAP) Observations: Cosmological Parameter Results}},
  \href{https://doi.org/10.1088/0067-0049/208/2/19}{\emph{Astrophys. J. Suppl.}
  {\bfseries 208} (2013) 19} [\href{https://arxiv.org/abs/1212.5226}{{\ttfamily
  1212.5226}}].

\bibitem{Aghanim:2018eyx}
{\scshape Planck} collaboration, \emph{{Planck 2018 results. VI. Cosmological
  parameters}},
  \href{https://doi.org/10.1051/0004-6361/201833910}{\emph{Astron. Astrophys.}
  {\bfseries 641} (2020) A6}
  [\href{https://arxiv.org/abs/1807.06209}{{\ttfamily 1807.06209}}].

\bibitem{Bertone:2004pz}
G.~Bertone, D.~Hooper and J.~Silk, \emph{{Particle dark matter: Evidence,
  candidates and constraints}},
  \href{https://doi.org/10.1016/j.physrep.2004.08.031}{\emph{Phys. Rept.}
  {\bfseries 405} (2005) 279}
  [\href{https://arxiv.org/abs/hep-ph/0404175}{{\ttfamily hep-ph/0404175}}].

\bibitem{Minkowski:1977sc}
P.~Minkowski, \emph{{$\mu \to e\gamma$ at a Rate of One Out of $10^{9}$ Muon
  Decays?}}, \href{https://doi.org/10.1016/0370-2693(77)90435-X}{\emph{Phys.
  Lett. B} {\bfseries 67} (1977) 421}.

\bibitem{Gell-Mann:1979vob}
M.~Gell-Mann, P.~Ramond and R.~Slansky, \emph{{Complex Spinors and Unified
  Theories}}, {\emph{Conf. Proc. C} {\bfseries 790927} (1979) 315}
  [\href{https://arxiv.org/abs/1306.4669}{{\ttfamily 1306.4669}}].

\bibitem{Mohapatra:1979ia}
R.N.~Mohapatra and G.~Senjanovic, \emph{{Neutrino Mass and Spontaneous Parity
  Violation}}, \href{https://doi.org/10.1103/PhysRevLett.44.912}{\emph{Phys.
  Rev. Lett.} {\bfseries 44} (1980) 912}.

\bibitem{Schechter:1981cv}
J.~Schechter and J.W.F.~Valle, \emph{{Neutrino Decay and Spontaneous Violation
  of Lepton Number}},
  \href{https://doi.org/10.1103/PhysRevD.25.774}{\emph{Phys. Rev. D} {\bfseries
  25} (1982) 774}.

\bibitem{Schechter:1980gr}
J.~Schechter and J.W.F.~Valle, \emph{{Neutrino Masses in SU(2) x U(1)
  Theories}}, \href{https://doi.org/10.1103/PhysRevD.22.2227}{\emph{Phys. Rev.}
  {\bfseries D22} (1980) 2227}.

\bibitem{King:2015sfk}
S.F.~King, \emph{{Neutrino Mass and Mixing in the Seesaw Playground}},
  \href{https://doi.org/10.1016/j.nuclphysb.2015.12.015}{\emph{Nucl. Phys. B}
  {\bfseries 908} (2016) 456}
  [\href{https://arxiv.org/abs/1511.03831}{{\ttfamily 1511.03831}}].

\bibitem{Xing:2020ald}
Z.-z.~Xing and Z.-h.~Zhao, \emph{{The minimal seesaw and leptogenesis models}},
  \href{https://doi.org/10.1088/1361-6633/abf086}{\emph{Rept. Prog. Phys.}
  {\bfseries 84} (2021) 066201}
  [\href{https://arxiv.org/abs/2008.12090}{{\ttfamily 2008.12090}}].

\bibitem{King:1999mb}
S.F.~King, \emph{{Large mixing angle MSW and atmospheric neutrinos from single
  right-handed neutrino dominance and U(1) family symmetry}},
  \href{https://doi.org/10.1016/S0550-3213(00)00109-7}{\emph{Nucl. Phys. B}
  {\bfseries 576} (2000) 85}
  [\href{https://arxiv.org/abs/hep-ph/9912492}{{\ttfamily hep-ph/9912492}}].

\bibitem{King:2002nf}
S.F.~King, \emph{{Constructing the large mixing angle MNS matrix in seesaw
  models with right-handed neutrino dominance}},
  \href{https://doi.org/10.1088/1126-6708/2002/09/011}{\emph{JHEP} {\bfseries
  09} (2002) 011} [\href{https://arxiv.org/abs/hep-ph/0204360}{{\ttfamily
  hep-ph/0204360}}].

\bibitem{Morisi:2012fg}
S.~Morisi and J.W.F.~Valle, \emph{{Neutrino masses and mixing: a flavour
  symmetry roadmap}},
  \href{https://doi.org/10.1002/prop.201200125}{\emph{Fortsch. Phys.}
  {\bfseries 61} (2013) 466} [\href{https://arxiv.org/abs/1206.6678}{{\ttfamily
  1206.6678}}].

\bibitem{Ishimori:2010au}
H.~Ishimori, T.~Kobayashi, H.~Ohki, Y.~Shimizu, H.~Okada and M.~Tanimoto,
  \emph{{Non-Abelian Discrete Symmetries in Particle Physics}},
  \href{https://doi.org/10.1143/PTPS.183.1}{\emph{Prog. Theor. Phys. Suppl.}
  {\bfseries 183} (2010) 1} [\href{https://arxiv.org/abs/1003.3552}{{\ttfamily
  1003.3552}}].

\bibitem{Altarelli:2010gt}
G.~Altarelli and F.~Feruglio, \emph{{Discrete Flavor Symmetries and Models of
  Neutrino Mixing}},
  \href{https://doi.org/10.1103/RevModPhys.82.2701}{\emph{Rev. Mod. Phys.}
  {\bfseries 82} (2010) 2701}
  [\href{https://arxiv.org/abs/1002.0211}{{\ttfamily 1002.0211}}].

\bibitem{Feruglio:2019ybq}
F.~Feruglio and A.~Romanino, \emph{{Lepton flavor symmetries}},
  \href{https://doi.org/10.1103/RevModPhys.93.015007}{\emph{Rev. Mod. Phys.}
  {\bfseries 93} (2021) 015007}
  [\href{https://arxiv.org/abs/1912.06028}{{\ttfamily 1912.06028}}].

\bibitem{King:2013eh}
S.F.~King and C.~Luhn, \emph{{Neutrino Mass and Mixing with Discrete
  Symmetry}}, \href{https://doi.org/10.1088/0034-4885/76/5/056201}{\emph{Rept.
  Prog. Phys.} {\bfseries 76} (2013) 056201}
  [\href{https://arxiv.org/abs/1301.1340}{{\ttfamily 1301.1340}}].

\bibitem{Petcov:2017ggy}
S.T.~Petcov, \emph{{Discrete Flavour Symmetries, Neutrino Mixing and Leptonic
  CP Violation}},
  \href{https://doi.org/10.1140/epjc/s10052-018-6158-5}{\emph{Eur. Phys. J. C}
  {\bfseries 78} (2018) 709}
  [\href{https://arxiv.org/abs/1711.10806}{{\ttfamily 1711.10806}}].

\bibitem{Xing:2020ijf}
Z.-z.~Xing, \emph{{Flavor structures of charged fermions and massive
  neutrinos}}, \href{https://doi.org/10.1016/j.physrep.2020.02.001}{\emph{Phys.
  Rept.} {\bfseries 854} (2020) 1}
  [\href{https://arxiv.org/abs/1909.09610}{{\ttfamily 1909.09610}}].

\bibitem{Chauhan:2022gkz}
G.~Chauhan, P.S.B.~Dev, B.~Dziewit, W.~Flieger, J.~Gluza, K.~Grzanka et~al.,
  \emph{{Discrete Flavor Symmetries and Lepton Masses and Mixings}},  in
  \emph{{2022 Snowmass Summer Study}}, 3, 2022
  [\href{https://arxiv.org/abs/2203.08105}{{\ttfamily 2203.08105}}].

\bibitem{Vissani:1997pa}
F.~Vissani, \emph{{A Study of the scenario with nearly degenerate Majorana
  neutrinos}},  \href{https://arxiv.org/abs/hep-ph/9708483}{{\ttfamily
  hep-ph/9708483}}.

\bibitem{Barger:1998ta}
V.D.~Barger, S.~Pakvasa, T.J.~Weiler and K.~Whisnant, \emph{{Bimaximal mixing
  of three neutrinos}},
  \href{https://doi.org/10.1016/S0370-2693(98)00880-6}{\emph{Phys. Lett. B}
  {\bfseries 437} (1998) 107}
  [\href{https://arxiv.org/abs/hep-ph/9806387}{{\ttfamily hep-ph/9806387}}].

\bibitem{Datta:2003qg}
A.~Datta, F.-S.~Ling and P.~Ramond, \emph{{Correlated hierarchy, Dirac masses
  and large mixing angles}},
  \href{https://doi.org/10.1016/j.nuclphysb.2003.08.026}{\emph{Nucl. Phys. B}
  {\bfseries 671} (2003) 383}
  [\href{https://arxiv.org/abs/hep-ph/0306002}{{\ttfamily hep-ph/0306002}}].

\bibitem{Kajiyama:2007gx}
Y.~Kajiyama, M.~Raidal and A.~Strumia, \emph{{The Golden ratio prediction for
  the solar neutrino mixing}},
  \href{https://doi.org/10.1103/PhysRevD.76.117301}{\emph{Phys. Rev. D}
  {\bfseries 76} (2007) 117301}
  [\href{https://arxiv.org/abs/0705.4559}{{\ttfamily 0705.4559}}].

\bibitem{Albright:2010ap}
C.H.~Albright, A.~Dueck and W.~Rodejohann, \emph{{Possible Alternatives to
  Tri-bimaximal Mixing}},
  \href{https://doi.org/10.1140/epjc/s10052-010-1492-2}{\emph{Eur. Phys. J. C}
  {\bfseries 70} (2010) 1099}
  [\href{https://arxiv.org/abs/1004.2798}{{\ttfamily 1004.2798}}].

\bibitem{Harrison:2002er}
P.F.~Harrison, D.H.~Perkins and W.G.~Scott, \emph{{Tri-bimaximal mixing and the
  neutrino oscillation data}},
  \href{https://doi.org/10.1016/S0370-2693(02)01336-9}{\emph{Phys. Lett.}
  {\bfseries B530} (2002) 167}
  [\href{https://arxiv.org/abs/hep-ph/0202074}{{\ttfamily hep-ph/0202074}}].

\bibitem{Harrison:2002kp}
P.F.~Harrison and W.G.~Scott, \emph{{Symmetries and generalizations of tri -
  bimaximal neutrino mixing}},
  \href{https://doi.org/10.1016/S0370-2693(02)01753-7}{\emph{Phys. Lett.}
  {\bfseries B535} (2002) 163}
  [\href{https://arxiv.org/abs/hep-ph/0203209}{{\ttfamily hep-ph/0203209}}].

\bibitem{Chang:2004wy}
S.~Chang, S.K.~Kang and K.~Siyeon, \emph{{Minimal seesaw model with
  tri/bi-maximal mixing and leptogenesis}},
  \href{https://doi.org/10.1016/j.physletb.2004.06.104}{\emph{Phys. Lett. B}
  {\bfseries 597} (2004) 78}
  [\href{https://arxiv.org/abs/hep-ph/0404187}{{\ttfamily hep-ph/0404187}}].

\bibitem{Park:2011zt}
N.W.~Park, K.H.~Nam and K.~Siyeon, \emph{{Discrete flavor symmetry and minimal
  seesaw mechanism}},
  \href{https://doi.org/10.1103/PhysRevD.83.056013}{\emph{Phys. Rev. D}
  {\bfseries 83} (2011) 056013}
  [\href{https://arxiv.org/abs/1101.4134}{{\ttfamily 1101.4134}}].

\bibitem{Zhao:2011pv}
Z.-h.~Zhao, \emph{{Realizing Tri-bimaximal Mixing in Minimal Seesaw Model with
  S4 Family Symmetry}},
  \href{https://doi.org/10.1016/j.physletb.2011.06.050}{\emph{Phys. Lett. B}
  {\bfseries 701} (2011) 609}
  [\href{https://arxiv.org/abs/1106.2715}{{\ttfamily 1106.2715}}].

\bibitem{King:1999cm}
S.F.~King, \emph{{Atmospheric and solar neutrinos from single right-handed
  neutrino dominance and U(1) family symmetry}},
  \href{https://doi.org/10.1016/S0550-3213(99)00542-8}{\emph{Nucl. Phys. B}
  {\bfseries 562} (1999) 57}
  [\href{https://arxiv.org/abs/hep-ph/9904210}{{\ttfamily hep-ph/9904210}}].

\bibitem{Ma:2001dn}
E.~Ma and G.~Rajasekaran, \emph{{Softly broken A(4) symmetry for nearly
  degenerate neutrino masses}},
  \href{https://doi.org/10.1103/PhysRevD.64.113012}{\emph{Phys. Rev.}
  {\bfseries D64} (2001) 113012}
  [\href{https://arxiv.org/abs/hep-ph/0106291}{{\ttfamily hep-ph/0106291}}].

\bibitem{Ma:2004zv}
E.~Ma, \emph{{A(4) symmetry and neutrinos with very different masses}},
  \href{https://doi.org/10.1103/PhysRevD.70.031901}{\emph{Phys. Rev. D}
  {\bfseries 70} (2004) 031901}
  [\href{https://arxiv.org/abs/hep-ph/0404199}{{\ttfamily hep-ph/0404199}}].

\bibitem{Ma:2002yp}
E.~Ma, \emph{{Quark mass matrices in the A(4) model}},
  \href{https://doi.org/10.1142/S0217732302006722}{\emph{Mod. Phys. Lett.}
  {\bfseries A17} (2002) 627}
  [\href{https://arxiv.org/abs/hep-ph/0203238}{{\ttfamily hep-ph/0203238}}].

\bibitem{Altarelli:2005yp}
G.~Altarelli and F.~Feruglio, \emph{{Tri-bimaximal neutrino mixing from
  discrete symmetry in extra dimensions}},
  \href{https://doi.org/10.1016/j.nuclphysb.2005.05.005}{\emph{Nucl. Phys. B}
  {\bfseries 720} (2005) 64}
  [\href{https://arxiv.org/abs/hep-ph/0504165}{{\ttfamily hep-ph/0504165}}].

\bibitem{Altarelli:2005yx}
G.~Altarelli and F.~Feruglio, \emph{{Tri-bimaximal neutrino mixing, A(4) and
  the modular symmetry}},
  \href{https://doi.org/10.1016/j.nuclphysb.2006.02.015}{\emph{Nucl. Phys. B}
  {\bfseries 741} (2006) 215}
  [\href{https://arxiv.org/abs/hep-ph/0512103}{{\ttfamily hep-ph/0512103}}].

\bibitem{Babu:2002dz}
K.S.~Babu, E.~Ma and J.W.F.~Valle, \emph{{Underlying A(4) symmetry for the
  neutrino mass matrix and the quark mixing matrix}},
  \href{https://doi.org/10.1016/S0370-2693(02)03153-2}{\emph{Phys. Lett. B}
  {\bfseries 552} (2003) 207}
  [\href{https://arxiv.org/abs/hep-ph/0206292}{{\ttfamily hep-ph/0206292}}].

\bibitem{Abe:2011fz}
{\scshape Double Chooz} collaboration, \emph{{Indication for the disappearance
  of reactor electron antineutrinos in the Double Chooz experiment}},
  \href{https://doi.org/10.1103/PhysRevLett.108.131801}{\emph{Phys. Rev. Lett.}
  {\bfseries 108} (2012) 131801}
  [\href{https://arxiv.org/abs/1112.6353}{{\ttfamily 1112.6353}}].

\bibitem{An:2012eh}
{\scshape Daya Bay} collaboration, \emph{{Observation of electron-antineutrino
  disappearance at Daya Bay}},
  \href{https://doi.org/10.1103/PhysRevLett.108.171803}{\emph{Phys. Rev. Lett.}
  {\bfseries 108} (2012) 171803}
  [\href{https://arxiv.org/abs/1203.1669}{{\ttfamily 1203.1669}}].

\bibitem{Ahn:2012nd}
{\scshape RENO} collaboration, \emph{{Observation of Reactor Electron
  Antineutrino Disappearance in the RENO Experiment}},
  \href{https://doi.org/10.1103/PhysRevLett.108.191802}{\emph{Phys. Rev. Lett.}
  {\bfseries 108} (2012) 191802}
  [\href{https://arxiv.org/abs/1204.0626}{{\ttfamily 1204.0626}}].

\bibitem{T2K:2013ppw}
{\scshape T2K} collaboration, \emph{{Observation of Electron Neutrino
  Appearance in a Muon Neutrino Beam}},
  \href{https://doi.org/10.1103/PhysRevLett.112.061802}{\emph{Phys. Rev. Lett.}
  {\bfseries 112} (2014) 061802}
  [\href{https://arxiv.org/abs/1311.4750}{{\ttfamily 1311.4750}}].

\bibitem{MINOS:2013utc}
{\scshape MINOS} collaboration, \emph{{Measurement of Neutrino and Antineutrino
  Oscillations Using Beam and Atmospheric Data in MINOS}},
  \href{https://doi.org/10.1103/PhysRevLett.110.251801}{\emph{Phys. Rev. Lett.}
  {\bfseries 110} (2013) 251801}
  [\href{https://arxiv.org/abs/1304.6335}{{\ttfamily 1304.6335}}].

\bibitem{Adhikary:2008au}
B.~Adhikary and A.~Ghosal, \emph{{Nonzero U(e3), CP violation and leptogenesis
  in a see-saw type softly broken A(4) symmetric model}},
  \href{https://doi.org/10.1103/PhysRevD.78.073007}{\emph{Phys. Rev. D}
  {\bfseries 78} (2008) 073007}
  [\href{https://arxiv.org/abs/0803.3582}{{\ttfamily 0803.3582}}].

\bibitem{Brahmachari:2008fn}
B.~Brahmachari, S.~Choubey and M.~Mitra, \emph{{The A(4) flavor symmetry and
  neutrino phenomenology}},
  \href{https://doi.org/10.1103/PhysRevD.77.119901}{\emph{Phys. Rev. D}
  {\bfseries 77} (2008) 073008}
  [\href{https://arxiv.org/abs/0801.3554}{{\ttfamily 0801.3554}}].

\bibitem{King:2009qt}
S.F.~King, \emph{{Tri-bimaximal Neutrino Mixing and $\theta_{13}$}},
  \href{https://doi.org/10.1016/j.physletb.2009.04.031}{\emph{Phys. Lett. B}
  {\bfseries 675} (2009) 347}
  [\href{https://arxiv.org/abs/0903.3199}{{\ttfamily 0903.3199}}].

\bibitem{Branco:2009by}
G.~Branco, R.~Gonzalez~Felipe, M.~Rebelo and H.~Serodio, \emph{{Resonant
  leptogenesis and tribimaximal leptonic mixing with A(4) symmetry}},
  \href{https://doi.org/10.1103/PhysRevD.79.093008}{\emph{Phys. Rev. D}
  {\bfseries 79} (2009) 093008}
  [\href{https://arxiv.org/abs/0904.3076}{{\ttfamily 0904.3076}}].

\bibitem{AristizabalSierra:2009ex}
D.~Aristizabal~Sierra, F.~Bazzocchi, I.~de~Medeiros~Varzielas, L.~Merlo and
  S.~Morisi, \emph{{Tri-Bimaximal Lepton Mixing and Leptogenesis}},
  \href{https://doi.org/10.1016/j.nuclphysb.2009.10.009}{\emph{Nucl. Phys. B}
  {\bfseries 827} (2010) 34} [\href{https://arxiv.org/abs/0908.0907}{{\ttfamily
  0908.0907}}].

\bibitem{Morisi:2009sc}
S.~Morisi and E.~Peinado, \emph{{An A(4) model for lepton masses and mixings}},
  \href{https://doi.org/10.1103/PhysRevD.80.113011}{\emph{Phys. Rev. D}
  {\bfseries 80} (2009) 113011}
  [\href{https://arxiv.org/abs/0910.4389}{{\ttfamily 0910.4389}}].

\bibitem{Ahn:2011yj}
Y.H.~Ahn, H.-Y.~Cheng and S.~Oh, \emph{{Quark-lepton complementarity and
  tribimaximal neutrino mixing from discrete symmetry}},
  \href{https://doi.org/10.1103/PhysRevD.83.076012}{\emph{Phys. Rev. D}
  {\bfseries 83} (2011) 076012}
  [\href{https://arxiv.org/abs/1102.0879}{{\ttfamily 1102.0879}}].

\bibitem{Shimizu:2011xg}
Y.~Shimizu, M.~Tanimoto and A.~Watanabe, \emph{{Breaking Tri-bimaximal Mixing
  and Large $\theta_{13}$}},
  \href{https://doi.org/10.1143/PTP.126.81}{\emph{Prog. Theor. Phys.}
  {\bfseries 126} (2011) 81} [\href{https://arxiv.org/abs/1105.2929}{{\ttfamily
  1105.2929}}].

\bibitem{Ganguly:2020riw}
J.~Ganguly and R.S.~Hundi, \emph{{Neutrino Mixing by modifying the Yukawa
  coupling structure of constrained sequential dominance}},
  \href{https://doi.org/10.1103/PhysRevD.103.035007}{\emph{Phys. Rev. D}
  {\bfseries 103} (2021) 035007}
  [\href{https://arxiv.org/abs/2005.04023}{{\ttfamily 2005.04023}}].

\bibitem{Ganguly:2021nqx}
J.~Ganguly and R.S.~Hundi, \emph{{Deviation from tri-bimaximal mixing as a
  result of modification of Yukawa coupling structure of constrained sequential
  dominance}}, \href{https://doi.org/10.1088/1742-6596/2156/1/012183}{\emph{J.
  Phys. Conf. Ser.} {\bfseries 2156} (2021) 012183}.

\bibitem{Ahn:2011if}
Y.H.~Ahn, H.-Y.~Cheng and S.~Oh, \emph{{An extension of tribimaximal lepton
  mixing}}, \href{https://doi.org/10.1103/PhysRevD.84.113007}{\emph{Phys. Rev.
  D} {\bfseries 84} (2011) 113007}
  [\href{https://arxiv.org/abs/1107.4549}{{\ttfamily 1107.4549}}].

\bibitem{Antusch:2011ic}
S.~Antusch, S.F.~King, C.~Luhn and M.~Spinrath, \emph{{Trimaximal mixing with
  predicted $\theta_{13}$ from a new type of constrained sequential
  dominance}},
  \href{https://doi.org/10.1016/j.nuclphysb.2011.11.009}{\emph{Nucl. Phys. B}
  {\bfseries 856} (2012) 328}
  [\href{https://arxiv.org/abs/1108.4278}{{\ttfamily 1108.4278}}].

\bibitem{Borah:2019ldn}
D.~Borah, B.~Karmakar and D.~Nanda, \emph{{Planck scale origin of nonzero
  $\theta_{13}$ and super-WIMP dark matter}},
  \href{https://doi.org/10.1103/PhysRevD.100.055014}{\emph{Phys. Rev. D}
  {\bfseries 100} (2019) 055014}
  [\href{https://arxiv.org/abs/1906.02756}{{\ttfamily 1906.02756}}].

\bibitem{Ding:2011gt}
G.-J.~Ding and D.~Meloni, \emph{{A Model for Tri-bimaximal Mixing from a
  Completely Broken $A_4$}},
  \href{https://doi.org/10.1016/j.nuclphysb.2011.10.001}{\emph{Nucl. Phys. B}
  {\bfseries 855} (2012) 21} [\href{https://arxiv.org/abs/1108.2733}{{\ttfamily
  1108.2733}}].

\bibitem{King:2011ab}
S.F.~King and C.~Luhn, \emph{{A4 models of tri-bimaximal-reactor mixing}},
  \href{https://doi.org/10.1007/JHEP03(2012)036}{\emph{JHEP} {\bfseries 03}
  (2012) 036} [\href{https://arxiv.org/abs/1112.1959}{{\ttfamily 1112.1959}}].

\bibitem{Mukherjee:2015axj}
A.~Mukherjee and M.K.~Das, \emph{{Neutrino phenomenology and scalar Dark Matter
  with $A_{4}$ flavor symmetry in Inverse and type II seesaw}},
  \href{https://doi.org/10.1016/j.nuclphysb.2016.10.008}{\emph{Nucl. Phys. B}
  {\bfseries 913} (2016) 643}
  [\href{https://arxiv.org/abs/1512.02384}{{\ttfamily 1512.02384}}].

\bibitem{deMedeirosVarzielas:2010ppv}
I.~de~Medeiros~Varzielas and L.~Merlo, \emph{{Ultraviolet Completion of Flavour
  Models}}, \href{https://doi.org/10.1007/JHEP02(2011)062}{\emph{JHEP}
  {\bfseries 02} (2011) 062} [\href{https://arxiv.org/abs/1011.6662}{{\ttfamily
  1011.6662}}].

\bibitem{Ahn:2012cga}
Y.H.~Ahn and H.~Okada, \emph{{Non-zero $\theta_{13}$ linking to Dark Matter
  from Non-Abelian Discrete Flavor Model in Radiative Seesaw}},
  \href{https://doi.org/10.1103/PhysRevD.85.073010}{\emph{Phys. Rev. D}
  {\bfseries 85} (2012) 073010}
  [\href{https://arxiv.org/abs/1201.4436}{{\ttfamily 1201.4436}}].

\bibitem{Ahn:2012tv}
Y.H.~Ahn and S.K.~Kang, \emph{{Non-zero $\theta_{13}$ and CP violation in a
  model with $A_4$ flavor symmetry}},
  \href{https://doi.org/10.1103/PhysRevD.86.093003}{\emph{Phys. Rev. D}
  {\bfseries 86} (2012) 093003}
  [\href{https://arxiv.org/abs/1203.4185}{{\ttfamily 1203.4185}}].

\bibitem{BenTov:2012tg}
Y.~BenTov, X.-G.~He and A.~Zee, \emph{{An $A_{4}$ x $Z_{4}$ model for neutrino
  mixing}}, \href{https://doi.org/10.1007/JHEP12(2012)093}{\emph{JHEP}
  {\bfseries 12} (2012) 093} [\href{https://arxiv.org/abs/1208.1062}{{\ttfamily
  1208.1062}}].

\bibitem{Branco:2012vs}
G.C.~Branco, R.~Gonzalez~Felipe, F.R.~Joaquim and H.~Serodio,
  \emph{{Spontaneous leptonic CP violation and nonzero $\theta_{13}$}},
  \href{https://doi.org/10.1103/PhysRevD.86.076008}{\emph{Phys. Rev. D}
  {\bfseries 86} (2012) 076008}
  [\href{https://arxiv.org/abs/1203.2646}{{\ttfamily 1203.2646}}].

\bibitem{Borah:2017qdu}
D.~Borah, M.K.~Das and A.~Mukherjee, \emph{{Common origin of nonzero
  $\theta_{13}$ and baryon asymmetry of the Universe in a TeV scale seesaw
  model with $A_4$ flavor symmetry}},
  \href{https://doi.org/10.1103/PhysRevD.97.115009}{\emph{Phys. Rev. D}
  {\bfseries 97} (2018) 115009}
  [\href{https://arxiv.org/abs/1711.02445}{{\ttfamily 1711.02445}}].

\bibitem{CarcamoHernandez:2013yiy}
A.E.~Carcamo~Hernandez, I.~de~Medeiros~Varzielas, S.G.~Kovalenko, H.~P\"as and
  I.~Schmidt, \emph{{Lepton masses and mixings in an $A_4$ multi-Higgs model
  with a radiative seesaw mechanism}},
  \href{https://doi.org/10.1103/PhysRevD.88.076014}{\emph{Phys. Rev. D}
  {\bfseries 88} (2013) 076014}
  [\href{https://arxiv.org/abs/1307.6499}{{\ttfamily 1307.6499}}].

\bibitem{Bhattacharya:2016rqj}
S.~Bhattacharya, B.~Karmakar, N.~Sahu and A.~Sil, \emph{{Flavor origin of dark
  matter and its relation with leptonic nonzero $\theta_{13}$ and Dirac CP
  phase $\delta$}}, \href{https://doi.org/10.1007/JHEP05(2017)068}{\emph{JHEP}
  {\bfseries 05} (2017) 068}
  [\href{https://arxiv.org/abs/1611.07419}{{\ttfamily 1611.07419}}].

\bibitem{Barry:2010zk}
J.~Barry and W.~Rodejohann, \emph{{Deviations from tribimaximal mixing due to
  the vacuum expectation value misalignment in $A_4$ models}},
  \href{https://doi.org/10.1103/PhysRevD.81.119901}{\emph{Phys. Rev. D}
  {\bfseries 81} (2010) 093002}
  [\href{https://arxiv.org/abs/1003.2385}{{\ttfamily 1003.2385}}].

\bibitem{Chen:2012st}
M.-C.~Chen, J.~Huang, J.-M.~O'Bryan, A.M.~Wijangco and F.~Yu,
  \emph{{Compatibility of $\theta_{13}$ and the Type I Seesaw Model with $A_4$
  Symmetry}}, \href{https://doi.org/10.1007/JHEP02(2013)021}{\emph{JHEP}
  {\bfseries 02} (2013) 021} [\href{https://arxiv.org/abs/1210.6982}{{\ttfamily
  1210.6982}}].

\bibitem{Karmakar:2016cvb}
B.~Karmakar and A.~Sil, \emph{{An $A_4$ realization of inverse seesaw: neutrino
  masses, $\theta_{13}$ and leptonic non-unitarity}},
  \href{https://doi.org/10.1103/PhysRevD.96.015007}{\emph{Phys. Rev. D}
  {\bfseries 96} (2017) 015007}
  [\href{https://arxiv.org/abs/1610.01909}{{\ttfamily 1610.01909}}].

\bibitem{Zhao:2014yaa}
Z.-h.~Zhao, \emph{{Minimal modifications to the Tri-Bimaximal neutrino
  mixing}}, \href{https://doi.org/10.1007/JHEP11(2014)143}{\emph{JHEP}
  {\bfseries 11} (2014) 143} [\href{https://arxiv.org/abs/1405.3022}{{\ttfamily
  1405.3022}}].

\bibitem{Antusch:2013wn}
S.~Antusch, S.F.~King and M.~Spinrath, \emph{{Spontaneous CP violation in $A_4
  \times SU(5)$ with Constrained Sequential Dominance 2}},
  \href{https://doi.org/10.1103/PhysRevD.87.096018}{\emph{Phys. Rev. D}
  {\bfseries 87} (2013) 096018}
  [\href{https://arxiv.org/abs/1301.6764}{{\ttfamily 1301.6764}}].

\bibitem{Borah:2013upa}
M.~Borah, D.~Borah and M.K.~Das, \emph{{Radiative Generation of Non-zero
  $\theta_{13}$ in MSSM with broken $A_4$ Flavor Symmetry}},
  \href{https://doi.org/10.1016/j.nuclphysb.2014.05.023}{\emph{Nucl. Phys. B}
  {\bfseries 885} (2014) 76} [\href{https://arxiv.org/abs/1304.0164}{{\ttfamily
  1304.0164}}].

\bibitem{Ding:2013bpa}
G.-J.~Ding, S.F.~King and A.J.~Stuart, \emph{{Generalised CP and $A_4$ Family
  Symmetry}}, \href{https://doi.org/10.1007/JHEP12(2013)006}{\emph{JHEP}
  {\bfseries 12} (2013) 006} [\href{https://arxiv.org/abs/1307.4212}{{\ttfamily
  1307.4212}}].

\bibitem{Vien:2021eog}
V.V.~Vien, \emph{{Multiscalar $B-L$ extension with $A_4$ symmetry for fermion
  mass and mixing with co-bimaximal scheme}},
  \href{https://doi.org/10.1016/j.physletb.2021.136296}{\emph{Phys. Lett. B}
  {\bfseries 817} (2021) 136296}.

\bibitem{Ahn:2013mva}
Y.H.~Ahn, S.K.~Kang and C.S.~Kim, \emph{{Spontaneous CP Violation in $A_4$
  Flavor Symmetry and Leptogenesis}},
  \href{https://doi.org/10.1103/PhysRevD.87.113012}{\emph{Phys. Rev. D}
  {\bfseries 87} (2013) 113012}
  [\href{https://arxiv.org/abs/1304.0921}{{\ttfamily 1304.0921}}].

\bibitem{AristizabalSierra:2014zeq}
D.~Aristizabal~Sierra and I.~de~Medeiros~Varzielas, \emph{{Reactor mixing angle
  from hybrid neutrino masses}},
  \href{https://doi.org/10.1007/JHEP07(2014)042}{\emph{JHEP} {\bfseries 07}
  (2014) 042} [\href{https://arxiv.org/abs/1404.2529}{{\ttfamily 1404.2529}}].

\bibitem{Vien:2014pta}
V.V.~Vien and H.N.~Long, \emph{{Neutrino mixing with nonzero $\theta_{13}$ and
  CP violation in the 3-3-1 model based on $A_4$ flavor symmetry}},
  \href{https://doi.org/10.1142/S0217751X15501171}{\emph{Int. J. Mod. Phys. A}
  {\bfseries 30} (2015) 1550117}
  [\href{https://arxiv.org/abs/1405.4665}{{\ttfamily 1405.4665}}].

\bibitem{Vien:2020dlk}
V.V.~Vien, \emph{{Cobimaximal neutrino mixing in the $U(1)_{B-L}$ extension
  with $A_4$ symmetry}},
  \href{https://doi.org/10.1142/S0217732320503113}{\emph{Mod. Phys. Lett. A}
  {\bfseries 35} (2020) 2050311}.

\bibitem{CarcamoHernandez:2015rmj}
A.E.~C\'arcamo~Hern\'andez and R.~Martinez, \emph{{A predictive 3-3-1 model
  with $A_4$ flavor symmetry}},
  \href{https://doi.org/10.1016/j.nuclphysb.2016.02.025}{\emph{Nucl. Phys. B}
  {\bfseries 905} (2016) 337}
  [\href{https://arxiv.org/abs/1501.05937}{{\ttfamily 1501.05937}}].

\bibitem{Holthausen:2012wz}
M.~Holthausen, M.~Lindner and M.A.~Schmidt, \emph{{Lepton flavor at the
  electroweak scale: A complete $A_{4}$ model}},
  \href{https://doi.org/10.1103/PhysRevD.87.033006}{\emph{Phys. Rev. D}
  {\bfseries 87} (2013) 033006}
  [\href{https://arxiv.org/abs/1211.5143}{{\ttfamily 1211.5143}}].

\bibitem{Pramanick:2015qga}
S.~Pramanick and A.~Raychaudhuri, \emph{{A4-based seesaw model for realistic
  neutrino masses and mixing}},
  \href{https://doi.org/10.1103/PhysRevD.93.033007}{\emph{Phys. Rev. D}
  {\bfseries 93} (2016) 033007}
  [\href{https://arxiv.org/abs/1508.02330}{{\ttfamily 1508.02330}}].

\bibitem{Kalita:2015jaa}
R.~Kalita and D.~Borah, \emph{{Constraining a type I seesaw model with $A_4$
  flavor symmetry from neutrino data and leptogenesis}},
  \href{https://doi.org/10.1103/PhysRevD.92.055012}{\emph{Phys. Rev. D}
  {\bfseries 92} (2015) 055012}
  [\href{https://arxiv.org/abs/1508.05466}{{\ttfamily 1508.05466}}].

\bibitem{King:2013hj}
S.F.~King, S.~Morisi, E.~Peinado and J.W.F.~Valle, \emph{{Quark-Lepton Mass
  Relation in a Realistic $A_4$ Extension of the Standard Model}},
  \href{https://doi.org/10.1016/j.physletb.2013.05.067}{\emph{Phys. Lett. B}
  {\bfseries 724} (2013) 68} [\href{https://arxiv.org/abs/1301.7065}{{\ttfamily
  1301.7065}}].

\bibitem{Nomura:2016nfi}
T.~Nomura, Y.~Shimizu and T.~Yamada, \emph{{$A_4 \times U(1)_{PQ}$ model for
  the lepton flavor structure and the strong $CP$ problem}},
  \href{https://doi.org/10.1007/JHEP06(2016)125}{\emph{JHEP} {\bfseries 06}
  (2016) 125} [\href{https://arxiv.org/abs/1604.07650}{{\ttfamily
  1604.07650}}].

\bibitem{Borah:2013jia}
D.~Borah, \emph{{Deviations from Tri-Bimaximal Neutrino Mixing Using Type II
  Seesaw}}, \href{https://doi.org/10.1016/j.nuclphysb.2013.08.024}{\emph{Nucl.
  Phys. B} {\bfseries 876} (2013) 575}
  [\href{https://arxiv.org/abs/1307.2426}{{\ttfamily 1307.2426}}].

\bibitem{Memenga:2013vc}
N.~Memenga, W.~Rodejohann and H.~Zhang, \emph{{$A_4$ flavor symmetry model for
  Dirac neutrinos and sizable $U_{e3}$}},
  \href{https://doi.org/10.1103/PhysRevD.87.053021}{\emph{Phys. Rev. D}
  {\bfseries 87} (2013) 053021}
  [\href{https://arxiv.org/abs/1301.2963}{{\ttfamily 1301.2963}}].

\bibitem{Karmakar:2014dva}
B.~Karmakar and A.~Sil, \emph{{Nonzero $\theta_{13}$ and leptogenesis in a
  type-I seesaw model with $A_4$ symmetry}},
  \href{https://doi.org/10.1103/PhysRevD.91.013004}{\emph{Phys. Rev.}
  {\bfseries D91} (2015) 013004}
  [\href{https://arxiv.org/abs/1407.5826}{{\ttfamily 1407.5826}}].

\bibitem{Puyam:2022mej}
V.~Puyam, S.R.~Singh and N.N.~Singh, \emph{{Deviation from Tribimaximal mixing
  using A4 flavour model with five extra scalars}},
  \href{https://doi.org/10.1016/j.nuclphysb.2022.115932}{\emph{Nucl. Phys. B}
  {\bfseries 983} (2022) 115932}
  [\href{https://arxiv.org/abs/2204.10122}{{\ttfamily 2204.10122}}].

\bibitem{Xing:2006ms}
Z.-z.~Xing and S.~Zhou, \emph{{Tri-bimaximal Neutrino Mixing and
  Flavor-dependent Resonant Leptogenesis}},
  \href{https://doi.org/10.1016/j.physletb.2007.08.009}{\emph{Phys. Lett. B}
  {\bfseries 653} (2007) 278}
  [\href{https://arxiv.org/abs/hep-ph/0607302}{{\ttfamily hep-ph/0607302}}].

\bibitem{Grimus:2008tt}
W.~Grimus and L.~Lavoura, \emph{{A Model for trimaximal lepton mixing}},
  \href{https://doi.org/10.1088/1126-6708/2008/09/106}{\emph{JHEP} {\bfseries
  09} (2008) 106} [\href{https://arxiv.org/abs/0809.0226}{{\ttfamily
  0809.0226}}].

\bibitem{Ma:2006km}
E.~Ma, \emph{{Verifiable radiative seesaw mechanism of neutrino mass and dark
  matter}}, \href{https://doi.org/10.1103/PhysRevD.73.077301}{\emph{Phys. Rev.
  D} {\bfseries 73} (2006) 077301}
  [\href{https://arxiv.org/abs/hep-ph/0601225}{{\ttfamily hep-ph/0601225}}].

\bibitem{Zee:1980ai}
A.~Zee, \emph{{A Theory of Lepton Number Violation, Neutrino Majorana Mass, and
  Oscillation}},
  \href{https://doi.org/10.1016/0370-2693(80)90349-4}{\emph{Phys. Lett. B}
  {\bfseries 93} (1980) 389}.

\bibitem{Cheng:1980qt}
T.P.~Cheng and L.-F.~Li, \emph{{Neutrino Masses, Mixings and Oscillations in
  SU(2) x U(1) Models of Electroweak Interactions}},
  \href{https://doi.org/10.1103/PhysRevD.22.2860}{\emph{Phys. Rev. D}
  {\bfseries 22} (1980) 2860}.

\bibitem{Restrepo:2013aga}
D.~Restrepo, O.~Zapata and C.E.~Yaguna, \emph{{Models with radiative neutrino
  masses and viable dark matter candidates}},
  \href{https://doi.org/10.1007/JHEP11(2013)011}{\emph{JHEP} {\bfseries 11}
  (2013) 011} [\href{https://arxiv.org/abs/1308.3655}{{\ttfamily 1308.3655}}].

\bibitem{Babu:1988ki}
K.S.~Babu, \emph{{Model of 'Calculable' Majorana Neutrino Masses}},
  \href{https://doi.org/10.1016/0370-2693(88)91584-5}{\emph{Phys. Lett. B}
  {\bfseries 203} (1988) 132}.

\bibitem{Cai:2017jrq}
Y.~Cai, J.~Herrero-Garc\'\i{}a, M.A.~Schmidt, A.~Vicente and R.R.~Volkas,
  \emph{{From the trees to the forest: a review of radiative neutrino mass
  models}}, \href{https://doi.org/10.3389/fphy.2017.00063}{\emph{Front. in
  Phys.} {\bfseries 5} (2017) 63}
  [\href{https://arxiv.org/abs/1706.08524}{{\ttfamily 1706.08524}}].

\bibitem{Rojas:2018wym}
N.~Rojas, R.~Srivastava and J.W.F.~Valle, \emph{{Simplest Scoto-Seesaw
  Mechanism}},
  \href{https://doi.org/10.1016/j.physletb.2018.12.014}{\emph{Phys. Lett. B}
  {\bfseries 789} (2019) 132}
  [\href{https://arxiv.org/abs/1807.11447}{{\ttfamily 1807.11447}}].

\bibitem{Schechter:1981bd}
J.~Schechter and J.W.F.~Valle, \emph{{Neutrinoless Double beta Decay in SU(2) x
  U(1) Theories}}, \href{https://doi.org/10.1103/PhysRevD.25.2951}{\emph{Phys.
  Rev. D} {\bfseries 25} (1982) 2951}.

\bibitem{King:1998jw}
S.F.~King, \emph{{Atmospheric and solar neutrinos with a heavy singlet}},
  \href{https://doi.org/10.1016/S0370-2693(98)01055-7}{\emph{Phys. Lett. B}
  {\bfseries 439} (1998) 350}
  [\href{https://arxiv.org/abs/hep-ph/9806440}{{\ttfamily hep-ph/9806440}}].

\bibitem{Antusch:2010tf}
S.~Antusch, S.~Boudjemaa and S.F.~King, \emph{{Neutrino Mixing Angles in
  Sequential Dominance to NLO and NNLO}},
  \href{https://doi.org/10.1007/JHEP09(2010)096}{\emph{JHEP} {\bfseries 09}
  (2010) 096} [\href{https://arxiv.org/abs/1003.5498}{{\ttfamily 1003.5498}}].

\bibitem{Antusch:2004gf}
S.~Antusch and S.F.~King, \emph{{Sequential dominance}},
  \href{https://doi.org/10.1088/1367-2630/6/1/110}{\emph{New J. Phys.}
  {\bfseries 6} (2004) 110}
  [\href{https://arxiv.org/abs/hep-ph/0405272}{{\ttfamily hep-ph/0405272}}].

\bibitem{King:2005bj}
S.F.~King, \emph{{Predicting neutrino parameters from SO(3) family symmetry and
  quark-lepton unification}},
  \href{https://doi.org/10.1088/1126-6708/2005/08/105}{\emph{JHEP} {\bfseries
  08} (2005) 105} [\href{https://arxiv.org/abs/hep-ph/0506297}{{\ttfamily
  hep-ph/0506297}}].

\bibitem{Mandal:2021yph}
S.~Mandal, R.~Srivastava and J.W.F.~Valle, \emph{{The simplest scoto-seesaw
  model: WIMP dark matter phenomenology and Higgs vacuum stability}},
  \href{https://doi.org/10.1016/j.physletb.2021.136458}{\emph{Phys. Lett. B}
  {\bfseries 819} (2021) 136458}
  [\href{https://arxiv.org/abs/2104.13401}{{\ttfamily 2104.13401}}].

\bibitem{Barreiros:2020gxu}
D.M.~Barreiros, F.R.~Joaquim, R.~Srivastava and J.W.F.~Valle, \emph{{Minimal
  scoto-seesaw mechanism with spontaneous CP violation}},
  \href{https://doi.org/10.1007/JHEP04(2021)249}{\emph{JHEP} {\bfseries 04}
  (2021) 249} [\href{https://arxiv.org/abs/2012.05189}{{\ttfamily
  2012.05189}}].

\bibitem{Koide:2007sr}
Y.~Koide, \emph{{S(4) flavor symmetry embedded into SU(3) and lepton masses and
  mixing}}, \href{https://doi.org/10.1088/1126-6708/2007/08/086}{\emph{JHEP}
  {\bfseries 08} (2007) 086} [\href{https://arxiv.org/abs/0705.2275}{{\ttfamily
  0705.2275}}].

\bibitem{Adulpravitchai:2009kd}
A.~Adulpravitchai, A.~Blum and M.~Lindner, \emph{{Non-Abelian Discrete Groups
  from the Breaking of Continuous Flavor Symmetries}},
  \href{https://doi.org/10.1088/1126-6708/2009/09/018}{\emph{JHEP} {\bfseries
  09} (2009) 018} [\href{https://arxiv.org/abs/0907.2332}{{\ttfamily
  0907.2332}}].

\bibitem{Luhn:2011ip}
C.~Luhn, \emph{{Spontaneous breaking of SU(3) to finite family symmetries: a
  pedestrian's approach}},
  \href{https://doi.org/10.1007/JHEP03(2011)108}{\emph{JHEP} {\bfseries 03}
  (2011) 108} [\href{https://arxiv.org/abs/1101.2417}{{\ttfamily 1101.2417}}].

\bibitem{Merle:2011vy}
A.~Merle and R.~Zwicky, \emph{{Explicit and spontaneous breaking of SU(3) into
  its finite subgroups}},
  \href{https://doi.org/10.1007/JHEP02(2012)128}{\emph{JHEP} {\bfseries 02}
  (2012) 128} [\href{https://arxiv.org/abs/1110.4891}{{\ttfamily 1110.4891}}].

\bibitem{Rachlin:2017rvm}
B.L.~Rachlin and T.W.~Kephart, \emph{{Spontaneous Breaking of Gauge Groups to
  Discrete Symmetries}},
  \href{https://doi.org/10.1007/JHEP08(2017)110}{\emph{JHEP} {\bfseries 08}
  (2017) 110} [\href{https://arxiv.org/abs/1702.08073}{{\ttfamily
  1702.08073}}].

\bibitem{King:2018fke}
S.F.~King and Y.-L.~Zhou, \emph{{Spontaneous breaking of $SO(3)$ to finite
  family symmetries with supersymmetry - an $A_4$ model}},
  \href{https://doi.org/10.1007/JHEP11(2018)173}{\emph{JHEP} {\bfseries 11}
  (2018) 173} [\href{https://arxiv.org/abs/1809.10292}{{\ttfamily
  1809.10292}}].

\bibitem{Burrows:2009pi}
T.J.~Burrows and S.F.~King, \emph{{A(4) Family Symmetry from SU(5) SUSY GUTs in
  6d}}, \href{https://doi.org/10.1016/j.nuclphysb.2010.04.002}{\emph{Nucl.
  Phys. B} {\bfseries 835} (2010) 174}
  [\href{https://arxiv.org/abs/0909.1433}{{\ttfamily 0909.1433}}].

\bibitem{King:2017guk}
S.F.~King, \emph{{Unified Models of Neutrinos, Flavour and CP Violation}},
  \href{https://doi.org/10.1016/j.ppnp.2017.01.003}{\emph{Prog. Part. Nucl.
  Phys.} {\bfseries 94} (2017) 217}
  [\href{https://arxiv.org/abs/1701.04413}{{\ttfamily 1701.04413}}].

\bibitem{deAdelhartToorop:2011re}
R.~de~Adelhart~Toorop, F.~Feruglio and C.~Hagedorn, \emph{{Finite Modular
  Groups and Lepton Mixing}},
  \href{https://doi.org/10.1016/j.nuclphysb.2012.01.017}{\emph{Nucl. Phys. B}
  {\bfseries 858} (2012) 437}
  [\href{https://arxiv.org/abs/1112.1340}{{\ttfamily 1112.1340}}].

\bibitem{Feruglio:2017spp}
F.~Feruglio, \emph{{Are neutrino masses modular forms?}},  in \emph{{From My
  Vast Repertoire ...}: {Guido Altarelli's Legacy}}, A.~Levy, S.~Forte and
  G.~Ridolfi, eds., pp.~227--266 (2019),
  \href{https://doi.org/10.1142/9789813238053_0012}{DOI}
  [\href{https://arxiv.org/abs/1706.08749}{{\ttfamily 1706.08749}}].

\bibitem{deAnda:2018ecu}
F.J.~de~Anda, S.F.~King and E.~Perdomo, \emph{{$SU(5)$ grand unified theory
  with $A_4$ modular symmetry}},
  \href{https://doi.org/10.1103/PhysRevD.101.015028}{\emph{Phys. Rev. D}
  {\bfseries 101} (2020) 015028}
  [\href{https://arxiv.org/abs/1812.05620}{{\ttfamily 1812.05620}}].

\bibitem{Novichkov:2018yse}
P.P.~Novichkov, S.T.~Petcov and M.~Tanimoto, \emph{{Trimaximal Neutrino Mixing
  from Modular A4 Invariance with Residual Symmetries}},
  \href{https://doi.org/10.1016/j.physletb.2019.04.043}{\emph{Phys. Lett. B}
  {\bfseries 793} (2019) 247}
  [\href{https://arxiv.org/abs/1812.11289}{{\ttfamily 1812.11289}}].

\bibitem{Criado:2018thu}
J.C.~Criado and F.~Feruglio, \emph{{Modular Invariance Faces Precision Neutrino
  Data}}, \href{https://doi.org/10.21468/SciPostPhys.5.5.042}{\emph{SciPost
  Phys.} {\bfseries 5} (2018) 042}
  [\href{https://arxiv.org/abs/1807.01125}{{\ttfamily 1807.01125}}].

\bibitem{Kobayashi:2018wkl}
T.~Kobayashi, Y.~Shimizu, K.~Takagi, M.~Tanimoto, T.H.~Tatsuishi and H.~Uchida,
  \emph{{Finite modular subgroups for fermion mass matrices and baryon/lepton
  number violation}},
  \href{https://doi.org/10.1016/j.physletb.2019.05.034}{\emph{Phys. Lett. B}
  {\bfseries 794} (2019) 114}
  [\href{https://arxiv.org/abs/1812.11072}{{\ttfamily 1812.11072}}].

\bibitem{Penedo:2018nmg}
J.T.~Penedo and S.T.~Petcov, \emph{{Lepton Masses and Mixing from Modular $S_4$
  Symmetry}},
  \href{https://doi.org/10.1016/j.nuclphysb.2018.12.016}{\emph{Nucl. Phys. B}
  {\bfseries 939} (2019) 292}
  [\href{https://arxiv.org/abs/1806.11040}{{\ttfamily 1806.11040}}].

\bibitem{Ding:2019xna}
G.-J.~Ding, S.F.~King and X.-G.~Liu, \emph{{Neutrino mass and mixing with $A_5$
  modular symmetry}},
  \href{https://doi.org/10.1103/PhysRevD.100.115005}{\emph{Phys. Rev. D}
  {\bfseries 100} (2019) 115005}
  [\href{https://arxiv.org/abs/1903.12588}{{\ttfamily 1903.12588}}].

\bibitem{Borah:2017dmk}
D.~Borah and B.~Karmakar, \emph{{$A_4$ flavour model for Dirac neutrinos: Type
  I and inverse seesaw}},
  \href{https://doi.org/10.1016/j.physletb.2018.03.047}{\emph{Phys. Lett.}
  {\bfseries B780} (2018) 461}
  [\href{https://arxiv.org/abs/1712.06407}{{\ttfamily 1712.06407}}].

\bibitem{Borah:2018nvu}
D.~Borah and B.~Karmakar, \emph{{Linear seesaw for Dirac neutrinos with $A_4$
  flavour symmetry}},
  \href{https://doi.org/10.1016/j.physletb.2018.12.006}{\emph{Phys. Lett. B}
  {\bfseries 789} (2019) 59}
  [\href{https://arxiv.org/abs/1806.10685}{{\ttfamily 1806.10685}}].

\bibitem{Borah:2018gjk}
D.~Borah, B.~Karmakar and D.~Nanda, \emph{{Common Origin of Dirac Neutrino Mass
  and Freeze-in Massive Particle Dark Matter}},
  \href{https://doi.org/10.1088/1475-7516/2018/07/039}{\emph{JCAP} {\bfseries
  07} (2018) 039} [\href{https://arxiv.org/abs/1805.11115}{{\ttfamily
  1805.11115}}].

\bibitem{He:2006dk}
X.-G.~He, Y.-Y.~Keum and R.R.~Volkas, \emph{{A(4) flavor symmetry breaking
  scheme for understanding quark and neutrino mixing angles}},
  \href{https://doi.org/10.1088/1126-6708/2006/04/039}{\emph{JHEP} {\bfseries
  04} (2006) 039} [\href{https://arxiv.org/abs/hep-ph/0601001}{{\ttfamily
  hep-ph/0601001}}].

\bibitem{Lin:2008aj}
Y.~Lin, \emph{{A Predictive A(4) model, Charged Lepton Hierarchy and
  Tri-bimaximal Sum Rule}},
  \href{https://doi.org/10.1016/j.nuclphysb.2008.12.025}{\emph{Nucl. Phys. B}
  {\bfseries 813} (2009) 91} [\href{https://arxiv.org/abs/0804.2867}{{\ttfamily
  0804.2867}}].

\bibitem{Chen:2009um}
M.-C.~Chen and S.F.~King, \emph{{A4 See-Saw Models and Form Dominance}},
  \href{https://doi.org/10.1088/1126-6708/2009/06/072}{\emph{JHEP} {\bfseries
  06} (2009) 072} [\href{https://arxiv.org/abs/0903.0125}{{\ttfamily
  0903.0125}}].

\bibitem{Karmakar:2015jza}
B.~Karmakar and A.~Sil, \emph{{Spontaneous CP violation in lepton-sector: A
  common origin for $\theta_{13}$, the Dirac CP phase, and leptogenesis}},
  \href{https://doi.org/10.1103/PhysRevD.93.013006}{\emph{Phys. Rev. D}
  {\bfseries 93} (2016) 013006}
  [\href{https://arxiv.org/abs/1509.07090}{{\ttfamily 1509.07090}}].

\bibitem{Bhattacharya:2016lts}
S.~Bhattacharya, B.~Karmakar, N.~Sahu and A.~Sil, \emph{{Unifying the flavor
  origin of dark matter with leptonic nonzero $\theta_{13}$}},
  \href{https://doi.org/10.1103/PhysRevD.93.115041}{\emph{Phys. Rev. D}
  {\bfseries 93} (2016) 115041}
  [\href{https://arxiv.org/abs/1603.04776}{{\ttfamily 1603.04776}}].

\bibitem{Molinaro:2009lud}
E.~Molinaro and S.T.~Petcov, \emph{{The Interplay Between the 'Low' and 'High'
  Energy CP-Violation in Leptogenesis}},
  \href{https://doi.org/10.1140/epjc/s10052-009-0985-3}{\emph{Eur. Phys. J. C}
  {\bfseries 61} (2009) 93} [\href{https://arxiv.org/abs/0803.4120}{{\ttfamily
  0803.4120}}].

\bibitem{Branco:2011iw}
G.C.~Branco, P.M.~Ferreira, L.~Lavoura, M.N.~Rebelo, M.~Sher and J.P.~Silva,
  \emph{{Theory and phenomenology of two-Higgs-doublet models}},
  \href{https://doi.org/10.1016/j.physrep.2012.02.002}{\emph{Phys. Rept.}
  {\bfseries 516} (2012) 1} [\href{https://arxiv.org/abs/1106.0034}{{\ttfamily
  1106.0034}}].

\bibitem{King:2011zj}
S.F.~King and C.~Luhn, \emph{{Trimaximal neutrino mixing from vacuum alignment
  in A4 and S4 models}},
  \href{https://doi.org/10.1007/JHEP09(2011)042}{\emph{JHEP} {\bfseries 09}
  (2011) 042} [\href{https://arxiv.org/abs/1107.5332}{{\ttfamily 1107.5332}}].

\bibitem{ParticleDataGroup:2020ssz}
{\scshape Particle Data Group} collaboration, \emph{{Review of Particle
  Physics}}, \href{https://doi.org/10.1093/ptep/ptaa104}{\emph{PTEP} {\bfseries
  2020} (2020) 083C01}.

\bibitem{Ma:2012ez}
E.~Ma, A.~Natale and A.~Rashed, \emph{{Scotogenic $A_4$ Neutrino Model for
  Nonzero $\theta_{13}$ and Large $\delta_{CP}$}},
  \href{https://doi.org/10.1142/S0217751X12501345}{\emph{Int. J. Mod. Phys. A}
  {\bfseries 27} (2012) 1250134}
  [\href{https://arxiv.org/abs/1206.1570}{{\ttfamily 1206.1570}}].

\bibitem{Hernandez:2012ra}
D.~Hernandez and A.Y.~Smirnov, \emph{{Lepton mixing and discrete symmetries}},
  \href{https://doi.org/10.1103/PhysRevD.86.053014}{\emph{Phys. Rev. D}
  {\bfseries 86} (2012) 053014}
  [\href{https://arxiv.org/abs/1204.0445}{{\ttfamily 1204.0445}}].

\bibitem{Tanimoto:2015nfa}
M.~Tanimoto, \emph{{Neutrinos and flavor symmetries}},
  \href{https://doi.org/10.1063/1.4915578}{\emph{AIP Conf. Proc.} {\bfseries
  1666} (2015) 120002}.

\bibitem{Shimizu:2014ria}
Y.~Shimizu, M.~Tanimoto and K.~Yamamoto, \emph{{Predicting CP violation in
  Deviation from Tri-bimaximal mixing of Neutrinos}},
  \href{https://doi.org/10.1142/S0217732315500029}{\emph{Mod. Phys. Lett. A}
  {\bfseries 30} (2015) 1550002}
  [\href{https://arxiv.org/abs/1405.1521}{{\ttfamily 1405.1521}}].

\bibitem{Esteban:2020cvm}
I.~Esteban, M.~Gonzalez-Garcia, M.~Maltoni, T.~Schwetz and A.~Zhou, \emph{{The
  fate of hints: updated global analysis of three-flavor neutrino
  oscillations}}, \href{https://doi.org/10.1007/JHEP09(2020)178}{\emph{JHEP}
  {\bfseries 09} (2020) 178}
  [\href{https://arxiv.org/abs/2007.14792}{{\ttfamily 2007.14792}}].

\bibitem{Capozzi:2021fjo}
F.~Capozzi, E.~Di~Valentino, E.~Lisi, A.~Marrone, A.~Melchiorri and A.~Palazzo,
  \emph{{Unfinished fabric of the three neutrino paradigm}},
  \href{https://doi.org/10.1103/PhysRevD.104.083031}{\emph{Phys. Rev. D}
  {\bfseries 104} (2021) 083031}
  [\href{https://arxiv.org/abs/2107.00532}{{\ttfamily 2107.00532}}].

\bibitem{Rodejohann:2011vc}
W.~Rodejohann and J.W.F.~Valle, \emph{{Symmetrical Parametrizations of the
  Lepton Mixing Matrix}},
  \href{https://doi.org/10.1103/PhysRevD.84.073011}{\emph{Phys. Rev. D}
  {\bfseries 84} (2011) 073011}
  [\href{https://arxiv.org/abs/1108.3484}{{\ttfamily 1108.3484}}].

\bibitem{Zyla:2020zbs}
{\scshape Particle Data Group} collaboration, \emph{{Review of Particle
  Physics}}, \href{https://doi.org/10.1093/ptep/ptaa104}{\emph{PTEP} {\bfseries
  2020} (2020) 083C01}.

\bibitem{KamLAND-Zen:2016pfg}
{\scshape KamLAND-Zen} collaboration, \emph{{Search for Majorana Neutrinos near
  the Inverted Mass Hierarchy Region with KamLAND-Zen}},
  \href{https://doi.org/10.1103/PhysRevLett.117.082503}{\emph{Phys. Rev. Lett.}
  {\bfseries 117} (2016) 082503}
  [\href{https://arxiv.org/abs/1605.02889}{{\ttfamily 1605.02889}}].

\bibitem{GERDA:2018pmc}
{\scshape GERDA} collaboration, \emph{{Improved Limit on Neutrinoless
  Double-$\beta$ Decay of $^{76}$Ge from GERDA Phase II}},
  \href{https://doi.org/10.1103/PhysRevLett.120.132503}{\emph{Phys. Rev. Lett.}
  {\bfseries 120} (2018) 132503}
  [\href{https://arxiv.org/abs/1803.11100}{{\ttfamily 1803.11100}}].

\bibitem{LEGEND:2021bnm}
{\scshape LEGEND} collaboration, \emph{{The Large Enriched Germanium Experiment
  for Neutrinoless $\beta\beta$ Decay}: {LEGEND-1000 Preconceptual Design
  Report}},  \href{https://arxiv.org/abs/2107.11462}{{\ttfamily 2107.11462}}.

\bibitem{nEXO:2021ujk}
{\scshape nEXO} collaboration, \emph{{nEXO: neutrinoless double beta decay
  search beyond 10$^{28}$ year half-life sensitivity}},
  \href{https://doi.org/10.1088/1361-6471/ac3631}{\emph{J. Phys. G} {\bfseries
  49} (2022) 015104} [\href{https://arxiv.org/abs/2106.16243}{{\ttfamily
  2106.16243}}].

\bibitem{Toma:2013zsa}
T.~Toma and A.~Vicente, \emph{{Lepton Flavor Violation in the Scotogenic
  Model}}, \href{https://doi.org/10.1007/JHEP01(2014)160}{\emph{JHEP}
  {\bfseries 01} (2014) 160} [\href{https://arxiv.org/abs/1312.2840}{{\ttfamily
  1312.2840}}].

\bibitem{Vicente:2014wga}
A.~Vicente and C.E.~Yaguna, \emph{{Probing the scotogenic model with lepton
  flavor violating processes}},
  \href{https://doi.org/10.1007/JHEP02(2015)144}{\emph{JHEP} {\bfseries 02}
  (2015) 144} [\href{https://arxiv.org/abs/1412.2545}{{\ttfamily 1412.2545}}].

\bibitem{Hagedorn:2018spx}
C.~Hagedorn, J.~Herrero-Garc\'\i{}a, E.~Molinaro and M.A.~Schmidt,
  \emph{{Phenomenology of the Generalised Scotogenic Model with Fermionic Dark
  Matter}}, \href{https://doi.org/10.1007/JHEP11(2018)103}{\emph{JHEP}
  {\bfseries 11} (2018) 103}
  [\href{https://arxiv.org/abs/1804.04117}{{\ttfamily 1804.04117}}].

\bibitem{Aushev:2010bq}
T.~Aushev et~al., \emph{{Physics at Super B Factory}},
  \href{https://arxiv.org/abs/1002.5012}{{\ttfamily 1002.5012}}.

\bibitem{Ilakovac:1994kj}
A.~Ilakovac and A.~Pilaftsis, \emph{{Flavor violating charged lepton decays in
  seesaw-type models}},
  \href{https://doi.org/10.1016/0550-3213(94)00567-X}{\emph{Nucl. Phys. B}
  {\bfseries 437} (1995) 491}
  [\href{https://arxiv.org/abs/hep-ph/9403398}{{\ttfamily hep-ph/9403398}}].

\bibitem{Tommasini:1995ii}
D.~Tommasini, G.~Barenboim, J.~Bernabeu and C.~Jarlskog, \emph{{Nondecoupling
  of heavy neutrinos and lepton flavor violation}},
  \href{https://doi.org/10.1016/0550-3213(95)00201-3}{\emph{Nucl. Phys. B}
  {\bfseries 444} (1995) 451}
  [\href{https://arxiv.org/abs/hep-ph/9503228}{{\ttfamily hep-ph/9503228}}].

\bibitem{Dinh:2012bp}
D.N.~Dinh, A.~Ibarra, E.~Molinaro and S.T.~Petcov, \emph{{The $\mu - e$
  Conversion in Nuclei, $\mu \to e \gamma, \mu \to 3e$ Decays and TeV Scale
  See-Saw Scenarios of Neutrino Mass Generation}},
  \href{https://doi.org/10.1007/JHEP08(2012)125}{\emph{JHEP} {\bfseries 08}
  (2012) 125} [\href{https://arxiv.org/abs/1205.4671}{{\ttfamily 1205.4671}}].

\bibitem{Bambhaniya:2016rbb}
G.~Bambhaniya, P.S.~Bhupal~Dev, S.~Goswami, S.~Khan and W.~Rodejohann,
  \emph{{Naturalness, Vacuum Stability and Leptogenesis in the Minimal Seesaw
  Model}}, \href{https://doi.org/10.1103/PhysRevD.95.095016}{\emph{Phys. Rev.
  D} {\bfseries 95} (2017) 095016}
  [\href{https://arxiv.org/abs/1611.03827}{{\ttfamily 1611.03827}}].

\bibitem{Ghosh:2017fmr}
P.~Ghosh, A.K.~Saha and A.~Sil, \emph{{Study of Electroweak Vacuum Stability
  from Extended Higgs Portal of Dark Matter and Neutrinos}},
  \href{https://doi.org/10.1103/PhysRevD.97.075034}{\emph{Phys. Rev. D}
  {\bfseries 97} (2018) 075034}
  [\href{https://arxiv.org/abs/1706.04931}{{\ttfamily 1706.04931}}].

\bibitem{MEG:2016leq}
{\scshape MEG} collaboration, \emph{{Search for the lepton flavour violating
  decay $\mu ^+ \rightarrow \mathrm {e}^+ \gamma $ with the full dataset of the
  MEG experiment}},
  \href{https://doi.org/10.1140/epjc/s10052-016-4271-x}{\emph{Eur. Phys. J. C}
  {\bfseries 76} (2016) 434}
  [\href{https://arxiv.org/abs/1605.05081}{{\ttfamily 1605.05081}}].

\bibitem{MEGII:2018kmf}
{\scshape MEG II} collaboration, \emph{{The design of the MEG II experiment}},
  \href{https://doi.org/10.1140/epjc/s10052-018-5845-6}{\emph{Eur. Phys. J. C}
  {\bfseries 78} (2018) 380}
  [\href{https://arxiv.org/abs/1801.04688}{{\ttfamily 1801.04688}}].

\bibitem{Dolle:2009fn}
E.M.~Dolle and S.~Su, \emph{{The Inert Dark Matter}},
  \href{https://doi.org/10.1103/PhysRevD.80.055012}{\emph{Phys. Rev. D}
  {\bfseries 80} (2009) 055012}
  [\href{https://arxiv.org/abs/0906.1609}{{\ttfamily 0906.1609}}].

\bibitem{Planck:2018vyg}
{\scshape Planck} collaboration, \emph{{Planck 2018 results. VI. Cosmological
  parameters}},
  \href{https://doi.org/10.1051/0004-6361/201833910}{\emph{Astron. Astrophys.}
  {\bfseries 641} (2020) A6}
  [\href{https://arxiv.org/abs/1807.06209}{{\ttfamily 1807.06209}}].

\bibitem{CMS:2022qva}
{\scshape CMS} collaboration, \emph{{Search for invisible decays of the Higgs
  boson produced via vector boson fusion in proton-proton collisions at
  $\sqrt{s} =$ 13 TeV}},
  \href{https://doi.org/10.1103/PhysRevD.105.092007}{\emph{Phys. Rev. D}
  {\bfseries 105} (2022) 092007}
  [\href{https://arxiv.org/abs/2201.11585}{{\ttfamily 2201.11585}}].

\end{thebibliography}\endgroup
\bibliographystyle{JHEP}
\end{document}